\documentclass[final,authoryear,5p,times,twocolumn]{elsarticle}
\usepackage{graphicx}
\usepackage{amssymb}
\usepackage{amsmath}
\usepackage{natbib}
\usepackage{threeparttable}
\journal{New Astronomy Reviews}

\newcommand\araa{ARA\&A}
\newcommand\apj{ApJ}
\newcommand\apjl{ApJ}
\newcommand\apjs{ApJS}
\newcommand\aap{A\&A}
\newcommand\pasj{PASJ}

\newcommand\mnras{MNRAS}
\newcommand\nat{Nature}
\newcommand\apss{Ap\&SS}

\def\beq{\begin{eqnarray}}
\def\eeq{\end{eqnarray}}

\begin{document}

\begin{frontmatter}

\title{Neutrino-dominated accretion flows as the central engine of gamma-ray bursts}

\author[ad1,ad2]{Tong Liu}
\address[ad1]{Department of Astronomy, Xiamen University, Xiamen, Fujian 361005, China}
\address[ad2]{Department of Physics and Astronomy, University of Nevada, Las Vegas, NV 89154, USA}
\ead{tongliu@xmu.edu.cn}
\author[ad1]{Wei-Min Gu}
\author[ad2,ad3,ad4]{Bing Zhang}
\address[ad3]{Department of Astronomy, School of Physics, Peking University, Beijing 100871, China}
\address[ad4]{Kavli Institute of Astronomy and Astrophysics, Peking University, Beijing 100871, China}

\begin{abstract}
Neutrino-dominated accretion flows (NDAFs) around rotating stellar-mass black holes (BHs) are plausible candidates for the central engines of gamma-ray bursts (GRBs). NDAFs are hyperaccretion disks with accretion rates in the range of around 0.001-10 $M_\odot ~\rm s^{-1}$, which have high density and temperature and therefore are extremely optically thick and geometrically slim or even thick. We review the theoretical progresses in studying the properties of NDAFs as well as their applications to the GRB phenomenology. The topics include: the steady radial and vertical structure of NDAFs and the implications for calculating neutrino luminosity and annihilation luminosity, jet power due to neutrino-antineutrino annihilation and Blandford-Znajek mechanism and their dependences on parameters such as BH mass, spin, and accretion rate, time evolution of NDAFs, effect of magnetic fields, applications of NDAF theories to the GRB phenomenology such as lightcurve variability, extended emission, X-ray flares, kilonovae, etc., as well as probing NDAFs using multi-messenger signals such as MeV neutrinos and gravitational waves.
\end{abstract}

\begin{keyword}

accretion, accretion disks; black hole physics; gamma-ray burst: general; gravitational waves; neutrinos.

\end{keyword}

\end{frontmatter}

\section{Introduction}

In astrophysics, accretion is a process that matter falls to a central object, which prompts part of gravitational binding energy of the infalling matter converted into heat and radiation due to viscous dissipation. The conservation of angular momentum, however, forces the accreted matter to form an accretion disk around the central object. Accretion disks widely exist in astrophysical systems, such as cataclysmic variable stars, X-ray binaries, protoplanetary disks, active galactic nuclei (AGNs), and gamma-ray bursts (GRBs).

As a fundamental physical model, black hole (BH) accretion disks have been widely studied \citep[see reviews by][]{Frank2002,Kato2008,Abramowicz2013,Blaes2014,Yuan2014}. Three classic accretion disk models, namely the Shakura-Sunyaev disk \citep[SSD,][]{Shakura1973}, the slim disk \citep{Abramowicz1988}, and the advection-dominated accretion disk \citep[ADAF,][]{Narayan1994,Abramowicz1995}, have been successfully applied to different systems. The SSD model is geometrically thin, optically thick, and Keplerianly rotating, where the viscous heating is balanced by the radiative cooling. Such a model is very successful in interpreting the high/soft state of X-ray binaries and can be even used to measure the spin of the BH \citep[e.g.,][]{Zhang1997}. It is also widely applied to high luminosity AGNs such as quasars. The slim disk model was introduced mainly for systems with super-Eddington accretion, where the disk is geometrically slim and optically thick. The fundamental difference from the SSD model is that the large amount of photons generated by the viscosity cannot escape from the disk. Most of the photons are carried by the accretion flow and finally fall into the BH. In other words, the main cooling mechanism is advection rather than radiation. The slim disk model is often applied to super-Eddington systems such as ultraluminous X-ray sources and narrow line Seyfert 1 galaxies, which is also applied to study cosmology \citep{Wang2013}. Different from the above two models, an ADAF is optically thin and has extremely high temperature. The main cooling mechanism is also advection rather than radiation. The difference from the slim disk is that, the energy advection in an ADAF is related to the internal energy of the flow instead of the photons. The ADAF model has been successfully applied to the low/hard and quiescent states of X-ray binaries and low luminosity AGNs.

Apart from the above three classic accretion models, there are some significant progresses in this field. The advection-dominated inflow outflow solution \citep[ADIOS,][]{Blandford1999} shows that the outflows may have an essential effect on the structure and radiation of the flow. The convection-dominated accretion flow \citep[CDAF,][]{Narayan2000} includes energy and angular momentum transfers by radially convective process. A luminous hot accretion flow \citep[LHAF,][]{Yuan2001} can provide both high luminosity and hard photons.

A well-known unified description of different accretion models is in the $\dot M$-$\Sigma$ parameter space (Figure 1), where $\dot M$ is the mass accretion rate and $\Sigma$ is the mass surface density. By including the thermally unstable Shapiro-Lightman-Eardley disk \citep[SLE,][]{Shapiro1976}, \citet{Abramowicz1995} first presented a unified picture of the thermal equilibrium solutions containing the above three classic models. Such a picture was improved by \citet{Yuan2003}, where the LHAF solution is included.

All the above mentioned accretion flows are photon radiation dominated and neutrino radiation is negligible. A main ingredient of this review, on the other hand, we will focus on the neutrino radiation-dominated disk which may be related to GRBs.

GRBs are extremely energetic transient events in the Universe and isotropically distributed over the sky (see e.g., \citealt{Meszaros2002,Zhang2004,Zhang2007a,Gehrels2009,Kumar2015,Wang2015} for reviews), which are sorted into two categories, i.e., short- and long-duration GRBs \citep[SGRBs and LGRBs, see][]{Kouveliotou1993} or Type I and II GRBs \citep{Zhang2006,Zhang2007b}. Milliseconds variability requires that GRBs are the stellar scale events, and they may be related to the stellar evolution in the external galaxies as well as cosmology. Moreover, their typically isotropic energy is about $10^{50}-10^{52}$ $\rm erg$, which asks for efficiently released energy approach. That is, a compact object should exist in the center of GRBs. SGRBs and LGRBs are generally considered to be mergers of two compact objects, i.e., two neutron stars (NSs) or a BH and an NS \citep[for reviews, see, e.g.,][]{Nakar2007,Berger2014}, and collapses of massive stars (e.g., \citealt{Woosley2006} for reviews), respectively. The popular models on the central engines of GRBs are either a rotating stellar BH surrounded by a hyperaccretion disk \citep[e.g.,][]{Paczynski1991,Narayan1992,MacFadyen1999} or a quickly rotating magnetar \citep[or protomagnetar, e.g.,][]{Usov1992,Dai1998,Zhang2001,Dai2006,Metzger2011,Lv2014,Lv2015,Lia2016}.

\begin{figure}
\centering
\includegraphics[angle=0,scale=0.5]{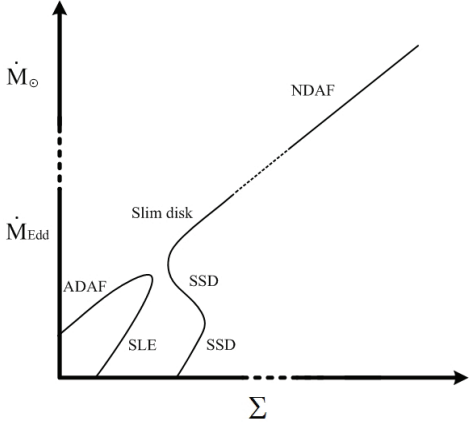}
\caption{Unified description of the thermal equilibrium solutions of different accretion models, where the horizontal axis is the mass surface density $\Sigma$ and the vertical axis is the mass accretion rate $\dot M$.}
\label{11}
\end{figure}

In the inner region of a hyperaccretion disk the density and temperature are so high ($\rho \sim$ $10^{10}-10^{13}$ g~cm$^{-3}$, $T \sim 10^{10}-10^{11}$ K) that photons are completely trapped. A large amount of energetic neutrinos are emitted from the surface of the disk, carrying away the viscous dissipation energy of accreted gas via the reactions that neutrinos participate. Annihilation of a fraction of neutrinos and antineutrinos produces a relativistic electron-positron pair dominated outflow, which can power a GRB. The hyperaccretion disk is referred as the neutrino-dominated accretion disk (NDAF). The properties of NDAFs were first investigated in details by \citet{Popham1999}. They proposed that NDAFs around stellar-mass rotating BHs are a plausible candidate for the central engine of GRBs.

\begin{figure}
\centering
\includegraphics[angle=0,scale=0.47]{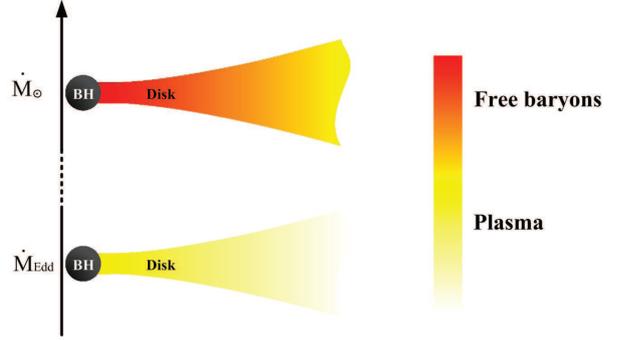}
\caption{Components of accretion disks.}
\label{12}
\end{figure}

Figure \ref{11} shows the unified description of the thermal equilibrium solutions of different accretion models, where the NDAF model is also included at extremely high accretion rates. This figure can be regarded as an extension of the unified picture of \citet{Abramowicz1995}, where the SLE disk is included whereas the LHAF is not considered. As the figure shows, slim disks and NDAFs have no separation boundary, and NDAFs can be considered as the naturally extended branch of slim disks for very high accretion rates. Certainly, NDAFs are extremely optically thick due to the high accretion rates and the corresponding high surface density. Figure \ref{12} shows the very different components between the classic accretion disks and NDAFs, which result from the significant difference in the mass accretion rates. The inner region of NDAFs is dominated by free baryons, which will be discussed in detail when in the results of the radial and vertical directions of NDAFs in Subsections 2.1 and 4.5.

According to the observations on GRBs and their afterglows, NDAF as a plausible and central engine model of GRBs must satisfy the following requirements: First, the released energy from NDAFs should be adequate for GRBs, and NDAFs should be able to launch an ultra-relativistic outflow with an opening angle of $\sim$ 0.1 and very low baryon contamination. NDAFs should be able to produce GRB emission with diverse temporal behaviors from single pulse to multi-pulses. These have been reviewed in Sections 2-5. Second, NDAFs should be able to reactivate to produce X-ray flares and extended X-ray emission after the prompt emission phase. These are reviewed in Section 6. Third, NDAFs should be produce diverse GRBs with different jet compositions, including both matter dominated fireball (neutrino-anti-neutrino annihilations) and Poynting flux dominated outflow [the Blandford-Znajek (BZ) mechanism \citep{Blandford1977}]. These are discussed in Sections 3 and 6.

In this review, we summarize the theoretical progresses and applications of NDAFs, and their possible detectable effects. In Sections 2 and 4, we describe the radial and vertical dynamics and radiation of NDAFs. In Section 3, magnetized NDAF models are discussed. Some simulations of NDAFs are briefly introduced in Section 5. In Section 6, numerical applications to GRB observations are presented. We review the validation of the existence of NDAFs in Section 7. Summary is given in Section 8.

\section{Radial dynamics and radiation of NDAFs}

\subsection{Radial dynamics of NDAFs}

The steady radial structure and neutrino luminosity of BH NDAFs are most widely studied \citep[see e.g,][]{Popham1999,Narayan2001,Di Matteo2002,Kohri2002,Kohri2005,Lee2005,Gu2006,Chen2007,Liu2007,Janiuk2007,Rossi2008,Janiuk2010,Kawanaka2007,Kawanaka2013b,Li2013,Luo2013b,Xue2013}. After the first work of \citet{Popham1999}, \citet{Di Matteo2002} recalculated the NDAF radial structure by taking into account neutrino opacity with Newtonian potential. They found that neutrinos are sufficiently trapped in the inner region of the disk, which may suggest that NDAFs may be inadequate to power bright GRBs. \citet{Gu2006} argued that when the general relativistic effects are considered, NDAFs can be powerful enough to be the central engine for GRBs. In another discussion, \citet{Kohri2002} pointed out that electron degeneracy can suppress neutrino emission. Kohri et al. (2005) started to introduce detailed neutrino physics into NDAF model calculations, which brings significant improvement to the NDAF model.

As most detailed study of NDAFs, \citet{Xue2013} investigated the relativistic one-dimensional (1D) steady global solutions of NDAFs by taking into account detailed neutrino physics, balance of chemical potentials, photodisintegration, and nuclear statistical equilibrium. This is mainly introduced below.

\subsubsection{Relativistic hydrodynamics}

The hydrodynamical model of disks is based on the ADAF model of \citet{Abramowicz1996}, the slim disk model of \citet{Sadowski2009} and \citet{Abramowicz2013}, and the NDAF model of \citet{Popham1999} and \citet{Xue2013}, which are all research on the 1D global solutions of accretion disks in Kerr metric. We describe in the units of $G=c=M=1$ ($M$ is the BH mass) just in this subsection.

The basic equations in Kerr geometry include:

(I) The equation of mass conservation (or the continuity equation) is
\beq \dot{M}=-4\pi\rho H \Delta^{1/2}\frac{v_r}{\sqrt{1-{v_r}^2}},\eeq
where $\dot M$ is the rest-mass accretion rate, $\rho$ is the rest-mass density, $H$ is the half thickness of the disk, $v_r$ is the radial velocity measured in the corotating frame, $\Delta\equiv r^2-2 r+a^2$ is a function of the Boyer-Lindquist radial coordinate $r$, and $a$ is the total specific angular momentum of the BH.

(II) The equation of radial momentum conservation is
\beq \frac{v_r}{1-{v_r}^2}\frac{d{v_r}}{dr}=\frac{\mathcal{A}}{r}-(1-{v_r}^2)\frac{1}{\lambda\rho}\frac{dp}{dr}, \eeq
where $p$ is pressure, $\lambda\equiv(\rho+p+u)/\rho$ is relativistic enthalpy, $u$ is specific internal energy, and the $\mathcal{A}$ term combines the effects of gravity and rotation,
\beq \mathcal{A}\equiv-\frac{A}{r^3\Delta\Omega^+_K\Omega^-_K}\frac{(\Omega-\Omega^+_K)(\Omega-\Omega^-_K)}{1-\tilde{\Omega}^2\tilde{R}^2},\eeq
where $\Omega$ is the angular velocity with respect to the stationary observer, $\Omega^{\pm}_K\equiv\pm(r^{3/2}\pm a)^{-1}$ are the angular velocities of the corotating and counter rotating Keplerian orbits, $\tilde{\Omega}\equiv\Omega-2ar/A$ is the angular velocity with respect to the local inertial observer, $\tilde{R}\equiv A/(r^2\Delta^{1/2})$ is the radius of gyration, and $A\equiv r^4+r^2 a^2+2 r a^2$.

(III) The equation of angular momentum conservation is
\beq \dot{M}(\mathcal{L}-\mathcal{L}_{\rm{in}})=\frac{4\pi p H A^{1/2}\Delta^{1/2}\gamma}{r}, \eeq
where $\mathcal{L}$ is the specific angular momentum of the disk, $\mathcal{L}_{\rm{in}}$ is the specific angular momentum at the inner edge of the disk, and the Lorentz factor $\gamma$ is written as
\beq \gamma=\sqrt{\frac{1}{1-{v_r}^2}+\frac{\mathcal{L}^2r^2}{A}}.\eeq

(IV) The equation of vertical mechanical equilibrium \citep{Abramowicz1997} is
\beq \frac{p}{\lambda\rho H^2}=\frac{\mathcal{L}^2-a^2(\epsilon^2-1)}{r^4}, \eeq
where $\epsilon$ is the energy at infinity, which is expressed as
\beq \epsilon=-\gamma\frac{r\Delta^{1/2}}{A^{1/2}}-\frac{2ar}{A}\mathcal{L}.\eeq

(V) The equation of energy conservation is
\beq -\frac{\dot{M}}{2\pi r^2}\left(\frac{u}{\rho}\frac{d\ln u}{d\ln r}-\frac{p}{\rho}\frac{d\ln\rho}{d\ln r}\right)=-\frac{2\alpha p H A \gamma^2}{r^3}\frac{d\Omega}{dr}-2Q^-,\eeq
where $\alpha$ is the viscosity parameter, and $Q^-$ is the total cooling rate (per unit area of a half-disk above or below the equator). The term on the left hand side is the advective cooling rate $Q_{\rm adv}$ (per unit area of a whole disk), and the first term on the right represents the viscous heating rate $Q_{\rm vis}$ (per unit area of a whole disk).

\subsubsection{Neutrino physics}

The cooling mechanism is the main difference between NDAFs and typical accretion disks. Neutrino radiation is dominantly in NDAFs, so microphysics, especially neutrino physics should be considered. In great detail, the microphysics in NDAFs is extended from that in NSs \citep[e.g.,][]{Shapiro1986} because of close resemblance.

The cooling process of NDAFs is much more complicated than that of classic accretion disks. In most previous works, neutrino physics is simplified through parameterizing the electron fraction $Y_{\rm e}=0.5$ or considering the effect of $^4$He. Below we present a detailed description of neutrino physics in NDAFs.

We first introduce the total optical depth for neutrinos
\beq \tau_{\nu_i}=\tau_{s, \nu_i} + \tau_{a,\nu_i}, \eeq
where $\tau_{s,\nu_i}$ and $\tau_{a,\nu_i}$ are the neutrino optical depth from scattering and absorption, respectively, the subscript ``$i$'' runs for the three species of neutrinos $\nu_{\rm e}$ , $\nu_{\rm \mu}$ , and $\nu_{\rm \tau}$.

Specifically, the optical depth for neutrinos through scattering off electrons and nucleons $\tau_{s, \nu_i}$ is given by
\beq \tau_{s,\nu_i} \approx H \displaystyle(\sigma_{{\rm e}, \nu_i} n_{\rm e} + {\sum_{j}} \sigma_{j,\nu_i} n_j), \eeq
where $\sigma_{{\rm e}, \nu_i}$, $\sigma_{j, \nu_i}$, $n_{\rm e}$ and $n_j$ ($j=1$, 2,...) are the cross sections of electron and nucleons ($n_1$ and $n_2$ are the number density of free protons and free neutrons), and the number density of electrons and nucleons ($j \geq 3$), respectively \citep[e.g.,][]{Kohri2005,Chen2007,Kawanaka2007,Liu2007,Liu2012a,Xue2013}. The major cross sections from scattering off electrons, free protons, free neutrons and other elements particles are given by \citep{Burrow2004,Chen2007}
\beq \sigma_{{\rm e},\nu_i}\approx \frac{3 k_{\rm B} T \sigma_0 e_{\nu_i}}{8 m_{\rm e} c^2} (1+ \frac{\eta_{\rm e}}{4})[(C_{V,\nu_i}+C_{A,\nu_i})^2 \\ \nonumber+ \frac{1}{3} (C_{V,\nu_i}-C_{A,\nu_i})^2],\eeq
\beq \sigma_{{\rm n_1},\nu_i}\approx\frac{\sigma_0 e_{\nu_i}^2}{4}[(C_{V,\nu_i}-1)^2+3 g_A^2 (C_{A,\nu_i}-1)^2] ,\eeq
\beq \sigma_{{\rm n_2},\nu_i}\approx\frac{\sigma_0 e_{\nu_i}^2}{4} \frac{1+3 g_A^2}{4}, \eeq
\beq \sigma_{n_j,\nu_i}\approx \frac{\sigma_0}{16} e^2_{\nu_i} (Z_j + N_j) [1-\frac{2Z_j}{Z_j+N_j}(1-2 \sin^2 \theta_W)]^2,\eeq
where $k_{\rm B}$ is the the Boltzmann constant and $\eta_{\rm e}$ is electron degeneracy, $\sigma_0= 4G^2_F(m_{\rm e}c^2)^2/\pi(\hbar c)^4 \approx 1.71 \times 10^{- 44} {\rm cm^2}$, $G_F \approx 1.436 \times 10^{-49} {\rm erg~cm^3}$, $e_{\nu_i}$ is the mean energy of neutrinos in units of $(m_{\rm e} c^2)$, $g_A \approx 1.26$, $\sin^2 \theta_W \approx0.23$, $Z_j$ and $N_j$ are defined as the number of protons and neutrons of a nucleus $X_j$, and $C_{V,\nu_e}=1/2 + 2 \sin^2 \theta_W$ , $C_{V,\nu_\mu} =C_{V,\nu_\tau}= - 1/2 + 2\sin^2 \theta_W$ , $ C_{A,\nu_e}= C_{A,\overline{\nu}_\mu}=C_{A,\overline{\nu}_\tau}=1/2$, $C_{A,\overline{\nu}_e} = C_{A,\nu_\mu}=C_{A,\nu_\tau}=- 1/2$. The electron degeneracy is an important physical parameter that affects electron fraction, degeneracy pressure, and neutrino cooling \citep{Kohri2005,Chen2007,Liu2007,Kawanaka2007}.

The number density of electrons and positrons can be calculated by the Fermi-Dirac integration \citep[see, e.g.,][]{Kohri2005,Kawanaka2007,Liu2007,Xue2013},
\beq n_{\rm e^{\mp}}= \frac{1}{\hbar^3 \pi^2} \int_0^\infty d p~p^2 \frac{1}{e^{({\sqrt{p^2 c^2+{m_{\rm e}}^2 c^4} \mp {\mu_{\rm e}})/k_{\rm B} T}}+1}, \eeq
where $\mu_{\rm e}=\eta_{\rm e} k_{\rm B} T$ is the chemical potential of electrons.

The absorption depth for neutrinos $\tau_{a, \nu_i}$ is defined by
\beq \tau_{a, \nu_i}= \frac{q_{\nu_i} H}{4 (7/8) \sigma T^4},\eeq
where $q_{\nu_i}$ is the total neutrino cooling rate (per unit volume) and is the sum of four terms,
\beq q_{\nu_i}=q_{\rm Urca}+q_{{\rm e^{-}+e^{+}} \rightarrow \nu_i+ \overline {\nu}_i}+q_{{\rm n + n \rightarrow  n + n +} \nu_i+\overline{\nu}_i}+q_{\tilde{\gamma} \rightarrow \nu_i+\overline \nu_i}.\eeq

Urca processes have been included in the proton-rich nuclear statistical equilibrium \citep[NSE,][]{Seitenzahl2008,Liu2013,Xue2013}. The neutrino cooling rate due to the Urca processes $q_{\rm Urca}$ relates only to electron neutrino and antineutrino. There are four major terms by electrons, positrons, free protons, free neutrons and nucleons \citep{Liu2007,Kawanaka2007}, which is expressed by
\beq q_{\rm Urca}=q_{\rm p+e^{-}\rightarrow n+\nu_{\rm e}}+q_{\rm n+e^{+}\rightarrow p+\overline{\rm \nu}_{\rm e}}+q_{\rm n \rightarrow p+e^{-} +\overline{\nu}_{\rm e}}\\ \nonumber+q_{ X_j+{\rm e^{-}}\rightarrow X'_j+\nu_{\rm e}}, \eeq
with
\beq &&q_{\rm p+e^{-}\rightarrow n+\nu_{\rm e}}= \frac{G_F^2 \cos^2 \theta_c}{2 \pi^2 \hbar^3 c^2}(1+3 g_A^2) n_{\rm 1} \nonumber\\ &&\times \int_Q^\infty d E_{\rm e}~E_{\rm e}~\sqrt{{E_{\rm e}}^2-{m_{\rm e}}^2 c^4}(E_{\rm e}-Q)^3 f_{{\rm e}^{-}} ,\eeq
\beq &&q_{\rm n+e^{+}\rightarrow p+\overline{\nu}_{\rm e}}=\frac{G_F^2 \cos^2 \theta_c}{2 \pi^2 \hbar^3 c^2}(1+3 g_A^2) n_{\rm 2} \nonumber\\&&\times\int_{m_{\rm e} c^2}^\infty d E_{\rm e}~E_{\rm e}~\sqrt{{E_{\rm e}}^2-{m_{\rm e}}^2 c^4}(E_{\rm e}+Q)^3 f_{{\rm e}^{+}}, \eeq
\beq &&q_{\rm n \rightarrow p+e^{-}+\overline{\nu}_{\rm e}}=\frac{G_F^2 \cos^2 \theta_c}{2 \pi^2 \hbar^3 c^2}(1+3 g_A^2) n_{\rm 2}\nonumber \\&&\times\int_{m_{\rm e} c^2}^Q d E_{\rm e}~E_{\rm e}~\sqrt{{E_{\rm e}}^2-{m_{\rm e}}^2 c^4}(Q-E_{\rm e})^3 (1-f_{{\rm e}^{-}}), \eeq
\beq &&q_{ X_j+{\rm e^{-}}\rightarrow X'_j+\nu_{\rm e}}= \frac{G_F^2 \cos^2 \theta_c}{2 \pi^2 \hbar^3 c^2}g_A^2 \frac{2}{7} N_p(Z_j) N_h(N_j)n_j \nonumber\\&&\times\int_{Q'}^\infty d E_{\rm e}~E_{\rm e}~\sqrt{{E_{\rm e}}^2-{m_{\rm e}}^2 c^4}(E_{\rm e}-Q')^3 f_{{\rm e}^{-}},\eeq
where $\cos^2 \theta_c \approx 0.947$, $Q = (m_{\rm n} - m_{\rm p})c^2$, $Q'\approx \mu'_{\rm n}-\mu'_{\rm p}+\Delta$, $\mu'_{\rm n}$ and $\mu'_{\rm p}$ are the chemical potential of protons and neutrons in their own nuclei, $\Delta\approx3 \rm MeV$ is the energy of the neutron 1$f_{5/2}$ state above the ground state, and $f_{{\rm e}^{\mp}}=\{\exp[(E_{\rm e} \mp \mu_{\rm e})/k_{\rm B} T]+1\}^{-1}$ is the Fermi-Dirac function. $N_p(Z_j)$ and $N_h(N_j)$ are satisfied with
\beq N_p(Z_j) =\begin{cases} 0, ~~Z_j<20,\\Z_j-20, ~~20<Z_j<28, \\ 8, ~~Z_j>28, \end{cases}\eeq
\beq N_h(N_j) =\begin{cases} 6, ~~N_j<34,\\40-N_j, ~~34<N_j<40, \\ 0, ~~N_j>40. \end{cases}\eeq

The electron-positron pair annihilation rate into neutrinos $q_{{\rm e^{-}+e^{+}} \rightarrow \nu_i+ \overline \nu_i}$ is \citep[e.g.,][]{Itoh1989,Yakovlev2001,Janiuk2007}
\beq q_{{\rm e^{-}+e^{+}} \rightarrow \nu_i+ \overline {\nu}_i}= \frac{Q_c}{36 \pi} \{(C^2_{V,\nu_i}+C^2_{A,\nu_i})^2[8(\Phi_1 U_2 + \Phi_2 U_1)\nonumber\\ -2(\Phi_{-1} U_2 + \Phi_2 U_{-1}) + 7(\Phi_0 U_1 + \Phi_1 U_0)]\} \nonumber\\+ \{[5(\Phi_0 U_{-1} + \Phi_{-1} U_0)]\nonumber\\+9(C^2_{V,\nu_i}-C^2_{A,\nu_i})^2 [\Phi_0(U_1+U_{-1})+(\Phi_{-1}+\Phi_1)U_0]\}, \eeq
where $Q_c = ({m_{\rm e} c}/{\hbar})^9 {G^2_F}/{\hbar} \approx 1.023 \times 10^{23} \rm erg~cm^{-3}~s^{-1}$, and the dimensionless functions $U_k$ and $\Phi_k$ ($k= -1, 0, 1, 2$) in the above equation can be expressed in terms of the Fermi-Dirac functions \citep{Kawanaka2007}. Once electrons are in degenerate state, this process can be ignored.

The nucleon-nucleon bremsstrahlung rate $q_{{\rm n + n \rightarrow  n + n +} \nu_i+\overline{\nu}_i}$ is the same for the three species of neutrinos \citep[e.g.,][]{Itoh1996,Di Matteo2002,Liu2007}, which can be simplified as
\beq q_{{\rm n + n \rightarrow  n + n +} \nu_i+\overline{\nu}_i} \approx 1.5 \times 10^{27} \rho_{10}^2 T_{11}^{5.5} ~{\rm erg~cm^{-3}~s^{-1}},\eeq
where ${\rho}_{10} \equiv {\rho}/10^{10}{\rm g~cm^{-3}}$ and $T_{11} \equiv T/10^{11}{\rm K}$.

The plasmon decay rate for the three species of neutrinos $q_{\tilde{\gamma} \rightarrow \nu_i+\overline \nu_i}$ also needs to be considered, where plasmons $\tilde{\gamma}$ are photons interacting with electrons \citep[e.g.,][]{Kawanaka2007,Xue2013},
\beq q_{\tilde{\gamma} \rightarrow \nu_{\rm e}+\overline \nu_{\rm e}} = \frac{\pi^4}{6 \alpha^*} C_{V,\nu_{\rm e}} \frac{\sigma_0 c}{(m_{\rm e}c^2)^2} \frac{(k_{\rm B} T)^9}{(2 \pi \hbar c)^6} \nonumber\\\times\gamma^6 (\gamma^2+2\gamma+2){\rm exp}(-\gamma),\eeq
\beq q_{\tilde{\gamma} \rightarrow \nu_\mu+\overline \nu_\mu} = q_{\tilde{\gamma} \rightarrow \nu_\tau+\overline \nu_\tau} = \frac{4 \pi^4}{6 \alpha^*} C_{V,\nu_\mu} \frac{\sigma_0 c}{(m_{\rm e}c^2)^2} \frac{(k_{\rm B} T)^9}{(2 \pi \hbar c)^6} \nonumber\\\times\gamma^6 (\gamma^2+2\gamma+2){\rm exp}(-\gamma),\eeq
where $\alpha^* \approx 1/137$ is the fine-structure constant, and $\gamma \approx 5.565 \times 10^{-2} {[(\pi^2 + 3 \eta_e^2)/3]}^{1/2}$. Here $q_{{\rm n + n \rightarrow  n + n +} \nu_i+\overline{\nu}_i}$ and $q_{\tilde{\gamma} \rightarrow \nu_i+\overline \nu_i}$ may become important only for very high electron degeneracy state.

Second, the electron fraction can be written as \citep{Liu2013,Xue2013}
\beq Y_{\rm e}=\frac{\sum\limits_{j} n_j Z_j} {\sum\limits_{j} n_j (Z_j+N_j)}. \eeq
Moreover, the condition of electrical neutrality naturally constrains $Y_{\rm e}$ \citep{Liu2007,Liu2013}, i.e.,
\beq \sum\limits_{j} n_j Z_j=\frac {\rho Y_{\rm e}}{m_u} = n_{\rm e^{-}}-n_{\rm e^{+}},\eeq
where $m_u$ is the mean mass of nucleus, and the mass fraction is considered to approximately equal the number density.

For NDAFs, in order to allow for a transition from the optically thin [$\mu_{\rm n}=\mu_{\rm p}+2\mu_{\rm e}$ \citep{Yuan2005}, where $\mu_{\rm n}$, $\mu_{\rm p}$, and $\mu_{\rm e}$ are the chemical potential of free neutrons, free protons, and electrons, respectively] to optically thick ($\mu_{\rm n}=\mu_{\rm p}+\mu_{\rm e}$) regimes, the bridging formula of free protons and neutrons can be established, which is given by \citep{Liu2007,Xue2013}
\beq \lg{\frac{n_{\rm 2}}{n_{\rm 1}}}= f(\tau_\nu) \frac{2 \mu_{\rm e}-Q}{k_{\rm B} T}+[1-f(\tau_\nu)] \frac{\mu_{\rm e}-Q}{k_{\rm B} T}, \eeq
where $f(\tau_\nu)=\rm exp(-\tau_{\nu_{\rm e}})$ is a weight factor. In addition, the bridging formula can be generally used in NDAFs if when nucleosynthesis is taken (see Subsection 2.1.3 below), because the outer region of the NDAF is optically thin.

If we assume that the heaviest nuclei is $^4 \rm He$, by combing with the above equations, the bridging formula can be simplified as \citep{Liu2007}
\beq Y_{\rm e}=\frac{1}{2}(1-X_{\rm nuc})+ \frac{X_{\rm nuc}}{1+\exp\{\frac{[1+f(\tau_\nu)] \mu_{\rm e}-Q}{k_{\rm B } T}\}}, \eeq
where $X_{\rm nuc}$ is the mass fraction of free nucleons \citep[e.g.,][]{Di Matteo2002,Kohri2005,Chen2007,Liu2007}, which can be reduced from the equation of NSE \citep[e.g.,][]{Meyer1994}.

\subsubsection{Nucleosynthesis}

As mentioned above, NSE is established by all nuclear reactions in the chemical equilibrium. \cite{Seitenzahl2008} developed a method of NSE especially for proton-rich material, which is applicable to almost the entire range of the electron fraction. The complicated and detailed balance has been included under the condition of chemical potential equilibrium. Although they focus on the proton-rich state of matter, it can be used in the description of all the state of matter. The conclusion is that, $\rm ^{56}Ni$ is favored under proton-rich conditions ($Y_{\rm e} \simeq 0.5$), which is very different from the case of Fe-peak nuclei domination with the largest binding energy per nucleon that have a proton to nucleon ratio close to the prescribed electron fraction \citep[e.g.,][]{Lattimer1991}. This method is suitable for the calculations of NDAFs whatever origins from compact objects mergers or collapsars.

One in particular is that the lower limit of the temperature in the NSE calculation is identified at about $2\times10^9$ K. Once the temperature is lower than this limit, their NSE solution is unreliable. Furthermore, \citet{Kawanaka2007} assumed that the inflowing gas is composed primarily of neutron-rich iron group nuclei as the outer boundary conditions, and the maximum electron fraction is less than 0.42. Anyway, it is predictably that the cooling process of the matter in the outflow of NDAFs can reveal the history of nucleosynthesis in very short timescale.

\subsubsection{Thermodynamics}

Distinguished from the classic accretion disks, the pressure from degenerate electrons and neutrinos are included in the equation of state (EoS), i.e.,
\beq  p=p_{\rm gas}+p_{\rm rad}+p_{\rm e}+p_\nu.\eeq

The gas pressure from free nucleons $p_{\rm gas}$ can be estimated by
\beq p_{\rm gas}= \sum\limits_{j} n_j k_{\rm B} T. \eeq

The disk is definitely optically thick for the photons, so photon radiation pressure $p_{\rm rad}$ can be given by
\beq p_{\rm rad}=\frac{1}{3}aT^4. \eeq

The electron pressure $p_{\rm e}$ is the sum of the electron and positron pressure, which is described by the exact Fermi-Dirac distribution \citep[e.g.,][]{Chen2007,Liu2007,Xue2013}. In NDAFs, electrons are neither nondegenerate nor strongly degenerate, and they are not ultrarelativistic at all radii. One has
\beq p_{\rm e}=p_{\rm e^{-}}+ p_{\rm e^{+}}, \eeq
where
\beq &&p_{\rm e^{\mp}}= \frac{1}{3 {\pi}^2 {\hbar}^3 c^3} \nonumber\\ && \times \int_0^{\infty} d p \frac{p^4}{\sqrt{p^2 c^2+{m_{\rm e}}^2 c^4}} \frac{1}{e^{({\sqrt{p^2 c^2+{m_{\rm e}}^2 c^4} \mp {\mu_{\rm e}})/k_{\rm B} T}}+1}. \eeq

The neutrino pressure $p_{\rm \nu}$ is
\beq p_{\rm \nu}=\frac{u_{\rm \nu}} {3}, \eeq
where ${u}_{\nu}$ is the energy density of the neutrinos, which is adopted by a bridging formula connecting the optically thin region with the optically thick region \citep[e.g.,][]{Di Matteo2002,Kohri2005,Liu2007,Xue2013},
\beq u_{\rm \nu}=\sum_{i} \frac{(7/8)a T^4 (\tau_{{\nu}_i}/2+1/ \sqrt{3})} {\tau_{{\nu}_i}/2+1/ \sqrt{3}+1/(3 \tau_{a,{\nu}_i})}. \eeq
In fact, this equation is derived from ADAF models via the approximation of the radiation transfer equations \citep[e.g.,][]{Hubeny1990,Narayan1995b,Popham1995,Artemova1996}.

\citet{Li2013} revisited various properties of the hot nuclear matter possible in the inner regions of GRBs and supernovae (SNe). They employed the microscopic Brueckner-Hartree-Fock approach to account for the strong interaction between nucleons, and calculated the nucleonic chemical potentials and the nucleonic EoS. A parameterized chemical potential equilibrium bridging between neutrino optically thin and thick regions was introduced. This method can update the above descriptions of EoS and neutrino physics in the inner regions of NDAFs.

A fraction of the viscous heating energy following inflows has been advected into the BH, and the rest part equals the cooling energy. The cooling rate $Q^{-}$ mentioned in Equation (8) is composed of the cooling rates of photodisintegration, neutrino emission, and photon radiation
\beq Q^{-}=Q_{\rm ph} +Q_{\nu}+Q_{\rm rad}. \eeq
However, cooling due to photon radiation can be ignored because NDAFs are optically thick.

The cooling rate by photodisintegration can be given by the NSE. For the case that the heaviest nucleus is $\alpha$-particles, $Q_{\rm ph}$ can be written as
\beq Q_{\rm ph} = 6.8 \times 10^{38} \rho_{10} v_{r,10} H \frac{d X_{\rm nuc}}{d r}~{\rm erg~s^{-1}}, \eeq
where $v_{r,10}=v_r/(10^{10}~\rm cm~s^{-1})$. The cooling rate by disintegration of other heavy nuclei can be ignored because of the lower number density of these nuclei and the absolutely dominant advective cooling rate in the outer region.

The cooling rate due to neutrino loss $Q_\nu$ is expressed in accordance to the above bridging formula of ${u}_{\nu}$ \citep[e.g.,][]{Di Matteo2002,Kohri2005,Liu2007,Xue2013},
\beq Q_{\nu}=\sum_{i} \frac{(7/8) {\sigma} T^4}{(3/4)[\tau_{{\nu}_i}/2+1/ \sqrt{3}+1/(3 \tau_{a,{\nu}_i})]}. \eeq

\begin{figure}[t]
\centering
\includegraphics[angle=0,scale=0.75]{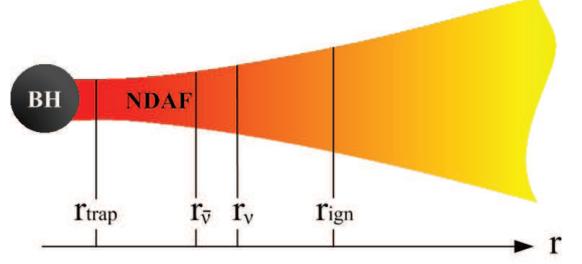}
\caption{Schematic picture of characteristic radii of NDAFs (adapted from Figure 10 in \citet{Chen2007}).}
\label{215}
\end{figure}

\subsubsection{Characteristic radii in NDAFs}

One can define three characteristic radii in NDAFs \citep{Chen2007} as shown in Figure \ref{215}.

(I) Neutrino trapping radius $r_{\rm trap}$

Photon trapping may occur in the supercritical accretion disk if the time of photon diffusion from the equatorial plane of the disk to the surface is longer than the accretion timescale \citep[e.g.,][]{Katz1977,Begelman1978,Ohsuga2002,Ohsuga2005}. Photons generated near the equatorial plane diffuse toward the disk surface at a speed of $\sim c/3 \tau$ \citep{Mihalas1984}, where $\tau$ is the total optical depth. The timescale of photon diffusion is about $t_{\rm diff} \sim H / (c/3 \tau)$. Similarly, neutrinos may also be trapped in NDAFs \citep{Chen2007,Liu2012,Xue2013}. The neutrino velocity $v_n$ replaces $c$ in the above equation of $t_{\rm diff}$, which can be estimated by $\sim (3.7 k_{\rm B} T c^2/0.07 \rm eV)^{1/2}$, where $\sim 3.7 k_{\rm B} T$ and $0.07 \rm eV$ roughly equals to the neutrino energy and the lower limit of neutrino rest-mass energy, respectively. The accretion timescale can be estimated by $t_{\rm acc} \sim -r/{v_r}$. One can then use $\tilde{t}=t_{\rm diff}/t_{\rm acc}=1$ to define the trapping radius,
\beq r_{\rm trap} \simeq -\frac{3 \tau_{\nu_{\rm e}}H v_r}{v_{\rm n}}. \eeq

Once neutrinos are trapped in the inner region, the neutrinos launched from the surface would decrease. As a result, neutrino trapping would greatly affect the neutrino luminosity and annihilation luminosity for high accretion rates ($\dot{M} \gtrsim 2~M_\odot~\rm s^{-1}$) and large spin ($a_* \sim 0.95$, where $a_*$ is the dimensionless spin parameter of the BH) \citep{Liu2012a,Xue2013}. Moreover, one can set a critical accretion rate $\dot{M}_{\rm trap}$ when $r_{\rm trap}$ appears in NDAFs.

(II) Ignition radius $r_{\rm ign}$

The region where $r<r_{\rm ign}$ meets that the neutrino emission switches on and neutrino cooling dominates. Accurately, $r_{\rm ign}$ is defined as the radius satisfied with $Q^-/Q_{\rm vis}=0.5$ \citep{Chen2007,Zalamea2011,Liu2012}. Once $r_{\rm ign}$ emerges, NDAFs are ignited, the corresponding $\dot{M}$ is defined as $\dot{M}_{\rm ign}$. For a rapidly rotating BH with $3~M_\odot$ surrounded by a NDAF with low viscosity, $\dot{M}_{\rm ign}$ is about 0.001 $M_\odot~\rm s^{-1}$. $\dot{M}_{\rm trap}$ and $\dot{M}_{\rm ign}$ mainly depend on the spin of BHs and the viscous parameter of disks in BH-NDAF systems.

(III) Neutrino opaque radii $r_\nu$ and $r_{\bar{\nu}}$

Radii $r_\nu$ and $r_{\bar{\nu}}$ locate where the disks become opaque for neutrinos and antineutrino \citep{Chen2007}, and the radiation becomes thermal. $r_\nu$ is usually larger than $r_{\bar{\nu}}$ due to the neutrino optical depth as shown in Subsection 2.1.2, and they are definitely less than $r_{\rm ign}$.

\subsubsection{Results}

\begin{figure}
\centering
\includegraphics[angle=0,scale=0.6]{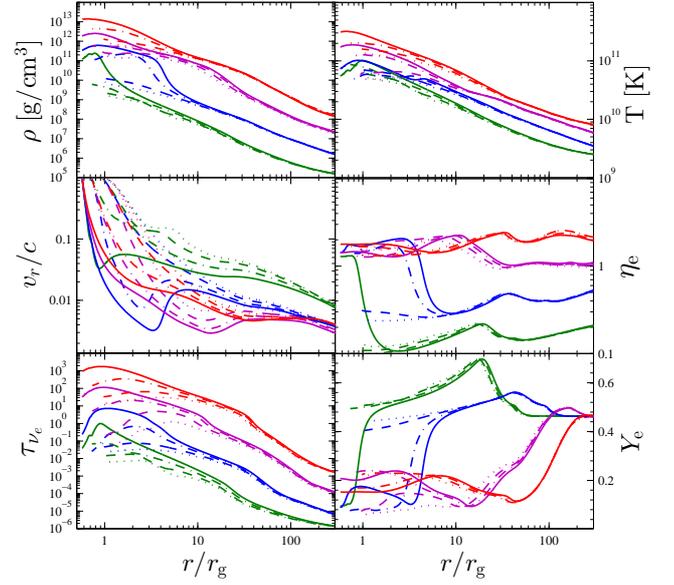}
\caption{The green, blue, purple, and red lines denote the different accretion rates $\dot{M}$ = $0.03$, $0.1$, $1$, and $10$ $M_\odot~\rm s^{-1}$, respectively. The dotted, dashed, dot-dashed, and solid denote the different BH spin values $a_*$ = $0$, $0.5$, $0.9$, and $0.99$. The six panels show the distributions of density $\rho$, temperature $T$, radial velocity $v_r$, electron degeneracy $\eta_{\rm e}$, optical depth of electron neutrino $\tau_{\nu_{\rm e}}$, and electron fraction $Y_{\rm e}$ with radius $r/r_{\rm g}$, respectively (adapted from Figure 1 in \citet{Xue2013}).}
\label{2161}
\end{figure}

\begin{figure}
\includegraphics[angle=0,scale=0.4]{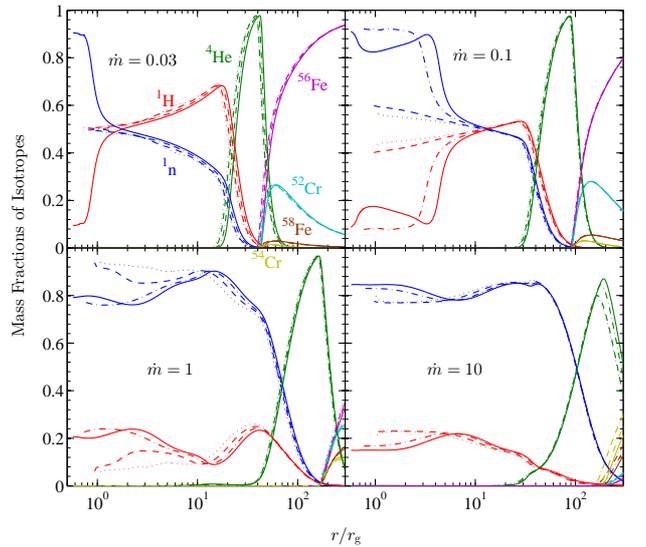}
\caption{The radial distributions of the mass fraction of seven major nucleons $\rm ^1n$, $\rm ^1H$, $\rm ^4He$, $\rm ^{52}Cr$, $\rm ^{54}Cr$, $\rm ^{56}Fe$. The meanings of different line-styles corresponding different $a_*$ are same with the ones in Fig. \ref{2161}, and $\dot{m}=\dot{M}/(M_\odot~\rm s^{-1})$ (adapted from Figure 4 in \citet{Xue2013}).}
\label{2162}
\end{figure}

Figure \ref{2161} shows the basic structure of NDAFs with $M=3M_\odot$ and $\alpha=0.1$. The six panels correspond with the variations of density $\rho$, temperature $T$, radial velocity $v_r$, electron degeneracy $\eta_{\rm e}$, optical depth of electron neutrino $\tau_{\nu_{\rm e}}$, and electron fraction $Y_{\rm e}$ with $r/r_{\rm g}$, where $r_{\rm g}=2GM/c^2$. From the outer to inner region of the disk, $\rho$, $T$, and $\tau_{\nu_{\rm e}}$ increase by about 6, 1, and 5 orders of magnitude. In the innermost region, they reach about $10^{13} ~\rm g~cm^{-3}$, $10^{12}~\rm K$, and $10^3$, respectively. The value of $Y_{\rm e}$ at the outer boundary of the disk in sixteen solutions all tend to about 0.46. There is a little different value $\sim 0.42$ in \citet{Kawanaka2007}, but this difference leads to hugely different nucleosynthesis products.

Figure \ref{2162} shows the radial distributions of the mass fractions of seven major nucleons cover almost $99\%$ mass of the flow, including $\rm ^1n$, $\rm ^1H$, $\rm ^4He$, $\rm ^{52}Cr$, $\rm ^{54}Cr$, $\rm ^{56}Fe$ and $\rm ^{58}Fe$. The mass fraction of $\rm ^{56}Fe$, $\rm ^4 He$, and free neutrons and protons dominate in the outer, middle, and inner regions of the disk for all accretion rates. It is an implication for the origin of heavy nuclei in GRBs accounting for the suspected detection of Fe K$\alpha$ X-ray lines and other emission lines \citep[e.g.,][]{Lazzati1999,Kallman2003,Gou2005,Butler2007}, which can play an important role in understanding the nature of GRBs, especially its central engine. For a comparison, \citet{Janiuk2014} obtained that $^{56}$Ni dominates out of about 500 $r_{\rm g}$ of the disk with $a_*=0.9$ by using simplified neutrino physics, which is similar to the vertical results in \citet{Liu2013}. We will detailed discuss nucleosynthesis of NDAFs in Subsection 6.4.

\subsection{Neutrino luminosity and annihilation luminosity}

Once the neutrino cooling rate $Q_\nu$ is obtained, the neutrino radiation luminosity before annihilation, $L_{\rm \nu}$, can be calculated by
\beq L_{\rm \nu}=4 \pi \int_{r_{\rm in}}^{r_{\rm out}} Q_{\rm \nu} r d r, \eeq
where $r_{\rm in}$ and $r_{\rm out}$ are the inner and outer edge of the disk. Here we sum two jets in the opposite direction. For comparison with the observations, the coefficient ``4$\pi$'' should be replaced by ``2$\pi$'' in the above equation. The same applies to the equation of the neutrino annihilation luminosity below.

\begin{figure}
\centering
\includegraphics[angle=0,scale=0.5]{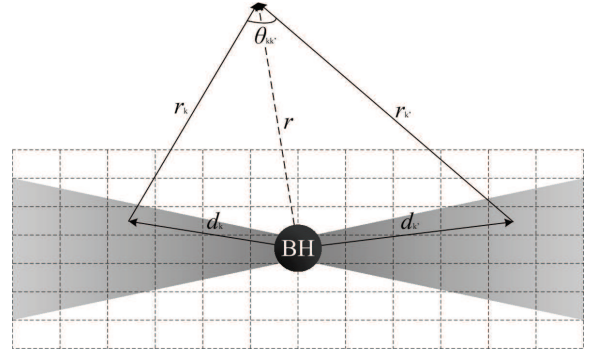}
\caption{Schematic picture of neutrino annihilation of NDAFs (adapted from Figure 3 in \citet{Rosswog2003}).}
\label{221}
\end{figure}

In the neutrino annihilation calculations, a Newtonian approach is introduced in \citet{Ruffert1997}, \citet{Popham1999}, and \citet{Rosswog2003}. Strictly, the general relativistic effects on the neutrino trajectory near the BH need to be considered \citep[e.g.,][]{Birkl2007,Zalamea2011}. The disk is modeled as a grid of cells in the equatorial plane. As shown in Figure \ref{221}, a cell $k$ has its mean neutrino energy $\varepsilon_{\nu_i}^k$, neutrino radiation luminosity $l_{\nu_i}^k$, and distance to a spatial point above (or below) the disk $d_k$. At any spatial point, the angle at which a neutrino from cell $k$ encounter an antineutrino from another cell $k'$ is denoted as $\theta_{kk'}$. Then the neutrino annihilation luminosity at that point is given by the summation over all pairs of cells,
\beq\ l_{\nu \overline{\nu}}=\sum_{i} A_{1,i} \sum_k \frac{l_{\nu_i}^k}{d_k^2} \sum_{k'} \frac{l_{\overline{\nu}_i}^{k'}}{d_{k'}^2} (\varepsilon_{\nu_i}^k + \varepsilon_{\overline{\nu}_i}^{k'}) {(1-\cos {\theta_{kk'}})}^2 \nonumber\\+ \sum_{i} A_{2,i} \sum_k \frac{l_{\nu_i}^k}{d_k^2} \sum_{k'} \frac{l_{\overline {\nu}_i}^{k'}}{d_{k'}^2} \frac{\varepsilon_{\nu_i}^k + \varepsilon_{\overline{\nu}_i}^{k'}}{\varepsilon_{\nu_i}^k \varepsilon_{\overline{\nu}_i}^{k'}} {(1-\cos {\theta_{kk'}})},\eeq
where $A_{1,i} = (1 / 12\pi^2)[\sigma_0/c{(m_{\rm e} c^2)}^2][{(C_{V,\nu_{i}}-C_{A,\nu_{i}})}^2 +{(C_{V,\nu_{i}}+C_{A,\nu_{i}})}^2]$, and $A_{2,i} = (1/6\pi^2) (\sigma_0/c)$ $ (2 C_{V,\nu_{i}}^2-C_{A,\nu_{i}}^2)$, with
$C_{V,\nu_{i}}$ and $C_{A,\nu_{i}}$ given below Equation (14). The total neutrino annihilation luminosity is the integration over the entire space outside the disk,
\beq L_{\nu \overline{\nu}}=4 \pi \int_{r_{\rm in}}^\infty \int_H^\infty l_{\nu \overline{\nu}} r d r d z. \eeq

The neutrino radiation luminosity and annihilation luminosity depend on with the BH spin and accretion rate. Two analytic formulae can be derived by fitting results in \citet{Xue2013},
\beq \log L_{\nu}~({\rm{erg\ s^{-1}}})\approx 52.5+1.17a_*+1.17\log\dot{m}, \eeq
\beq \log L_{\nu \bar{\nu}}~({\rm{erg\ s^{-1}}})\approx 49.5+2.45a_*+2.17\log\dot{m},\eeq
which are applicable for the accretion rate in the range of $0.01 \lesssim \dot{m} \lesssim 10$, where $\dot{m}=\dot{M}/(M_\odot~\rm s^{-1})$.

Including the effect of BH mass, the annihilation luminosity is approximated as \citep{Liu2016b}
\beq \log L_{\nu \bar{\nu}} ~({\rm{erg\ s^{-1}}}) \approx 52.98 + 3.88  a_* - 1.55  \log m \nonumber \\+ 5.0 \log \dot{m}, \eeq
where $m=M/M_\odot$. It should be emphasized that this formula is applicable for $0.01 \lesssim \dot{m} \lesssim 0.5$.

\begin{figure}
\centering
\includegraphics[angle=0,scale=0.32]{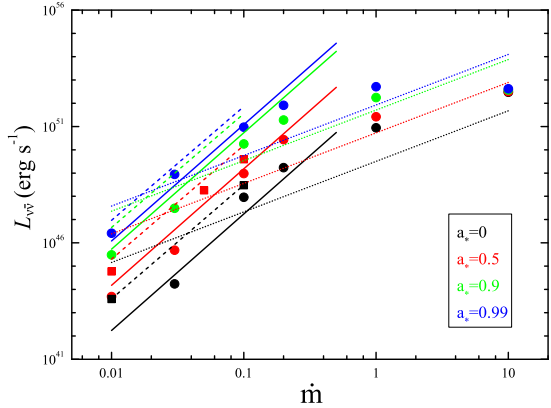}
\caption{Neutrino annihilation luminosity $L_{\nu \bar{\nu}}$ as a function of dimensionless accretion rate $\dot{m}$ for $a_*$ = 0, 0.5, 0.9, 0.99 and $M_{\rm BH}=3~M_\odot$. The circles represent the global solutions in \citep{Xue2013}, whereas the squares represent the solutions in \citet{Popham1999}. The solid, dashed, and dotted lines, represent the fitting lines of Equations (49), (48), and (50), respectively (adapted from Figure 2 in \citet{Liu2016b}).}
\label{222}
\end{figure}

\citet{Fryer1999} also displayed the approximate fit to the annihilation luminosity results of \citet{Popham1999}, i.e.,
\beq \log L_{\nu \bar{\nu}} ~({\rm{erg\ s^{-1}}}) \approx  53.4 + 3.4 a_* + 4.89 \log \dot{m}, \eeq
which is satisfied with $0.01\lesssim\dot{m}\lesssim0.1$. The comparison of the above three annihilation formulae is shown in Figure \ref{222}.

Another well-known analytic formula of neutrino annihilation luminosity is given by \citet{Zalamea2011}, which is expressed as
\beq &&L_{\nu\bar{\nu}}\approx5.7 \times 10^{52} ~x_{\rm ms}^{-4.8}~m^{-3/2} \nonumber \\ &&\times \Bigg \{
\begin{array}{ll}
0 & \hbox{for $\dot{m} < \dot {m}_{\rm ign}$}\\
\dot{m} ^{9/4} & \hbox{for $\dot {m}_{\rm ign} < \dot{m} < \dot{m}_{\rm trap}$}\\
\dot{m}_{\rm trap} ^{9/4} & \hbox{for $\dot{m} > \dot{m}_{\rm trap}$}\\
\end{array}
\Bigg \} ~\rm erg~s^{-1}, \eeq
where $x_{\rm ms}=r_{\rm ms}/r_{\rm g}$ is the dimensionless marginally stable orbit radius of the disk, $r_{\rm ms}$ is radius of the last marginally stable orbit, and $\dot {m}_{\rm ign}=\dot {M}_{\rm ign}/(M_\odot~\rm s^{-1})$, $\dot {m}_{\rm trap}=\dot {M}_{\rm trap}/(M_\odot~\rm s^{-1})$. Here $x_{\rm ms}=3+Z_{2}-\sqrt{(3-Z_{1})(3+Z_{1}+2Z_{2})}$, where $Z_{1}=1+(1-a_*^{2})^{1/3}[(1+a_*)^{1/3}+(1-a_*)^{1/3}]$ and $Z_{2}=\sqrt{3a_*^{2}+Z_{1}^{2}}$ for $0 < a_* <1$ \citep[e.g.,][]{Bardeen1972,Kato2008,Zalamea2011,Liu2015c}.

The reasons of the different forms include: (1) different levels of neutrino physics is considered, which cause the slight effects on the annihilation luminosity; (2) a different applying range of $\dot M$ is adopted; (3) we have fitted the numerical results directly for BH mass instead of introducing some analytical results as did in \cite{Zalamea2011}; (4) the difference between Newtonian approach or relativistic method also leads to the slight effects \citep{Yang2017}.

\begin{figure}
\centering
\includegraphics[angle=0,scale=0.35]{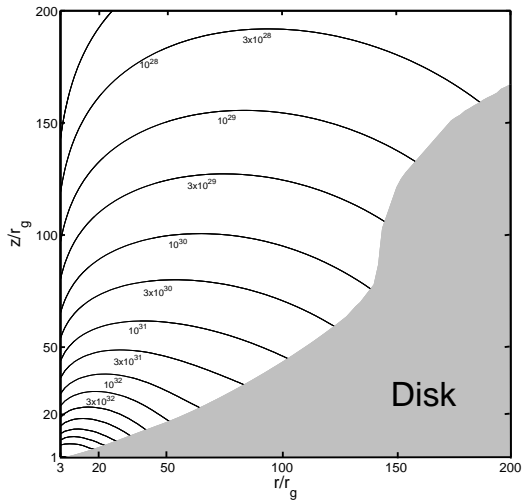}
\caption{Contours of the neutrino annihilation luminosity of a circle with cylindrical coordinates $r$ and $z$ for $\alpha=0.1$ and $\dot{m}=1$. The number attaching to each line is this luminosity in units of (${\rm erg~cm^{-2}~s^{-1}}$) (adapted from Figure 9 in \citet{Liu2007}).}
\label{223}
\end{figure}

In order to reveal the spatial distribution of neutrino annihilation luminosity, we plot in Figure \ref{223} contours of ($2\pi r  l_{\nu \overline{\nu}}$) in units of (${\rm erg~cm^{-2}~s^{-1}}$) with cylindrical coordinates $r$ and $z$. It is easy to find that most of the annihilation events occur in the region of $r \lesssim 20~r_{\rm g}$, and it varies more rapidly along the $z$ coordinate than along the $r$ coordinate \citep[also see,][]{Popham1999}. The results indicate that the annihilation luminosity is anisotropic and most of the annihilation energy escapes outward along the angular momentum axis of the disk.

\section{Magnetized NDAFs}
\subsection{Magnetized BH-NDAFs}

\citet{Blandford1977} suggested that the rotational energy of a BH can be efficiently extracted to power a Poynting jet via a large-scale poloidal magnetic field threading the horizon of the BH. Besides neutrino annihilation above the NDAF, magnetohydrodynamics (MHD) processes, such as the BZ mechanism \citep{Blandford1977} can also power a relativistic jet launched from the central engine of GRBs, even for accretion rates much lower than $\dot M_{\rm ign}$ \citep[e.g.,][]{Lee2000a,Lee2000b,Globus2014,Pan2015}.

The BZ luminosity can be estimated as \citep[e.g.,][]{Lei2005,Krolik2011,Kawanaka2013b,Liu2015b}
\beq L_{\rm BZ} = f(a_*)c r_{\rm g}^2 \frac{B_{\rm in}^2}{8 \pi}, \eeq
where $f (a_*)$ is a factor depending on the specific configuration of the magnetic field including the information of the field configuration \citep[e.g.,][]{Tchekhovskoy2008,Kawanaka2013b}, and $B_{\rm in}$ is the poloidal magnetic field strength near the horizon. The analytical expression of $f (a_*)$ was attempted \citep[e.g.,][]{Blandford1977,Tchekhovskoy2008}, which is an increasing function of $a_*$, and in the range from a small number to $\sim 1$ \citep{Hawley2006}. Many 2- or 3-dimensional (2D or 3D) MHD simulations have investigated on how a large-scale vertical magnetic field evolves with an accretion disk and which configurations may power a jet \citep[e.g.,][]{McKinney2004,Beckwith2008,Beckwith2009,McKinney2009}, yet the analytic form of $f (a_*)$ is still unclear. As simplification, $f (a_*)=1$ is adopted \citep{Kawanaka2013b,Liu2015b}, which is suitable for the fast-spinning BH in the center of GRBs.

The magnetic field energy can be estimated by the disk pressure near the horizon $p_{\rm in}$, which is presented as
\beq \beta_h \frac{B_{\rm in}^2}{8\pi}= p_{\rm in}, \eeq
where $\beta_h \sim 1$ is the ratio of the midplane pressure near the horizon of the BH to the magnetic pressure in the stretched horizon.

\citet{Cao2014} calculated the global solutions of the NDAFs considering the radial advection and diffusion \citep[e.g.,][]{Balbus1998,Rossi2008} of the large-scale magnetic field. Although the configuration of the magnetic field and the structure of NDAFs affect each other, they found that only the structure of the inner NDAF changes significantly once magnetic fields are considered. For the magnetic field with $10^{14}~ \rm G$, the BZ jet luminosity can reach $\sim 10^{53}-10^{54}~\rm erg ~s^{-1}$ for an extreme Kerr BH. Furthermore, the thermal instability\footnote{The instabilities in NDAFs will be discussed in Subsection 6.2.} in magnetized NDAFs has been studied to explain the GRB variability \citep{Janiuk2010,Xie2017}.

\citet{Lei2013} studied the baryon loading problem of a GRB jet launched by a NDAF under either the neutrino annihilation process and BZ mechanism. They argued that in no matter mechanism, more luminous jets tends to be more baryon poor. Comparing with jet produced by annihilation, the magnetically dominated jet is much cleaner. Thus they suggested that at least a good fraction of GRBs should have a magnetically dominated central engine. Furthermore, the relation between the Lorentz factor $\Gamma$ and isotropic luminosity $L_{\rm iso}$ of the jet can be interpreted for both two models.

\citet{Yuan2012} investigated the closed magnetic field lines that continuously emerge out of the hyperaccretion flow. Since the differential rotation of the accretion flow is shear and turbulent, the line may form flux ropes. Once the system loses its equilibrium, the flux rope is thrust outward and then an episodic jet launches. This mechanism can provide enormous amount of energy to trigger GRBs.

Alternatively, the magnetic coupling (MC) from the BH horizon to the inner region of the disk \citep{Li2000}, which can effectively transfer the angular momentum and rotational energy of the BH to heat the inner region of the disk, then radiate a larger number of neutrinos from the disk than those from non-magnetized NDAFs to produce the primordial fireball \citep[e.g.,][]{Lei2005,Lei2009,Luo2013}. The MC effect affects the EoS and angular momentum of NDAFs, and effectively heats the inner region of the disk. In addition, the inner region of MC-NDAF becomes thermally and viscously unstable, which may be invoked to interpret the GRB variability. \citet{Xie2016} studied magnetized NDAFs with non-zero boundary stresses. They also found that the boundary torque has strong effects on the properties of inner disk. The neutrino annihilation luminosity of such NDAFs may be powerful enough to account for most of GRBs including the ultra-LGRBs (ULGRBs), and viscously unstable emerges in the inner region of the disk, which may be interpreted the origin of GRBs variabilities.

\citet{Wu2013} presented that a BZ jet launched from the BH hyperaccretion system in the fall-back framework \citep{Kumar2008a,Kumar2008b} can explain the giant X-ray bump in GRB 121027A. \citet{Gao2016} investigated that the BZ jet and Blandford-Payne \citep[BP,][]{Blandford1982} outflow can interpret the ULGRB GRB 111209A and its associated SN 2011kl.

\subsection{NS-NDAF system}

After merger of a compact object binary or collapse of a massive star, a stellar-mass BH around a hyperaccretion disk may form to produce GRBs. Another possible product is an accreting NS or magnetar instead of a BH \citep[e.g.,][]{Duncan1992,Usov1992,Dai1998,Zhang2001,Gao2006}. \citet{Chevalier1996} first analyzed a system consisting of a neutrino-cooled accretion disk and a NS. In order to maintain the balance of neutrino cooling and viscous heating, the viscous parameter should be small ($\lesssim 2 \times 10^{-6}$). Alternatively, a possible option to relax neutrino cooling rate is advection.

\citet{Zhang2008,Zhang2009} studied the radial structure of a NDAF around a NS based on a two-region scenario, and calculated the corresponding neutrino luminosity and annihilation luminosity. As a result, they found in the weak field case ($\lesssim 10^{15}~\rm G$), the structure of the outer disk is very similar to that of a BH NDAF. A NS has a solid surface, which is different from a BH. Most of the advection energy has to be released in the inner region near the NS surface, so that an outflow may be produced. Moreover, the NS NDAF has a brighter neutrino luminosity compared with the BH NDAF caused by the additional neutrino emission from the NS surface boundary layer, and the neutrino annihilation efficiency of a NS-NDAF system should be higher than that of BH-NDAF system. The luminosity $L_s$ from the boundary layer at the NS surface is \citep{Frank2002,Zhang2009}
\beq L_s \simeq \frac{GM \dot{M}}{4 r_{\rm NS}}(1-\frac{\Omega_{\rm in}}{\Omega_{\rm NS}})^2, \eeq
where $r_{\rm NS}$ and $\Omega_{\rm NS}$ are the radius and angular velocity of the NS, respectively, $\Omega_{\rm in}$ is the angular velocity of the disk inner boundary.

For the strong field case ($\gtrsim 10^{15}~\rm G$) corresponding to a magnetar, the NDAF properties may be significantly changed. \citet{Zhang2010} found that the quantum effects (Landau levels) and the MC processes play two competitive roles in changing the disk properties, with the latter being the main factor to increase pressure, density and neutrino luminosity with increasing magnetic field strength. The strange star-NDAF model has also been studied \citep{Hao2013}.

\section{Vertical structure and luminosity of NDAFs}

\begin{figure}
\centering
\includegraphics[angle=0,scale=1.0]{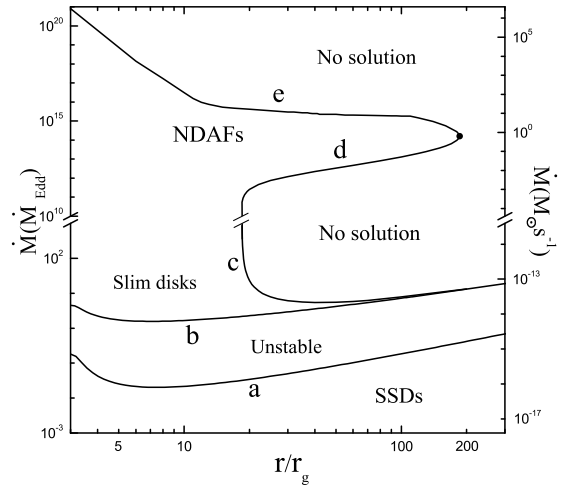}
\caption{Distribution of BH optically thick accretion disk solutions in the $\dot M$ - $r/r_{\rm g}$ plane, i.e., SSDs, slim disks, and NDAFs (adapted from Figure 1 in \citet{Liu2008}).}
\label{41}
\end{figure}

The simple well-known relationship ``$H = c_{\rm s}/\Omega_{\rm K}$" (or ``$H \Omega_{\rm K}/c_{\rm s} = {\rm constant}$") was widely adopted in the description of the vertical structure of accretion disks. Such a relationship can be derived from the simply vertical hydrostatic equilibrium
\beq \frac{\partial \psi}{\partial z} + \frac{1}{\rho} \frac{\partial p}{\partial z} =0, \eeq
with two additional assumptions in cylindrical coordinates ($r$, $\phi$, $z$) adopted, i.e., the approximated H\={o}shi form of gravitational potential \citep[e.g.,][]{Hoshi1977}
\beq \psi(r,z) \simeq \psi(r,0) + \Omega_{\rm K}^2 z^2/2, \eeq
and a one-zone approximation or a polytropic relation in the vertical direction \citep[e.g.,][]{Hoshi1977}
\beq p = {\mathcal K} \rho^{1+1/N}. \eeq
Obviously, the above assumptions work well for geometrically thin disks, but may be inaccurate when the mass accretion rate $\dot M$ approaches the Eddington rate $\dot M_{\rm Edd}$, for which the disk is likely not thin as shown in Figure 1. Consequently, the relationship ``$H = c_{\rm s}/\Omega_{\rm K}$" may be invalid for $\dot M \gtrsim \dot {M}_{\rm Edd}$ \citep{Gu2007,Gu2009,Liu2010a,Gu2012,Gu2016}, as well as $\dot M \gg \dot {M}_{\rm Edd}$ for NDAFs. The vertical structure of NDAFs should be revisited by using a more accurate description of the gravitational potential and vertically radiative transfer. The results would also affect the value of the neutrino luminosity.

\subsection{Gravitational potential}

\begin{figure*}
\centering
\includegraphics[angle=0,scale=0.496]{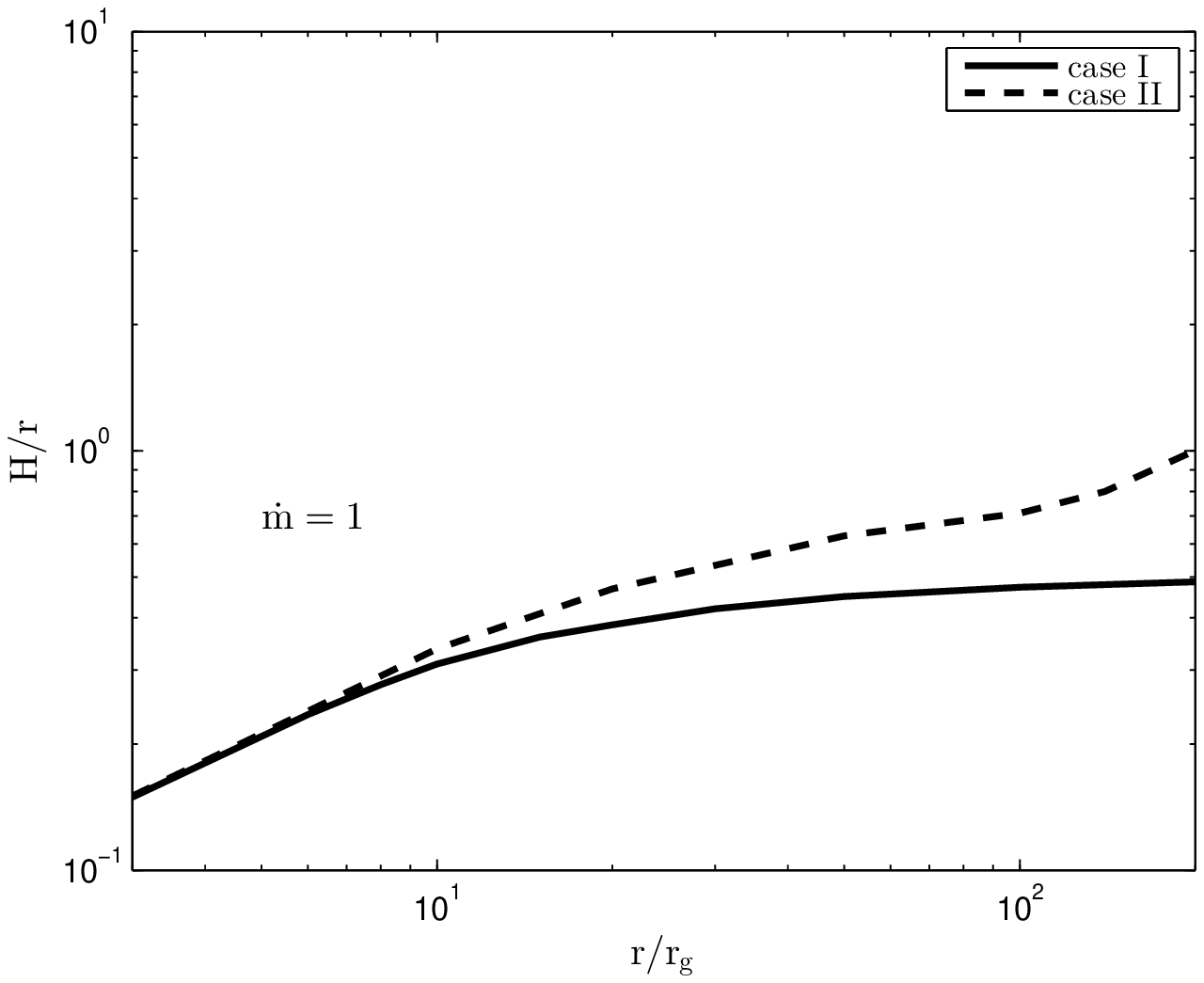}
\includegraphics[angle=0,scale=0.5]{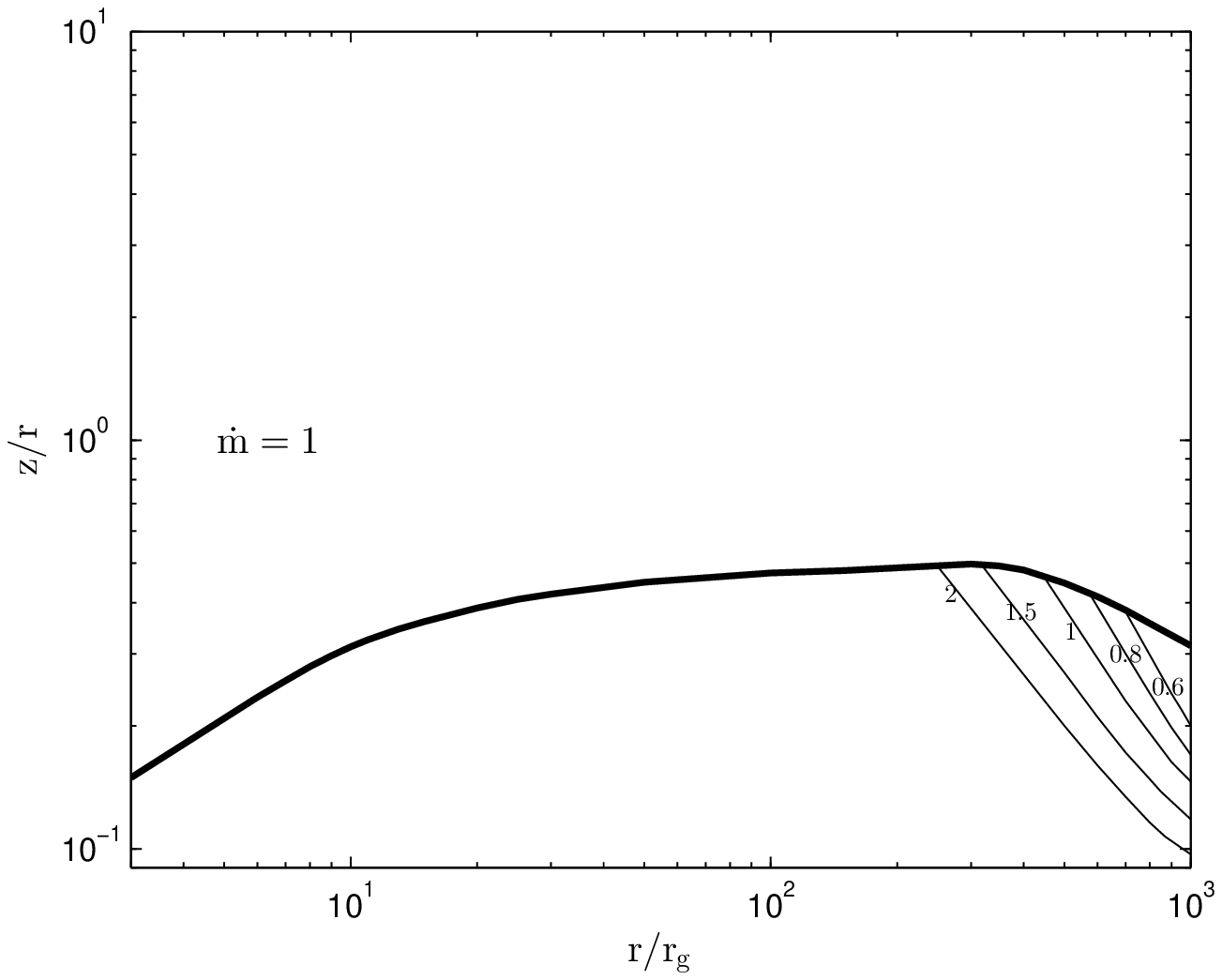}
\caption{Left panel: relative thickness $H/r$ as a function of $r$ for $\dot{m}=1$; Right panel: contours of Toomre parameter $Q_{\rm T}$ with cylindrical coordinates for $\dot{m}=1$ (adapted from Figures 3 and 5 in \citet{Liu2014b}).}
\label{42}
\end{figure*}

As discussion in Subsection 2.1.1, the hydrodynamics of NDAFs are expected to be similar to those of slim disks. If the angular velocity $\Omega=\Omega_{\rm K}$, the continuity and angular momentum equations of slim disks and NDAFs in cylindrical coordinates can be simplified as \citep{Liu2008}
\beq {\dot M} =-2 \pi r \Sigma v_r = {\rm constant}, \eeq
\beq {\dot M} (\Omega_{\rm K} r^2-j) = 2 \pi \alpha r^2 \Pi, \eeq
where $\Omega_{\rm K}=(GM/r)^{1/2}/(r-r_{\rm g})$ is the Keplerian angular velocity, $j=1.8c r_{\rm g}$ is an integration constant representing the specific angular momentum accreted by the BH, and $\Sigma$ and $\Pi$ are the surface density and vertically integrated pressure, respectively, which can be defined as
\beq \Sigma =2 \int_{0}^{\infty} {\rho} {\rm d} z,\eeq
\beq \Pi =2 \int_{0}^{\infty} {p} {\rm d} z.\eeq

The vertical hydrostatic equilibrium equation is provided by Equation (55), and here we adopt Paczy\'nski-Wiita potential \citep{Paczynski1980}
\beq \psi (r,z) = - \frac {G M}{\sqrt{r^2+z^2}-r_{\rm g}},\eeq
to replace the H\={o}shi form. This modification reverts the real effects of gravity in the vertical direction of the disk, which can bring significant change to the structure of the geometrically thick disks \citep{Liu2008}.

The equation of energy conservation reads
\beq Q_{\rm vis}=Q_{\rm adv}+2Q^-. \eeq
Here the viscous heating rate is given by
\beq Q_{\rm vis} =\frac{1}{2 \pi}{\dot M} \Omega_{\rm K} ^2 f g, \eeq
where $f = 1 - j/\Omega_{\rm K} r^2$, and $g = - {\rm d ln} \Omega_{\rm K}/{\rm d ln} r$. The advective cooling rate is
\beq Q_{\rm adv} =\frac{1}{2\pi} \frac{\xi {\dot M} {c_{\rm s}}^2}{r^2}, \eeq
with $\xi=3/2$ being a dimensionless quantity of the order of unity \citep[e.g.,][]{Kato2008,Liu2008}. The sound speed is further defined as $c_{\rm s} = (\Pi/\Sigma)^{1 / 2}$. The half thickness of the disk can be estimated via $H=\Sigma/2 \rho_0$, where $\rho_0$ is the mass density on the equatorial plane. Note that we can roughly use $Q_\nu$ instead of $Q^-$ here \citep{Liu2008}.

Figure \ref{41} shows thermal equilibria of SSDs, slim disks, and NDAFs at each radius $r$ with the corresponding accretion rate $\dot M$, which is plotted in units of the Eddington accretion rate ${\dot M}_{\rm Edd} = 64 \pi GM / c \kappa_{\rm es}$, where $\kappa_{\rm es} = 0.34 ~{\rm cm}^2 ~{\rm g}^{-1}$ is the electron scattering opacity, and $M_\odot$ s$^{-1}$ , respectively. The $\dot M$-$r$ plane can be divided into five regions: (1) The region below line $a$ corresponds to the stable, photon radiation-cooled, and gas pressure-supported SSDs; (2) the region between lines $a$ and $b$ corresponds to the unstable, photon radiation pressure-supported SSDs; (3) the region between lines $b$ and $c$ corresponds to the stable, advective cooling-dominated, and photon radiation pressure-supported slim disks \citep{Gu2007}; (4) line $c$ extends upward till line $d$, and the region between lines $d$ and $e$ corresponds to NDAFs; (5) the region on the right of lines $c$, $d$, and $e$ represents the `no solution' region.

Line $c$ represents a maximal possible accretion rate of slim disks for each radius, and lines $d$ and $e$ represent the lower and upper limits of $\dot M$ needed for NDAFs. They are defined by that no thermal equilibrium solutions exist because of viscous heating being always larger than total cooling. The filled circle at the connection between lines $d$ and $e$ at $r \approx 185$$r_{\rm g}$ defines the maximal possible outer boundary of an NDAF. The physical reason is that the correct BH's gravitational force in the vertical direction can only gather a limited amount of accreted gas. Once the pressure force is larger than the gravitational force, the balance is broken and outflows are produced. Outflows fall back to the BH, restart accretion process and may launch X-ray flares. We also notice that there is no boundary separating slim disks and NDAFs. It is easy to understand because both are very optically thick for photons. Along with the increase of the accretion rate, the neutrino emission processes operate and gradually become important. Thus the accretion flow changes from the slim disk form to the NDAF form. This picture is consistent with the results in the $\dot{M}$-$\Sigma$ plane in Figure 1. It should be noticed that the constraint on the accretion rate of NDAFs in this framework may be too tight because $H \sim r$ in cylindrical coordinates is considered.

\subsection{Self-gravity}

If the mass density of a disk becomes comparable to $M/r^3$, self-gravity becomes important and its resulting local instabilities may develop \citep[see, e.g.,][]{Paczynski1978a,Paczynski1978b,Abramowicz1984,Goodman1988}. The effects of self-gravity are generally important in some astrophysical processes, such as AGNs \citep[e.g.,][]{King2008,Hopkins2010} and protostars and protostellar disks \citep[e.g.,][]{Goodman2003,McKee2007,Rice2010}. Due to the high density of NDAFs, \citet{Liu2014b} argued that the self-gravity effect may be important to the structure of NDAF, and the neutrino luminosity.

First, we define the surface density of the disk in the range from the equatorial plane to a certain height $z$ ($z\leq H$),
\beq \Sigma_z = \int_{0}^{z} {\rho} {\rm d} z'. \eeq
We assume that it varies slowly with radius. A new term $4\pi G \Sigma_z$ should be added in the vertical equilibrium equation to represents the effect of self-gravity. Thus, Equation (55) can be rewritten as
\beq 4\pi G \Sigma_z + \frac{\partial \psi}{\partial z} + \frac{1}{\rho} \frac{\partial p}{\partial z} =0. \eeq
For comparison, we define cases I and II, which correspond to the vertical equilibrium forms with and without self-gravity, respectively.

The Toomre parameter is introduced to measure the local gravitational stability of the accretion disks, which is expressed as
\beq Q_{\rm T}=\frac{c_{\rm s} \Omega_{\rm K}}{\pi G \Sigma_z}, \eeq
where $Q_{\rm T}<1$ implies instability. If the effects of the self-gravity are considered in the vertical structure of NDAFs, the gravitational instability should be also reviewed in the framework.

As shown in the left panel of Figure \ref{42}, the effects of the self-gravity are mainly relevant in the outer region of the disk, especially for higher accretion rates. The right panel displays the equal $Q$ contours against the disk structure. One can see that $Q=1$ appears at the disk surface at $r/r_{\rm g}\sim250$ for an accretion rate $1M_\odot~\rm s^{-1}$. Since the self-gravity has limited effects on the structure of NDAF, there is also no significant influence on the neutrino luminosity \citep{Liu2014b}. The criterion, $Q<1$, indicates that the disk is gravitationally unstable, which may cause two classes of possible behaviors \citep[e.g.,][]{Perna2006}: (1) If the local cooling of the disk is rapid, the disk may fragment into two or more parts \citep[e.g.,][]{Nelson2000}. Since the fall back timescale is long enough, the accretion processes would restart, which may be related to the origin of late-time X-ray flares in GRBs. (2) The disk may evolve to a quasi-steady spiral structure transferring angular momentum outward and mass inward. This mode may drive long-duration, violent explosions if the disk mass is large enough \citep[e.g.,][]{Lodato2005}, which may give rise to SGRBs with extended emission \citep[e.g.,][]{Liu2012b,Cao2014} or X-ray flares in LGRBs.

\subsection{Vertical hydrostatic equilibrium}

\begin{figure}
\centering
\includegraphics[angle=0,scale=0.6]{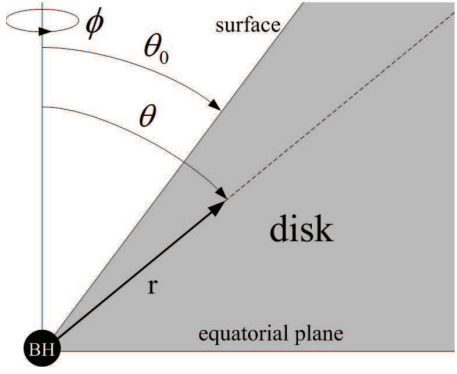}
\caption{Schematic picture of the system composed of a BH and an accretion disk in spherical coordinates (adapted from Figure 1 in \citet{Liu2011} and Figure 1 in \citet{Liu2013}).}
\label{431}
\end{figure}

The above work is based on the simple vertical hydrostatic equilibrium equation as shown in Equation (55), instead of the general form \citep{Abramowicz1997,Gu2009,Liu2010a},
\beq \frac{1}{\rho} \frac{\partial p}{\partial z} + \frac{\partial \psi}{\partial z} + v_r \frac{\partial v_z}{\partial r} + v_z \frac{\partial v_z}{\partial z} = 0, \eeq
where $v_z$ is the vertical velocity. Since $v_z$ is not negligible for geometrically thick disks, the solutions in \citet{Gu2007} and \citet{Liu2008} are still not self-consistent. To resolve this issue, we revisit the vertical structure of NDAFs in spherical coordinates ($r$, $\theta$, $\phi$) substituting cylindrical coordinates as shown in Figure \ref{431}.

We adopt the radial self-similar assumptions \citep{Narayan1995a} to simplify the basic equations of continuity and momentum \citep[see, e.g.,][]{Xue2005,Gu2009}, then obtain
\beq \frac{1}{2} {v_r}^2 + \frac{5}{2} {c_s}^2 + {v_\phi}^2 - r^2 {\Omega_{\rm K}}^2 = 0, \eeq
\beq \frac{1}{\rho} \frac{d p}{d \theta} = {v_\phi}^2 \cot \theta, \eeq
\beq v_r =-\frac{3}{2} \frac{\alpha {c_s}^2}{r \Omega_{\rm K}}, \eeq
where $v_r$ and $v_\phi$ are the radial and azimuthal components of the velocity ($v_\theta = 0$), and the sound speed $c_s$ is defined as $c_s^2 = p/\rho$, the Keplerian angular velocity is $\Omega_{\rm K} = (GM/r^3)^{1/2}$ here. The continuity equation can be adapted:
\beq \dot{M} = -4 \pi r^2 \int_{\theta_0}^{\frac{\pi}{2}} \rho v_r \sin \theta d \theta, \eeq
where $\theta_0$ is the polar angle of the surface.

\begin{figure*}
\centering
\includegraphics[angle=0,scale=0.2]{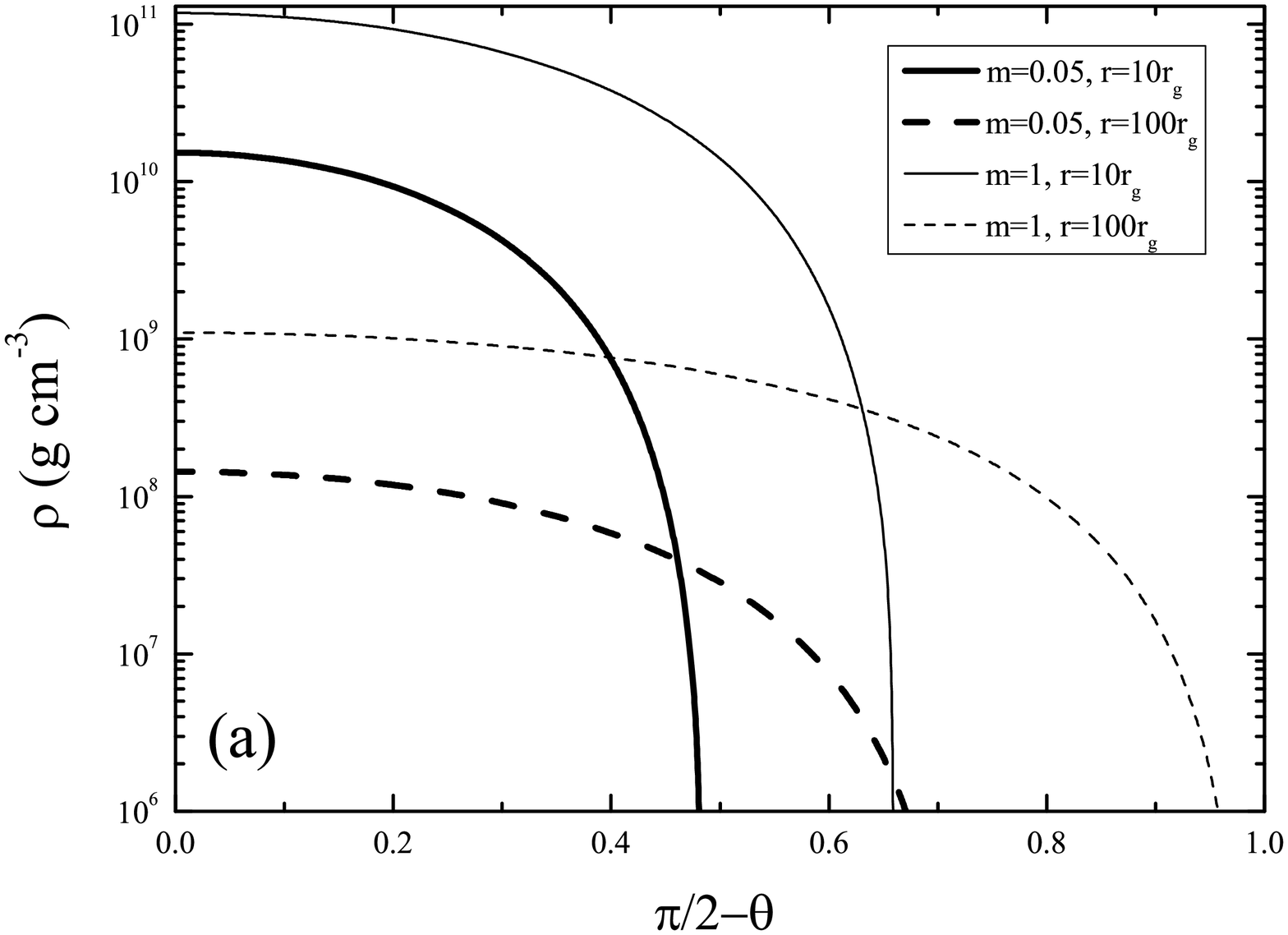}
\includegraphics[angle=0,scale=0.2]{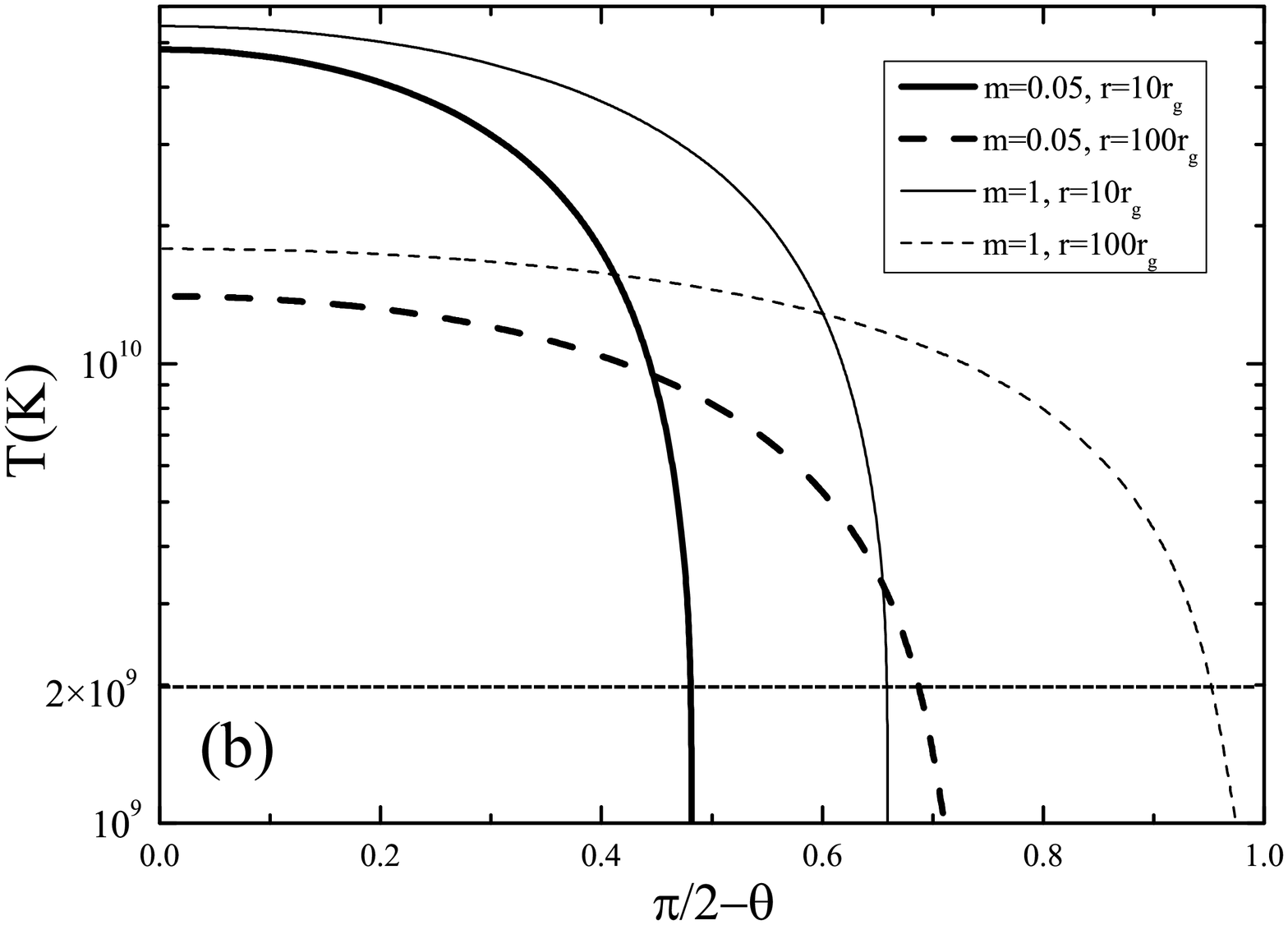}
\includegraphics[angle=0,scale=0.2]{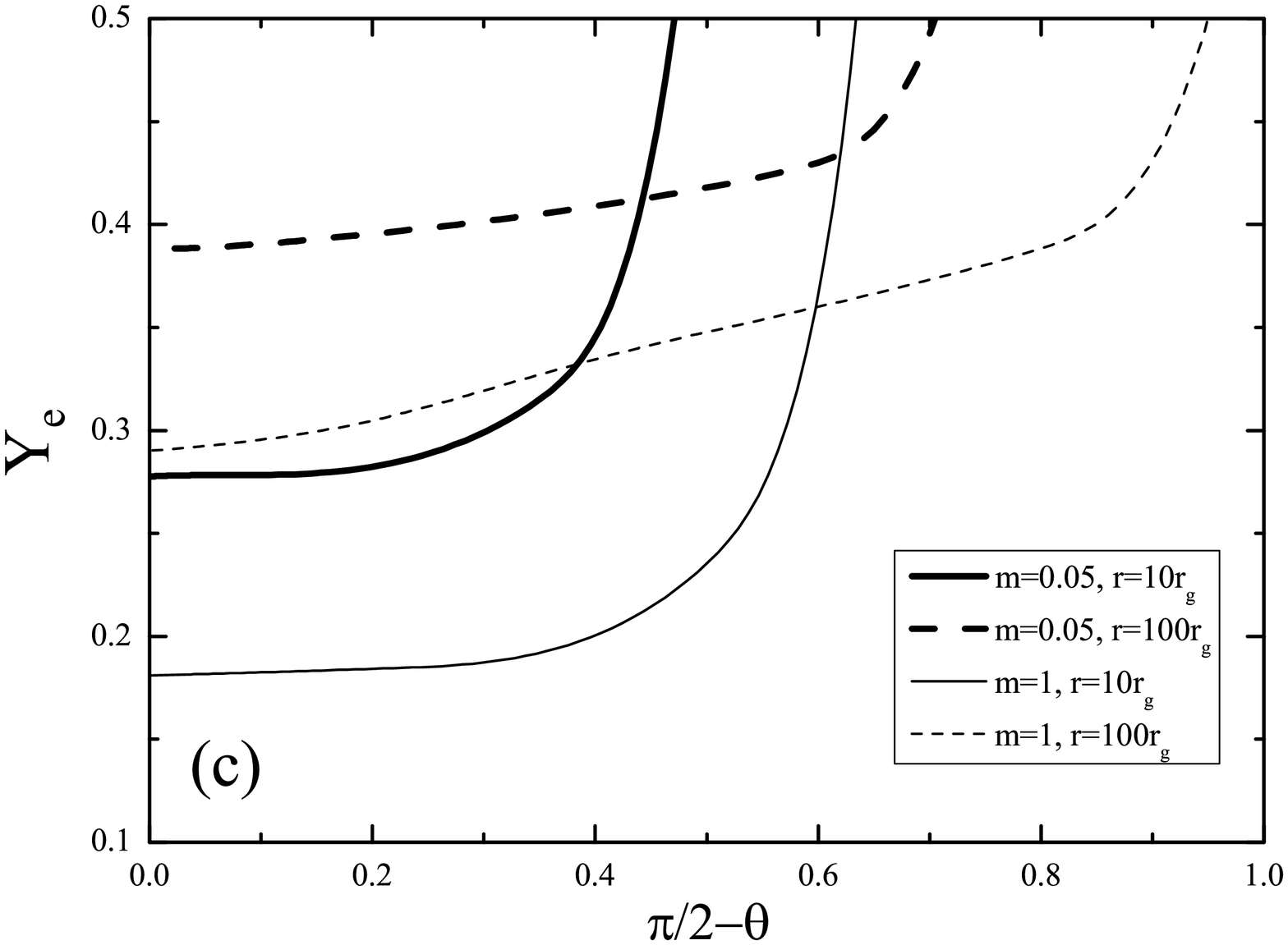}
\caption{Variations of density $\rho$, temperature $T$ and electron fraction $Y_{\rm e}$ with $\theta$ at $r=10r_g$ (solid lines) and $100r_g$ (dashed lines) for $m=0.05$ (thick lines) and $1$ (thin lines) (adapted from Figure 1 in \citet{Liu2013}).}
\label{432}
\end{figure*}

The EoS follows Equation (33), and the energy equation is written as
\beq Q_{\rm vis} = Q_{\rm adv} + Q_\nu. \eeq
where the cooling of photodisintegration of $\alpha$-particles is ignored \citep{Liu2010a,Liu2012a}. The viscous heating rate per unit volume $q_{\rm vis} = \nu \rho r^2 [\partial (v_{\phi}/r)/\partial r]^2$ ($\nu$ is kinematic coefficient of viscosity) and the advective cooling rate per unit volume $q_{\rm adv} = \rho v_r (\partial e/\partial r - (p/\rho^2) \partial \rho/\partial r)$ ($e$ is the internal energy per unit volume) are expressed, after self-similar simplification, we obtain
\beq q_{\rm vis} = \frac{9}{4} \frac{\alpha p v_{\phi}^2}{r^2 \Omega_{\rm K}}, \eeq
\beq q_{\rm adv} = - \frac{3}{2} \frac{(p-p_{\rm e}) v_r}{r}, \eeq
where the entropy of degenerate particles is neglected. Thus the vertical integration of $Q_{\rm vis}$ and $Q_{\rm adv}$ are the following \citep{Liu2010a,Liu2012a}:
\beq Q_{\rm vis} = 2 \int_{\theta_0}^{\frac{\pi}{2}} q_{\rm vis}  r \sin {\theta} d\theta \ , \eeq
\beq Q_{\rm adv} = 2 \int_{\theta_0}^{\frac{\pi}{2}} q_{\rm adv}  r \sin {\theta} d\theta . \eeq
The cooling rate due to neutrino radiation $Q_\nu$ can be defined as \citep{Lee2005,Liu2012a}
\beq Q_\nu = 2 \displaystyle{\sum_{k}}  \int_{\theta_0}^{\frac{\pi}{2}} q_{\nu_k} {\rm e}^{-\tau_{\nu_k}} r \sin {\theta} d\theta, \eeq
where $k$ represents different types of neutrinos and antineutrinos, and $q_{\nu_k}$ is the sum of the cooling rates per unit volume due to the Urca processes, electron-positron pair annihilation, nucleon-nucleon bremsstrahlung, and plasmon decay, as introduced in Subsection 2.1.2. The neutrino optical depth is calculated by using integral to $\theta$ instead of the forms by timing the half thickness of the disk $H$ in Equation (10).

Once gravity cannot hold the radiation pressure gradient, outflows may occur, so a boundary condition is essential. Analogous to the principle of Eddington luminosity, the mechanical equilibrium can be written as \citep{Liu2012a}
\beq p_{\rm rad} \mid _{\theta=\theta_0} \sigma_{\rm T} = \frac{2GMm_u}{r^2}\rm cot \theta_0, \eeq
combined with the photon radiation pressure, the surface temperature can be derived as \citep{Liu2012a}
\beq T \mid _{\theta=\theta_0}=(\frac{6 G M m_{u}}{a \sigma_{\rm T} r^2} \rm cot \theta_0)^{\frac{1}{4}}, \eeq
where $\sigma_{\rm T}$ is the Thompson scattering cross section.

Figure \ref{432} shows the variations of density $\rho$, temperature $T$ and electron fraction $Y_{\rm e}$ with $\pi/2-\theta$. The lower limit of the temperature is $\sim 2\times10^9 ~\rm K$ in NSE equations, which is marked in Figure \ref{432}(b). The values $\dot{M} = 0.05 ~M_\odot ~\rm s^{-1}$ and $\dot{M} = 1 ~M_\odot ~\rm s^{-1}$ correspond to $Y_{\rm e}$ around 0.49 and 0.47 near the disk surface, respectively. The half-opening angles of the disks have the positive correlation with accretion rate and radius.

\subsection{Outflow}

\begin{figure*}
\centering
\includegraphics[angle=0,scale=0.31]{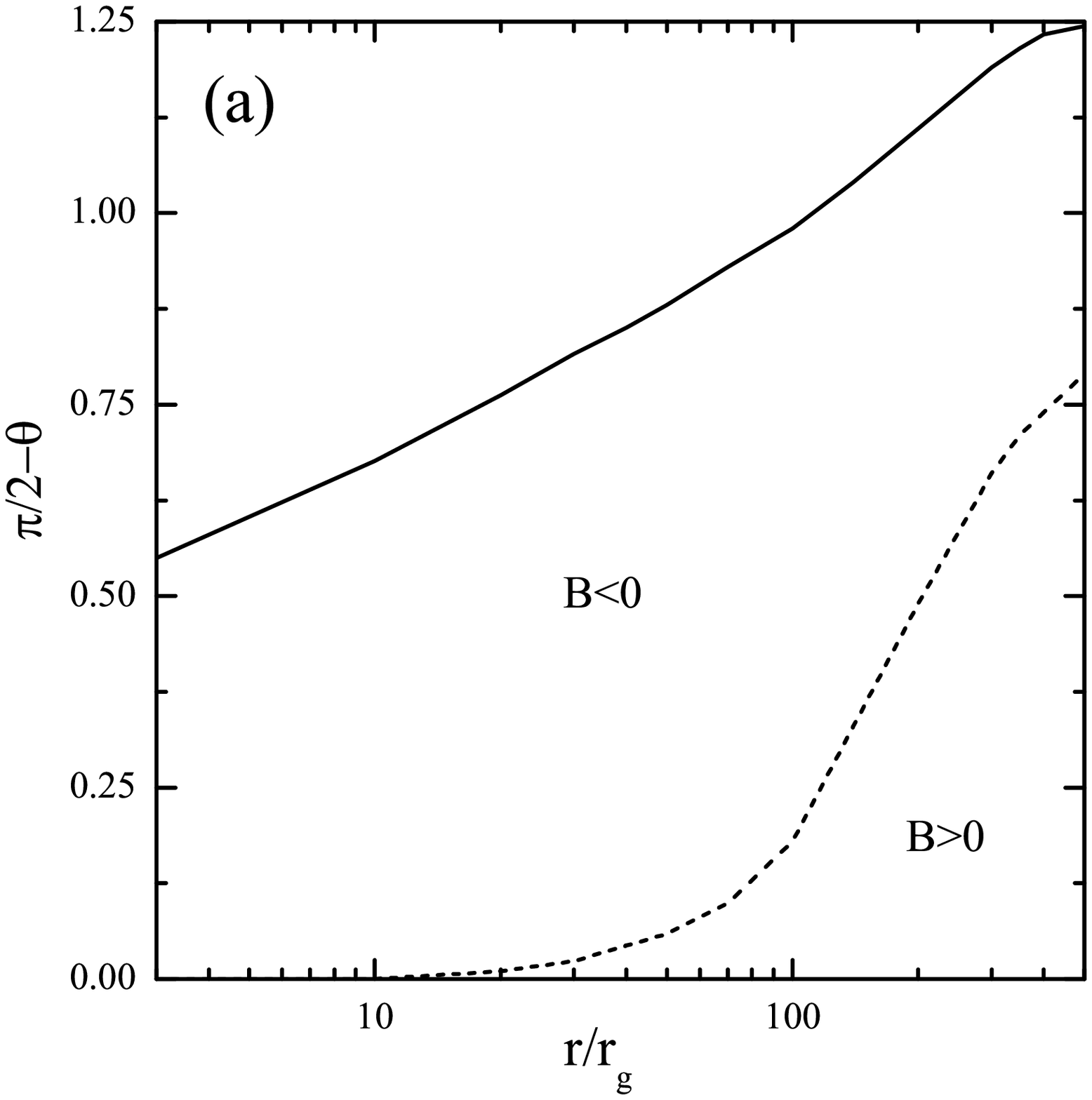}
\includegraphics[angle=0,scale=0.31]{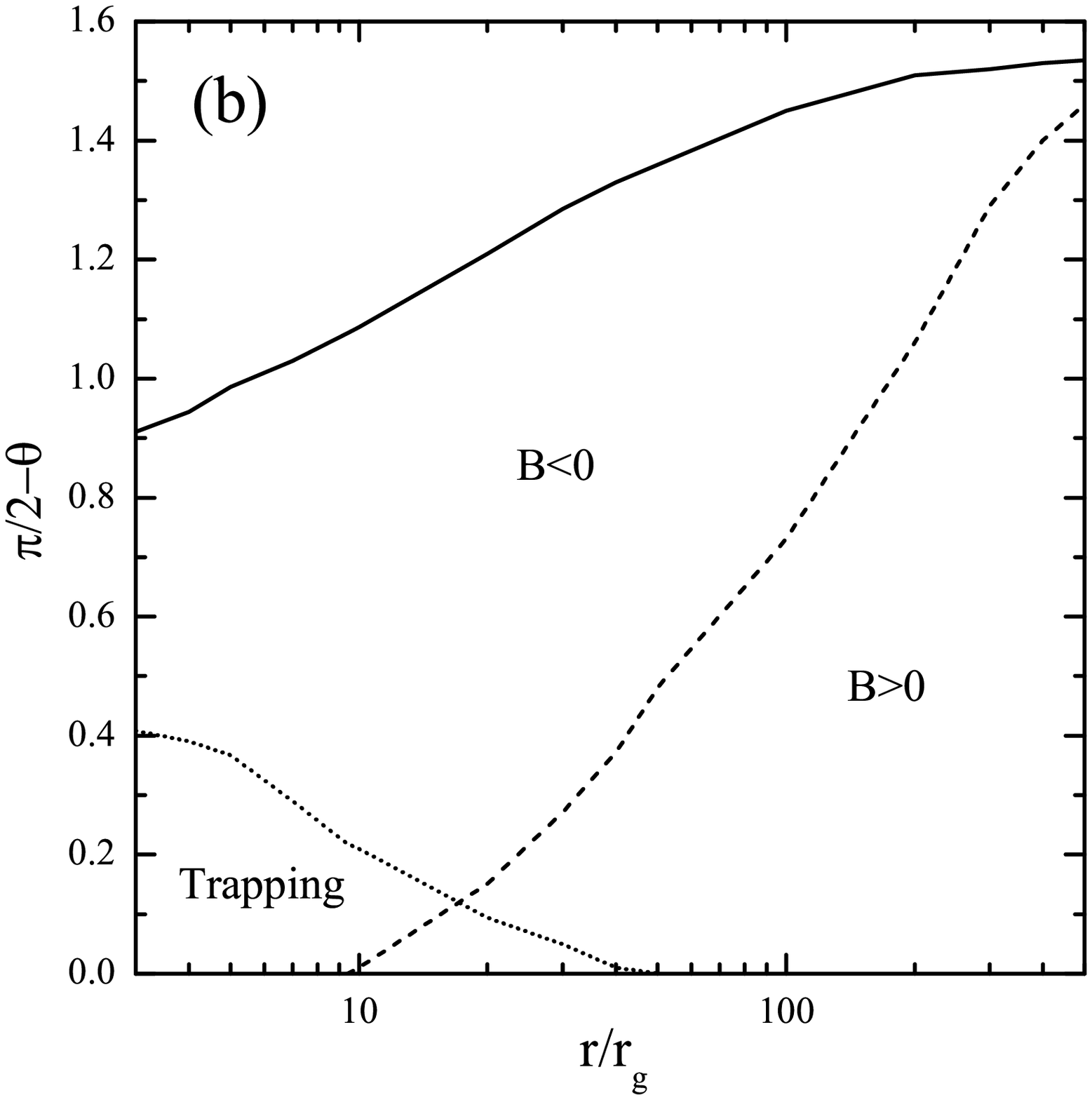}
\caption{Variations of the half-opening angles of the disk surface $(\pi/2-\theta_0)$ (solid lines), the $B=0$ surface (dashed lines), and $\tilde{t}=1$ (dotted line) with radius $r/r_g$ for (a) $\alpha=0.1$ and $\dot{M}=1~M_\odot ~\rm s^{-1}$; (b) $\alpha=0.01$ and $\dot{M}=10~M_\odot ~\rm s^{-1}$ (adapted from Figure 1 in \citet{Liu2012a}).}
\label{44}
\end{figure*}

The outflow from the critical accretion disk must be strong, which has been repeatedly demonstrated by theories \citep[e.g.,][]{Liu2008,Gu2015}, simulations \citep[e.g.,][]{Jiang2014,Sadowski2015}, and observations \citep[e.g.,][]{WangQ2013,Cheung2016,Parker2017}. In order to discuss the possible outflow from NDAFs, we introduce the Bernoulli parameter of the accreted matter, which is expressed as \citep[e.g.,][]{Narayan1995a}
\beq B=\frac{\gamma}{\gamma-1}{c_{\rm s}}^2 + \frac{1}{2}({v_r}^2+{v_\phi}^2)-\frac{GM}{r}. \eeq
This equation reflects the balance of energy, including the kinetic energy, potential energy and enthalpy of accreted matter. The region where $B>0$ is satisfied implies a possible outflow. Conversely, $B<0$ everywhere is a sufficient condition for the absence of an outflow \citep{Abramowicz2000}.

Figure \ref{44} shows the variation of the half-opening angles of the disk surface $(\pi/2-\theta_0)$, and the angle at which $B=0$ is satisfied, with radii. Figure \ref{44}(a) shows that for $\alpha$ = 0.1 and $\dot{M}$ = 1 $M_\odot~\rm s^{-1}$ the region with $B>0$ appears at $r \sim 10~r_{\rm g}$, and expands continuously with increasing radius. There is no region of neutrino trapping. In Figure \ref{44}(b), for $\alpha$ = 0.01 and $\dot{M}$ = 10 $M_\odot~\rm s^{-1}$ the region with $B>0$ appears at $\sim 9~r_{\rm g}$, and increases to near the surface of the disk at $\sim 500~r_{\rm g}$.

In the research on the vertical structure of NDAFs, we can further discuss the effects of neutrino trapping. Different from the trapping radius in radial structure of NDAFs, by equating the neutrino diffusing time and the accretion time, we can obtain a region of trapped neutrinos. Figure \ref{44}(b) shows the region of neutrino trapping occurs from inner area to $\sim 46~r_{\rm g}$, and its opening angle is from $\sim 0.4$ to $0$. We consider that the radius and the open angle of $B=0$ and $\tilde{t}=1$ are closely determined by the accretion rate and the viscous parameter.

\subsection{Vertical composition}

\begin{figure*}
\centering
\includegraphics[angle=0,scale=0.24]{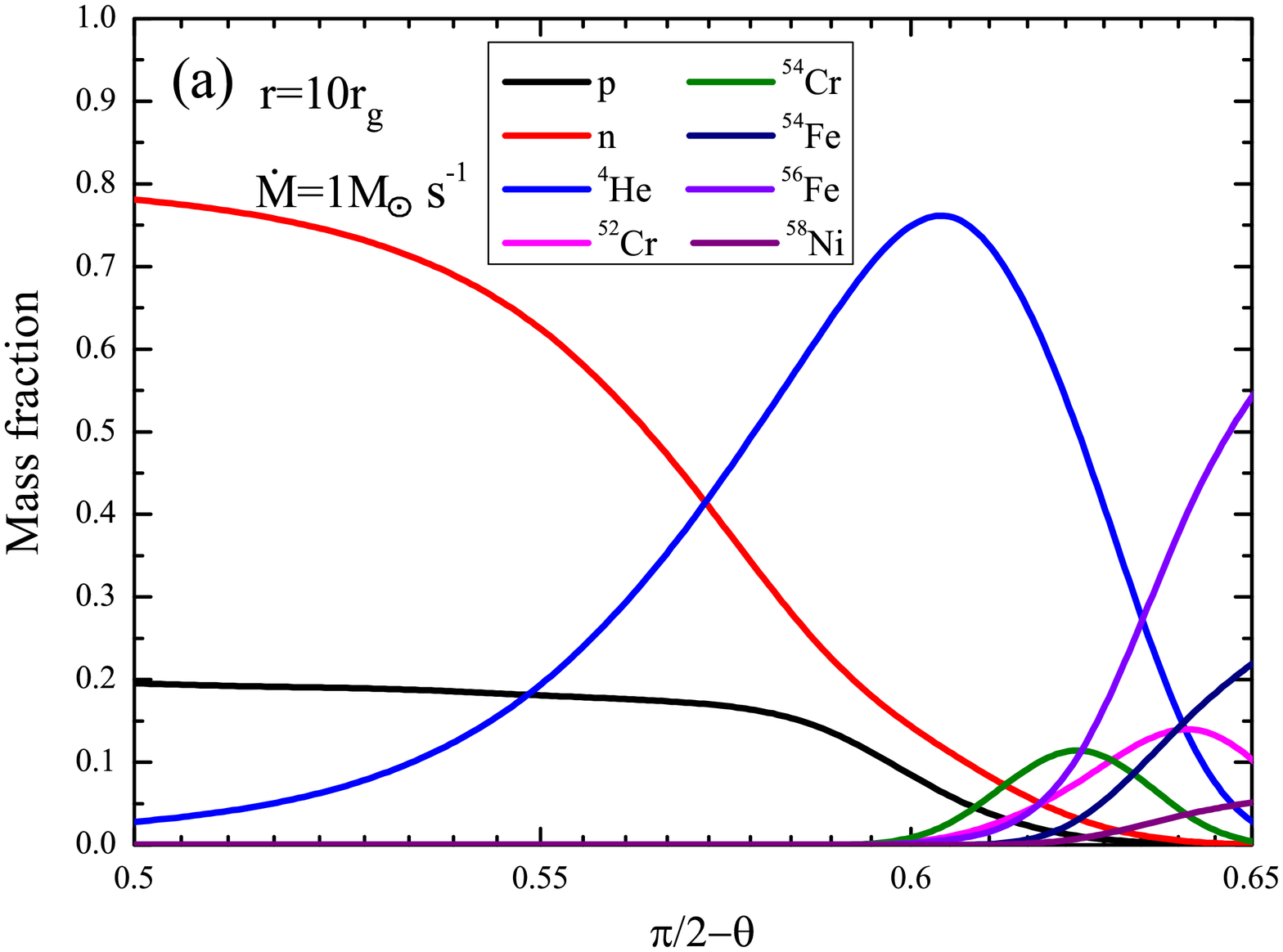}
\includegraphics[angle=0,scale=0.24]{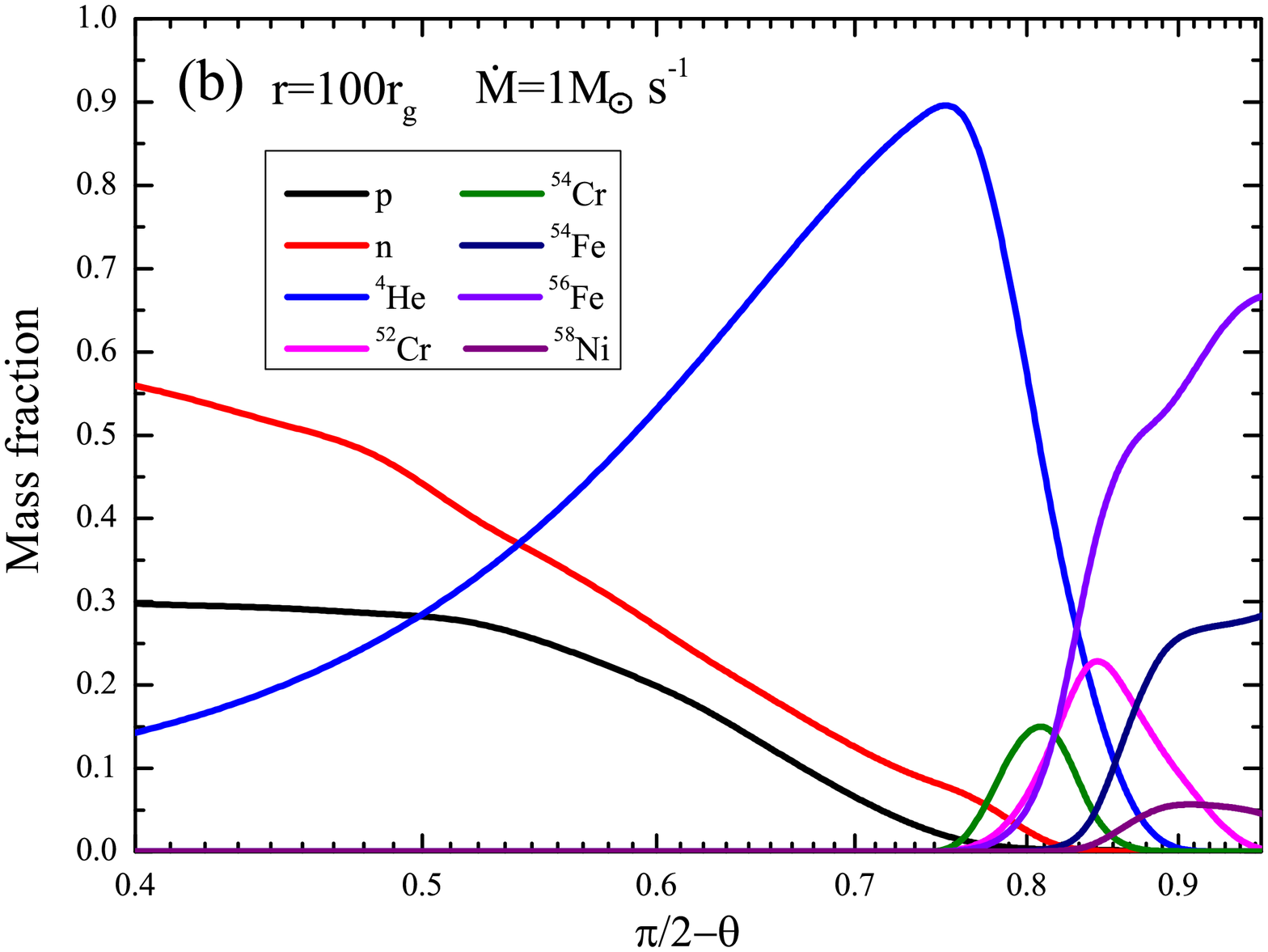}
\includegraphics[angle=0,scale=0.24]{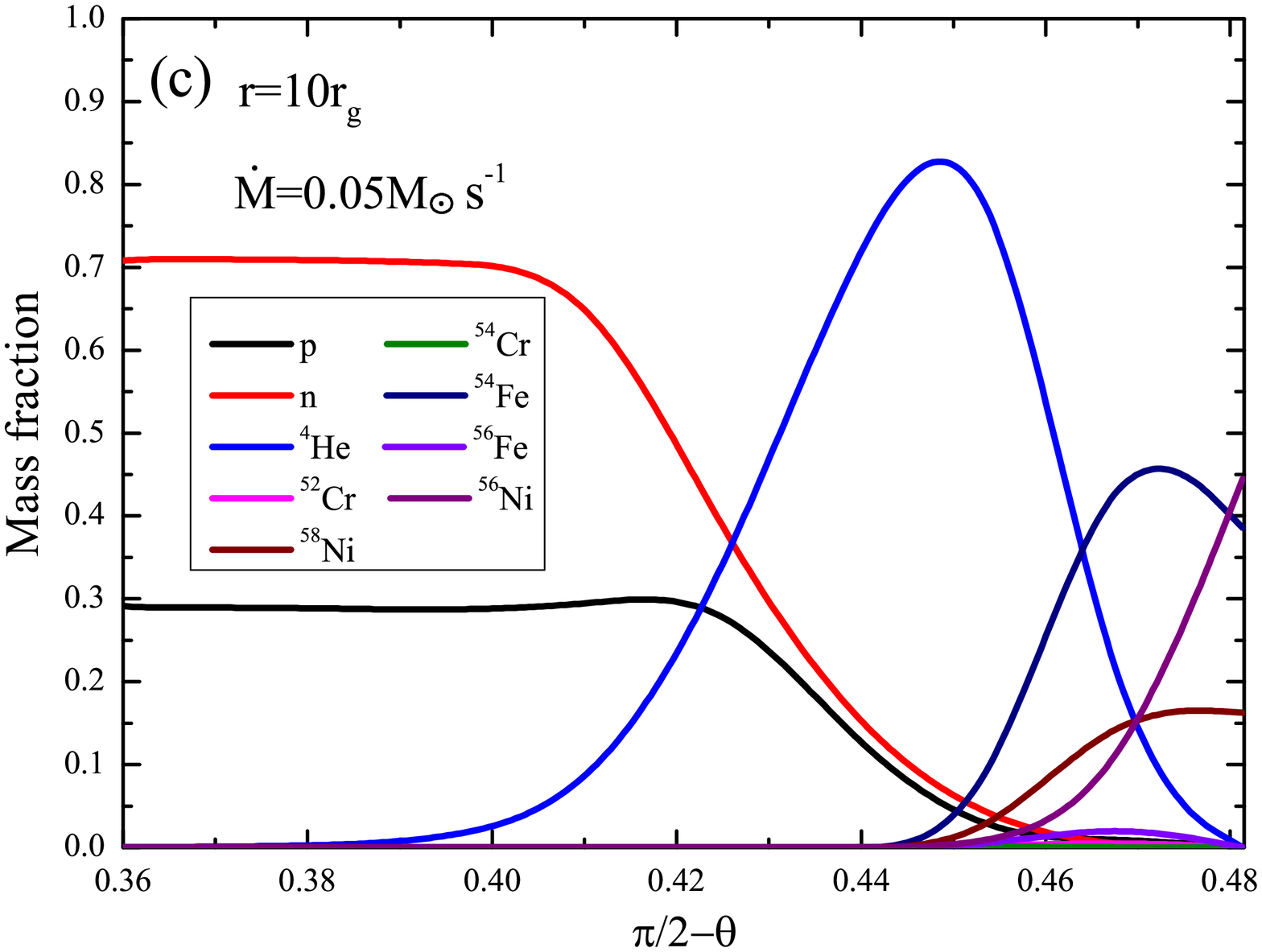}
\includegraphics[angle=0,scale=0.24]{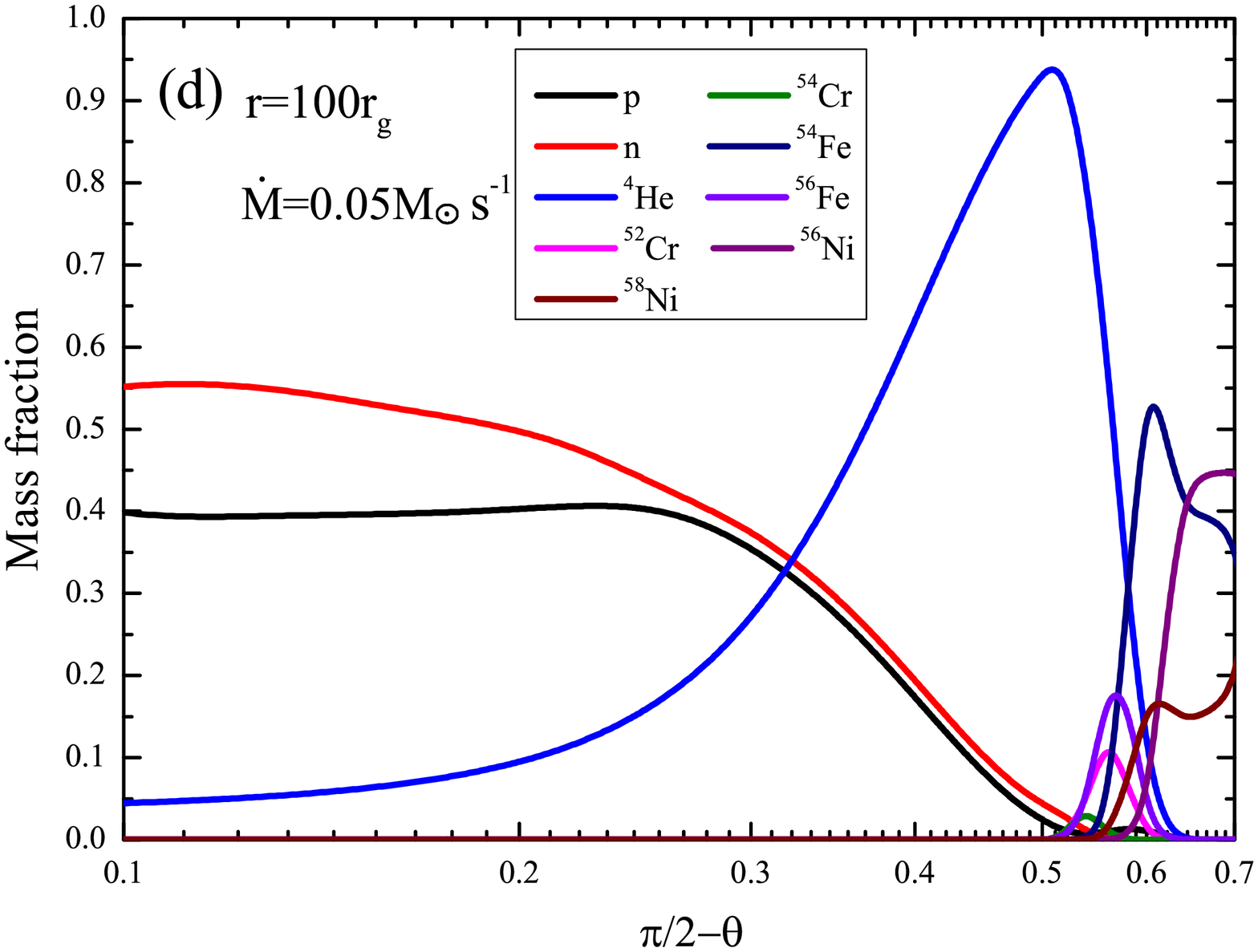}
\caption{Variations of the mass fraction of the main elements with $2\pi-\theta$ at $r=10r_g$ and $100r_g$ for $\dot{M} = 0.05 M_\odot ~\rm s^{-1}$ and $1 M_\odot ~\rm s^{-1}$ (adapted from Figure 2 in \citet{Liu2013}).}
\label{45}
\end{figure*}

In the NSE \citep{Seitenzahl2008,Liu2013,Xue2013}, the independent variables are the density $\rho$, temperature $T$ and electron fraction $Y_{\rm e}$, which are essential in the vertical NDAF model according to the description above. We thus obtain the vertical distribution of the mass fraction (also approximately equals the number density) of the free neutron and proton, and main elements (include $\rm ^4 He$, $\rm ^{52}Cr$, $\rm ^{54}Cr$, $\rm ^{54}Fe$, $\rm ^{56}Fe$, $\rm ^{56}Ni$, and $\rm ^{58}Ni$), as shown in Figure \ref{45}. $\rm ^{56}Ni$ dominates at the disk surface for $\dot{M} = 0.05 ~M_\odot ~\rm s^{-1}$, and $\rm ^{56}Fe$ dominates for $\dot{M} = 1 ~M_\odot ~\rm s^{-1}$, corresponding to $Y_{\rm e}$ around 0.49 and 0.47 as shown in Figure \ref{432}(c), respectively. The solutions show that the proportion of the nuclear matter increases with radius for the same accretion rate. The mass fraction of $\rm ^{56}Ni$ or $\rm ^{56}Fe$ near the surface increases with radius. In the middle region, $\rm ^4 He$ is dominant for all the accretion rates. The free neutrons and protons are dominant near the equatorial plane of the disk in the hot and dense state. Most of the free protons turn into free neutrons due to the Urca process \cite[see, e.g.,][]{Liu2007,Xue2013}, which causes the dominance of free neutrons and the decrease of electron fraction. In simple terms, the change of the electron fraction is inversely associated with the accretion rate and radius when the free baryons dominate.

\subsection{Convection}

\begin{figure}
\centering
\includegraphics[angle=0,scale=0.33]{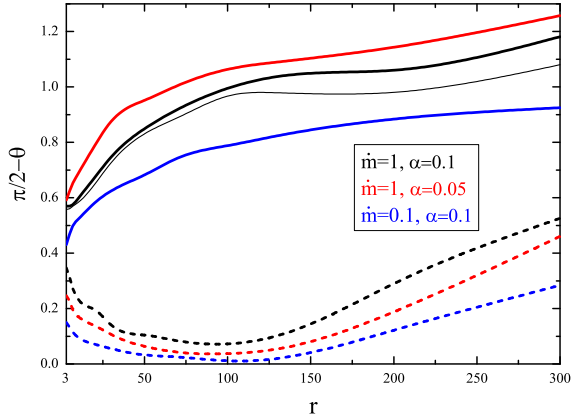}
\caption{Variations of the half-opening angle $(\pi /2 -\theta)$ with the dimensionless radius $r$ for the different cases. The black, red, and blue lines describe the cases that the given parameters ($\dot{m}$, $\alpha$) is (1, 0.1), (1, 0.05), and (0.1, 0.1), respectively. The solid and dashed lines correspond to $(\pi /2 -\theta_0)$ and $(\pi /2 -\theta_{\rm c})$. The thick and thin black solid lines correspond to the cases including and excluding the convection with $\dot{m}=1$ and $\alpha=0.1$ (adapted from Figure 3 in \citet{Liu2015a}).}
\label{461}
\end{figure}

In the vertical direction of the disk, the convective motion (or vertical advection) plays an important role, which can carry photons (or neutrinos) from the equatorial plane to the surface of the disk by magnetic buoyancy. For slim disk, the key issue is whether the vertical radiation transfer due to magnetic buoyancy is faster than the radial advection process, thus photons can escape from the surface before being advected into the BH. The results of \citet{Jiang2014} indicated that the radiation can be widely increased because of the vertical advection of radiation caused by magnetic buoyancy. We would like to examine whether the same mechanism exists in NDAFs to increase the efficiency of neutrino emission.

First of all, we estimate two typical timescales, the vertically convective timescale, $t_{\rm c}$, and the radially advective timescale, $t_{\rm adv}$. The vertical convective speed along the vertical direction can be approximatively expressed as \citep{Kawanaka2012}
\beq v_{\rm c} \simeq  - v_r \cos\theta. \eeq
This is the vertically ``infalling'' speed onto the equatorial plane due to the vertical component of the BH's gravity. Thus the convective timescale is
\beq t_{\rm c} \simeq \int_{\theta_0}^{\theta} \frac{r_0 d \cot \theta'}{v_{\rm c}} \simeq  \int_{\theta_0}^{\theta} \frac{r_0 d \theta'}{v_r \sin^2 \theta' \cos \theta'}, \eeq
where $r_0$ is the radius at the equatorial plane.

Meanwhile, the advection timescale (or accretion timescale $t_{\rm acc}$) can be expressed as
\beq t_{\rm adv} \simeq - \int_{3r_{\rm g}}^{r} \frac{d {r}'}{v'_r} - \frac{3r_{\rm g}}{v_r|_{r=3r_{\rm g}}}. \eeq

The vertical convection should exist if $t_{\rm c}<t_{\rm adv}$, i.e., the polar angle of the convective region should satisfy $\theta_0<\theta<\theta_{\rm c}$. Otherwise, convection would be destroyed by gravity. The critical polar angle $\theta_{\rm c}$ is defined by
\beq r \sin \theta_{\rm c} \int_{\theta_{\rm c}}^{\theta_0} \frac{d \theta}{v_r \sin^2 \theta \cos \theta} = \int_{3r_{\rm g}}^{r} \frac{d r'}{v'_r} + \frac{3r_{\rm g}}{v_r|_{r=3r_{\rm g}}}. \eeq

When vertical convection is included in the NDAF model and for $\theta_0 < \theta_{\rm c}$, Equation (78) should be rewritten as
\beq Q_{\rm adv} = 2 r \int_{\theta_{\rm c}}^{\pi/2} q_{\rm adv} \sin {\theta} d\theta, \eeq
which means that the advection has been suppressed.

Figure \ref{461} shows the variations of the half-opening angle $(\pi /2 -\theta)$ with the dimensionless radius $r$ for different cases. The half-opening angle of the disk in the case excluding vertical convection is similar to the solutions in \citet{Liu2012a,Liu2013}, which has a slight difference compared with the case including convection. As the cooling modes, advection and neutrino cooling dominate in the outer and inner region of the disk, respectively. Thus the effects of vertical convection mainly operate in the inner region until very near the BH. The vertically convective energy transfer can be effective to suppress radial advection in NDAFs, but the region dominated by vertical convection is deviated from the equatorial plane of the disk, thus it can be expected that the neutrino emission rate should increase slightly.

In the research of NDAFs in the vertical direction, the luminosity of neutrino annihilation $L_{\nu \bar{\nu}}$ can be roughly evaluated. We define the annihilation efficiency as $\eta \equiv L_{\nu \bar{\nu}}/L_\nu$, which satisfies $\eta \propto V_{\rm ann}^{-1}$ \citep[see, e.g.,][]{Mochkovitch1993,Liu2010a,Liu2015a}, where $V_{\rm ann}$ is the volume above the NDAF. We can estimate $V_{\rm ann}$ by integrating the region of $\theta < \theta_0$ and $r < r_{\rm out}$. Thus we can obtain the proportionality constant of the efficiency to the volume above the disk as a function of accretion rate from the vertically-integrated NDAF model where $\theta_0$ corresponds to $\pi/2$ \citep{Liu2007,Liu2015a}.

\begin{figure}
\centering
\includegraphics[angle=0,scale=0.33]{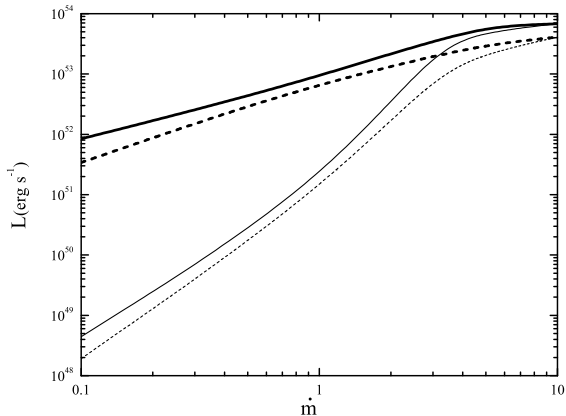}
\caption{Neutrino luminosity $L_\nu$ and annihilation luminosity $L_{\nu \bar{\nu}}$ as a function of dimensionless mass accretion rates $\dot{m}$. The solid and dashed lines correspond to the cases including and excluding the vertical convection, respectively. The thick and thin lines display the estimations of the neutrino luminosity and annihilation luminosity, respectively (adapted from Figure 4 in \citet{Liu2015a}).}
\label{462}
\end{figure}

Figure \ref{462} displays the neutrino luminosity $L_\nu$ and annihilation luminosity $L_{\nu \bar{\nu}}$ as a function of $\dot{m}$. The solid and dashed lines correspond to the cases including \citep{Liu2015a} and excluding  \citep{Liu2010a,Liu2012a} the vertical convection, respectively. The thick and thin lines correspond to the neutrino luminosity and annihilation luminosity. The certain increase of luminosity comes from the suppressed advection and correspondingly increased neutrino cooling. We also notice that for $\dot{M}\sim 5~M_\odot~{\rm s}^{-1}$, the density of radiated neutrino is so large that the annihilation efficiency is close to 1. The extremely thick disk restrains the ejection in a very narrow empty funnel along the rotation axis. Similar to the discussion in \citet{Liu2010a,Liu2015a}, thanks to the large opening angle of the disk and the sufficient output of energy, the neutrino annihilable ejection required by GRBs may be naturally realized by NDAFs.

\subsection{Radiative transfer}

At mentioned above, the polytropic relation should be substituted by the equations of neutrino radiative transfer. \citet{Sawyer2003} firstly proposed the two-stream approximation to solve the neutrino transport problem in the analysis of some accretion disks. \citet{Rossi2007} derived a very simple form of neutrino radiative transfer, which can be used in the solutions of NDAFs.

\citet{Pan2012a} calculated the 1D Boltzmann equation of neutrino transfer in the accretion disk and obtained the neutrino spectra by given the distribution of density, temperature and chemical components of the disk. They also verified that the approaches of \citet{Sawyer2003} is a good approximation. \citet{Pan2012b} further investigated the vertical structure, neutrino luminosity, and annihilation luminosity of NDAFs by considering neutrino radiative transfer, and they considered that the effects of the BH spin and magnetic field might be introduced in NDAF models to power GRBs.

\section{Numerical simulations of NDAFs}

The time-dependent 2D or 3D hydrodynamical simulations of hyperaccretion disks in BH-NS or NS-NS mergers and in collapsars have been widely studied \citep[e.g.,][]{Ruffert1999,Lee2004,Setiawan2006,Lee2009,Carballido2011,Sekiguchi2011,Caballero2012,Fernandez2013,Janiuk2013,Just2015,Fernandez2015,Foucart2015,Batta2016,Just2016}.

\citet{Janiuk2013} calculated the structure and violent evolution of a turbulent torus accreting onto a BH. In the 2D simulations, neutrino cooling makes the disk much denser, geometrically thinner and less magnetized. For the accretion rate $\dot{m}\sim$ 0.03-0.1, the neutrino luminosity reaches $10^{53}-10^{54}$ erg s$^{-1}$, which is 1-2 orders of magnitude larger than the BZ jet luminosity. The conclusion is similar to the results on the vertical structure and luminosity of NDAFs by \citet{Liu2010a,Liu2015a}. Moreover, the neutrino cooled torus launches a fast, rarefied wind that is responsible for a powerful mass outflow, the power of which is correlated with the net accretion rate. The neutrino cooling rates are similar for the inner $\sim$ 20-30 $r_{\rm g}$ in the 1D and 2D calculations. \citet{JaniukA2017} studied the general relativistic, MHD NDAF models including self-consistent nuclear EoS. The structure of inflows and outflows, neutrino radiation, and nucleosynthesis are studied. They argued that the central BH characteristics may be estimated by the observations on the decay of the radioactive elements in the inflows and outflows \citep[e.g.,][]{Surman2011}.

Different from the traditional simulation method in accretion disks, the 2D or 3D Lagrangian smooth particle hydrodynamics methods are also introduced in NDAFs \citep[e.g.,][]{Lee2009, Batta2016}. \citet{Batta2016} investigated the collapse and accretion onto BHs of spherically rotating envelopes by considering the angular momentum distribution. Contrary to the results obtained in previous 2D hydrodynamical simulations, they found that the collapsing gas with angular momentum between supercritical and subcritical causes the production of large quiescent times originating from the absence of an accretion disk near the BH. Moreover, the collapse of extremely subcritical materials on to the disk would result in a shutdown of the inner engine, if the critical angular momentum increases beyond the angular momentum of the materials in the disk before supercritical materials fill in.

\section{Applications to GRBs}

\subsection{Jet luminosity}

The most important motivation in study of NDAFs is to explain the source of GRB power. Two mechanisms, i.e., the neutrino annihilation process and BZ mechanism, have been widely discussed in the literatures. In the following, we confront both mechanisms against the GRB data.

\subsubsection{Neutrino annihilation for SGRBs}

\begin{figure}
\centering
\includegraphics[angle=0,scale=0.4]{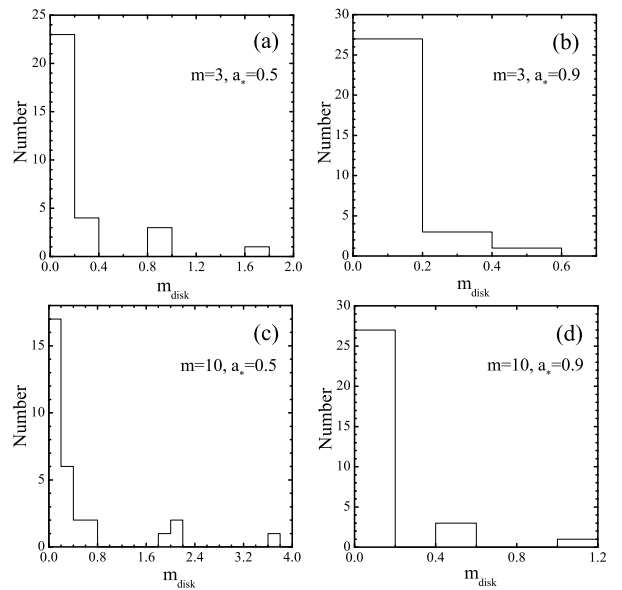}
\caption{Distributions of the disk masses $m_{\rm disk}$ for different typical BH masses and spins (adapted from Figure 2 in \citet{Liu2015b}).}
\label{611}
\end{figure}

As mentioned above, \citet{Popham1999} and \citet{Liu2007} investigated the spatial distribution of neutrino annihilation rate and found that most of the annihilation luminosity is ejected from the region $r \lesssim 20~ r_{\rm g}$. In the studies on the vertical structure of NDAF model, \citet{Liu2010a,Liu2012a,Liu2013}, found that the very large half-opening angle of the disk for a typical accretion rate can naturally constrain the neutrino annihilable ejection to produce the primary fireball of a GRB.

The observed fireball mean power outputting $\dot{E}$ from a central engine is a fraction of $L_{\nu\bar{\nu}}$, i.e.,
\beq \dot{E}=\varepsilon L_{\nu\bar{\nu}}, \eeq
where $\varepsilon$ is the conversion factor \citep[e.g.,][]{Aloy2005,Fan2011,Liu2012b}. The output power can be estimated from the observation data, i.e.,
\beq \dot{E} \approx \frac {(1+z)(E_{\rm \gamma,iso}+E_{\rm k,iso})\theta_{\rm jet}^{2}}{2 T_{90}}, \eeq
where $z$ is the redshift, $E_{\rm \gamma,iso}$ is the isotropic radiated energy in the prompt emission phase, $E_{\rm k,iso}$ is the isotropic kinetic energy of the afterglow, $T_{\rm 90}$ is the duration of GRBs, which can be roughly considered as the duration of the the central engine activity, and $\theta_{\rm jet}$ is the half opening angle of the ejecta.

Hence, for the cases of $\dot{m}_{\rm ign}<\dot{m}<\dot{m}_{\rm trap}$ in the analytic formula of \citet{Zalamea2011}, one has the mean accretion rate \citep{Fan2011,Liu2015b}
\beq \dot{m} \approx 0.12~[\frac{(1+z)(E_{\rm \gamma,iso,51}+E_{\rm k,iso,51})\theta_{\rm jet}^{2}}{\varepsilon T_{90,\rm s}}]^{4/9}~x_{\rm ms}^{2.1}~m^{2/3}, \eeq
where $E_{\rm k,iso,51}=E_{\rm k,iso}/(10^{51}~\rm erg)$, $E_{\rm \gamma,iso,51}=E_{\rm \gamma,iso}/(10^{51}~\rm erg)$, and $T_{90, \rm s}=T_{90}/(1~\rm s)$. The dimensionless mean disk mass derived from $m_{\rm disk}=\dot{m}T_{\rm 90,s}/(1+z)$, which reads \citep{Liu2015b}
\beq m_{\rm disk} \approx  0.12~[\frac{(E_{\rm \gamma,iso,51}+E_{\rm k,iso,51})\theta_{\rm jet}^{2}}{\varepsilon}]^{4/9}~\nonumber \\ \times (\frac{T_{90,\rm s}}{{1+z}})^{5/9}x_{\rm ms}^{2.1}~m^{2/3}.\eeq
Notice that $E_{\rm \gamma,iso}$ can be derived from the prompt emission data, and $E_{\rm k,iso}$ and $\theta_{\rm jet}$ can be deduced from afterglow modeling \citep[e.g.,][]{Sari1998,Sari1999,Frail2001,Panaitescu2002,Lloyd-Ronning2004,Zhang2007b,Fong2012,WangX2015}.

According to the above equations, one can estimate the disk mass by using the observational data. \citet{Liu2015b} collected 31 SGRBs with the authentic X-ray detections and known redshifts discovered by $\emph{Swift}$ and HETE-2. Figure \ref{611} displays the distributions of the disk masses $m_{\rm disk}$ with $\varepsilon=0.3$ for different typical BH masses and spins. One can see that the spin parameters are more effective than BH mass on the values of disk mass. The disk masses of most SGRBs are below $0.2-0.4~M_\odot$, and occasionally reach the limit of $0.5~M_\odot$. Even for the extreme case of $m=3$, $a_*=0.9$, there still exists one SGRB, whose disk mass is larger than $0.45~M_\odot$. These cases with massive disks may point towards an origin that invoke a massive star (Type II) rather than a compact stars merger (Type I) \citep[e.g.,][]{ZhangB2009,Lazzati2010}.

\subsubsection{Neutrino annihilation for LGRBs}

\begin{figure*}
\centering
\includegraphics[scale=1.2]{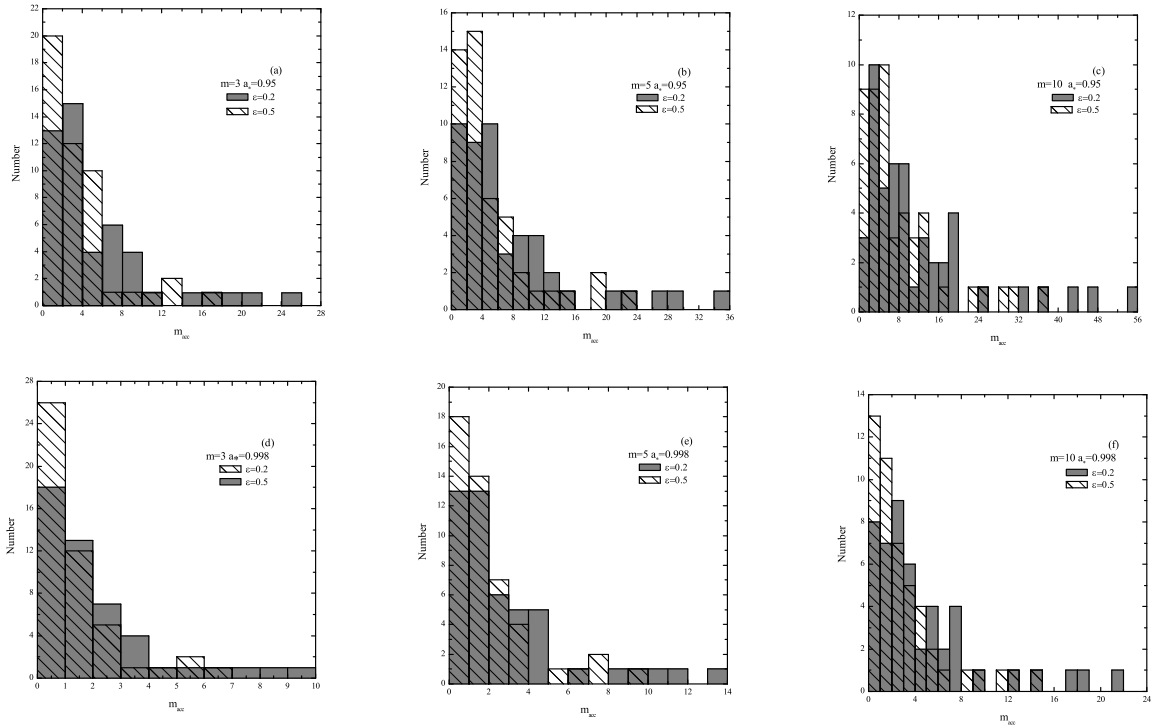}
\caption{Distributions of the accreted masses $m_{\rm acc}$ for different typical mean BH masses and spins and conversion factors (adapted from Figure 1 in \citet{Song2016}).}
\label{612}
\end{figure*}

\citet{Song2016} collected the isotropic gamma-ray radiation energy and jet kinetic energy of 48 LGRBs with known redshifts. Similar to \citet{Liu2015b}, they estimated the mean accreted masses $m_{\rm acc}$ (instead of $m_{\rm disk}$ in SGRBs) of LGRBs in the sample to investigate whether NDAFs can power LGRBs with reasonable BH parameters and conversion factor $\varepsilon$. Figure \ref{612} shows the distributions of the accreted masses for different typical mean BH masses and spins and conversion factors. Since the BHs in the centre of collapsars should be rotating very rapidly, the spin parameter is set to be larger than 0.9 \citep[e.g.,][]{MacFadyen1999,Popham1999,Zhang2003,Woosley2006}. If one considers $a_*$ = 0.998, $m=3$, and $\varepsilon$ = 0.5, all the values of the accreted masses are less than $7~M_\odot$. Obviously, if the accreted mass is larger than $7~M_\odot$, the mean BH mass must be much larger than $3~M_\odot$. As a result, most of the values of the accreted masses are less than that from the collapsar simulation, $\sim 5~M_\odot$, for extreme Kerr BHs and a high conversion factor. It suggests that the NDAFs may be suitable for most LGRBs except for some extremely energetic sources.

In the above studies, the effects of the outflow are ignored \citep{Liu2012a,Janiuk2013}. Including them may significant influence the disk mass of LGRBs even not so much for SGRBs. Furthermore, X-ray flares \citep[e.g.,][]{Burrows2005,Chincarini2007,Falcone2007} and the shallow decay phase \citep[e.g.,][]{Zhang2006a} as the common feature in GRB afterglows have not been considered in this framework, which may bring more challenges to NDAFs.

\subsubsection{Evolution of NDAFs for GRBs}

High accretion rate of NDAFs should trigger violent evolution of BH characteristics, which further leads to the evolution of the neutrino annihilation luminosity \citep{Janiuk2004,Song2015}. The evolution equations of a Kerr BH, based on the conservation of energy and angular momentum, can be expressed by \citep[e.g.,][]{Liu2012b,Song2015}
\beq \frac{d M}{dt}=\dot{M}e_{\rm ms}, \eeq
\beq \frac{d J}{dt}=\dot{M}l_{\rm ms}, \eeq
where $J=a_*GM^2/c$ is the angular momentum of the BH, $e_{\rm ms}$ and $l_{\rm ms}$ are the specific energy and angular momentum corresponding to the marginally stable orbit radius $r_{\rm ms}$ of the disk, i.e., \citep[e.g.,][]{Novikov1973,Wu2013,Hou2014b}
\beq e_{\rm ms}= \frac{1}{\sqrt{3 x_{\rm ms}}} (4- \frac{3 a_\ast}{\sqrt{x_{\rm ms}}}),\eeq
\beq l_{\rm ms}=2 \sqrt{3} \frac{G M}{c} (1-\frac{2 a_\ast}{3\sqrt{x_{\rm ms}}}),\eeq

Therefore the evolution of the BH spin is expressed by
\beq \frac{d{a}_\ast}{dt} =  2\sqrt{3} \frac{\dot{M}}{M} (1-\frac{a_\ast}{\sqrt{x_{\rm ms}}})^2. \eeq

According to the above equations, one can obtain the characteristics of the BH if the initial mass $M_0$ and spin $a_0$ of the BH are given. Similar to the above discussion, one can calculate along with the evolution of the BH mass and spin, the relations between the neutrino annihilation energy $E_{\nu\bar{\nu}}$ and $T_{90}$ by using the analytic formula of \citet{Zalamea2011} with $\dot{m}_{\rm ign}<\dot{m}<\dot{m}_{\rm trap}$, i.e.,
\beq E_{\nu\bar{\nu}}=1.59 \times 10^{54} ~\int_0^{T_{90,\rm s}/(1+z)} x_{\rm ms}^{-4.8}~m^{-3/2}~\dot{m}^{9/4}d t_{\rm s} ~\rm erg, \eeq
where $t_{\rm s}=t/(1~\rm s)$.

\begin{figure*}
\centering
\includegraphics[angle=0,scale=0.42]{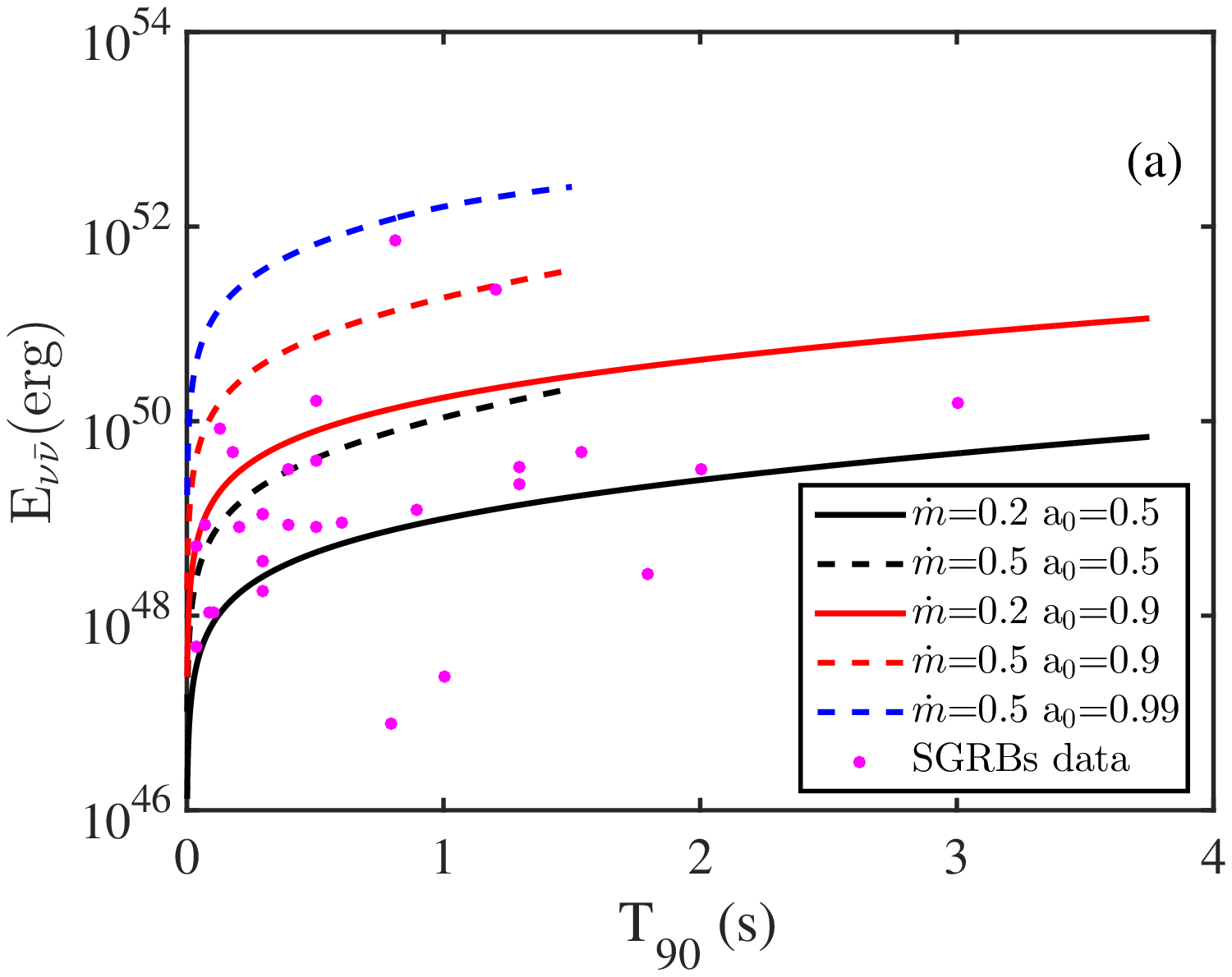}
\includegraphics[angle=0,scale=0.42]{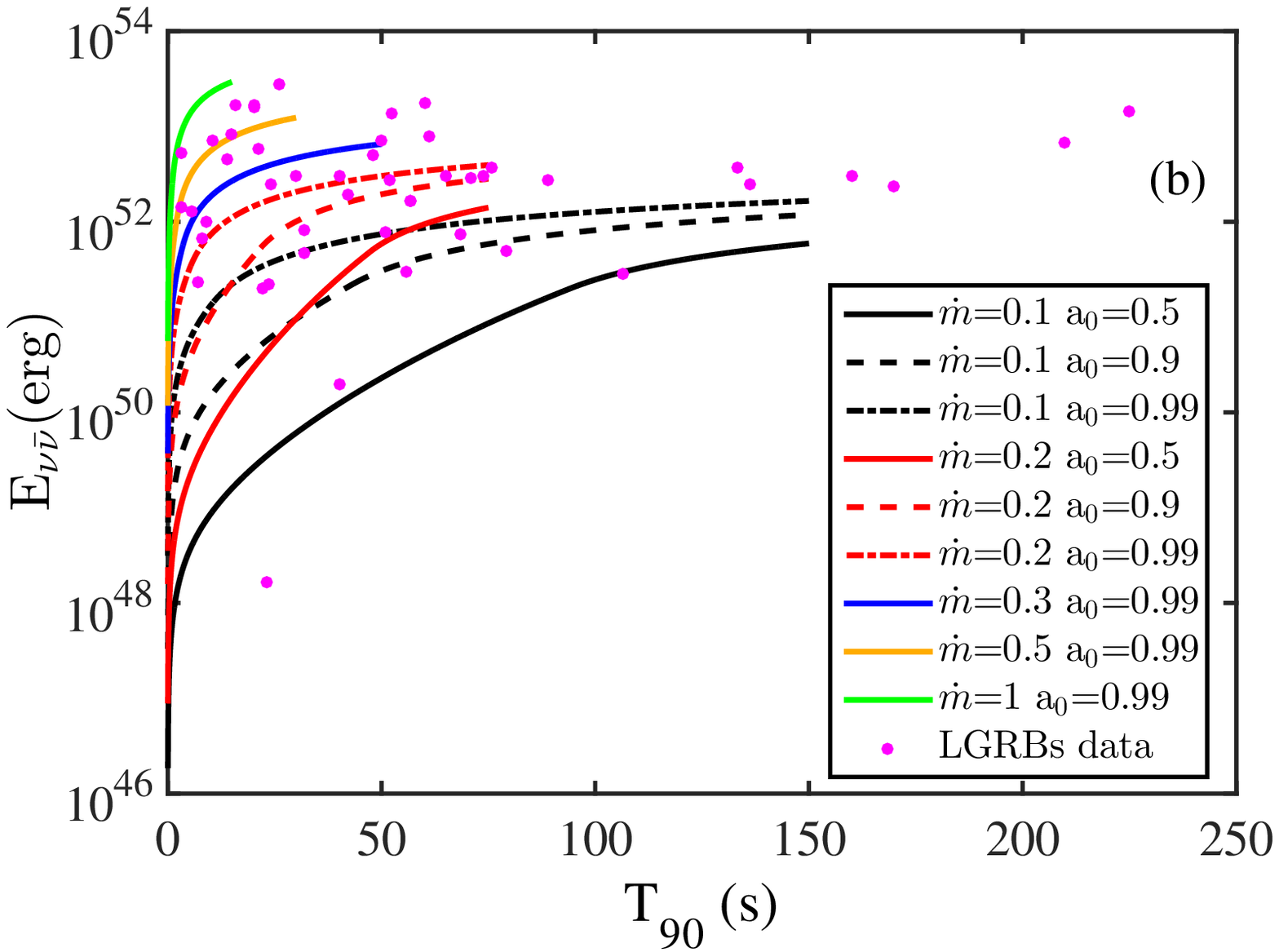}
\caption{Predictions of NDAF model compared with SGRB and LGRB observational data, corresponding to (a) and (b). Plane (a): the initial dimensionless value of BH mass $m_0=2.3$ and typical red shift $z=0.5$; Plane (b): $m_0=3$ and $z=2$ (adapted from Figures 2 and 5 in \citet{Song2015}).}
\label{613}
\end{figure*}

Figure \ref{613}(a) displays the individual neutrino annihilation energy of the SGRB data in \citet{Liu2015a} compared with the typical theoretical lines. All curves correspond to the initial value of BH mass $m_0=2.3$, conversion factor $\varepsilon=0.3$, and typical redshift $z=0.5$, and all the lines are truncated at the disk mass $m_{\rm disk}=0.5$. We notice that all the points are under our predicted lines, which means that NDAFs can interpret the profiles in SGRB sample even considering BH evolution.

Figure \ref{613}(b) shows a comparison between the predictions of NDAFs and the data of LGRBs in \citet{Nemmen2012} with parameters $m_0=3$, $\varepsilon=0.3$, and $z=2$. The duration of LGRBs shorter than 2 s or longer than 300 s are not considered \citep{Leng2014,Liu2015c}. All lines are truncated when $m_{\rm disk}=5$. In this figure, more than half of the LGRB data are below our predicted lines, which means that the neutrino annihilation processes can power these LGRBs with reasonable initial conditions. Other LGRB data still exceed the range of the lines, suggesting that alternative MHD processes may be required.

\subsubsection{Neutrino annihilation versus BZ mechanism}

Since the neutrino annihilation process cannot explain all the observed GRBs, BZ mechanism should be further studied. \citet{Kawanaka2013b} investigated the BZ jet luminosity and efficiency expected from BH-NDAF systems, and obtain the analytic descriptions of BZ luminosity and compared it with the neutrino annihilation luminosity.

\citet{Liu2015b} used the analytic formula of $p_{\rm in}$ from \citet{Xue2013},
\beq \log p_{\rm in}~({\rm{erg\ cm^{-3}}}) \approx 30.0+1.22a_*+1.00\log\dot{m}, \eeq
and combined with the BZ luminosity described by Equations (52) and (53) by ignoring the effects of the magnetic field configuration to estimate the BZ and neutrino annihilation luminosities as the functions of the disk masses and BH spin parameters, and contrasted the observational GRB jet luminosities. As the results, the BZ mechanism is more effective than the neutrino annihilation processes, especially for LGRBs. Actually, if the energy of afterglows and flares is included, the distinction between these two mechanisms is more significant. Mounting evidence suggests that at least for some GRBs outflow carries significant Poynting flux: missing or weak thermal components \citep{Zhangb2009}, strong linear polarization in gamma-rays \citep{Fan2005,Lai2015,Yonetoku2011} and early afterglows \citep{Steele2009,Mundell2013}, stringent upper limits of neutrino flux \citep{Zhang2013}, and bulk acceleration in GRBs and X-ray flares \citep{Uhm2016a,Uhm2016b}. Future GRB polarization observations by the POLAR detector may give more evidence and further distinguish these two mechanisms.

\begin{figure}
\centering
\includegraphics[angle=0,scale=0.7]{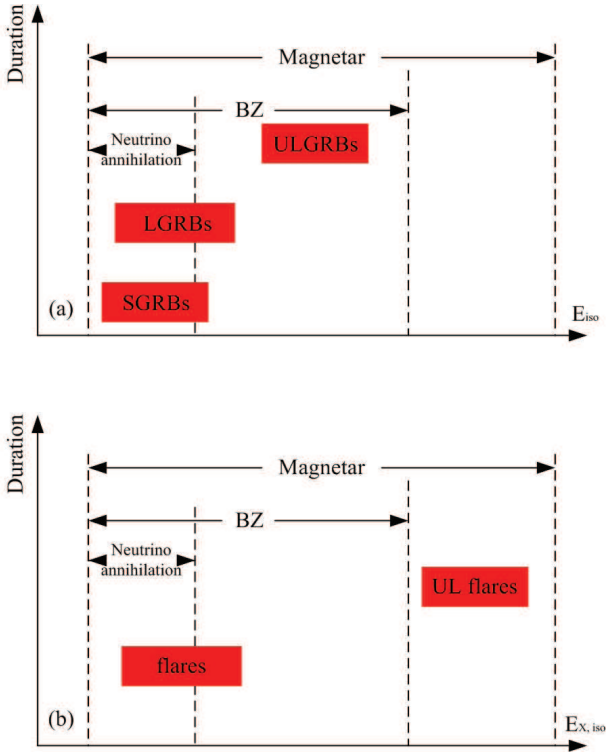}
\caption{Applications of the different mechanisms of the GRB central engine to the isotropic energy of GRBs and flares.}
\label{614}
\end{figure}

As the summary and comparison, Figure \ref{614} shows the applications of three central engine mechanisms, i.e., neutrino annihilation, BZ jet, and magnetar, to GRBs and flares. Most of SGRBs, about half of LGRBs, and non ULGRBs \citep[e.g.,][]{Virgili2013,Levan2014,Zhangbb2014} can be explained by the neutrino annihilation \citep[e.g.,][]{Nathanail2015}. If we consider that some flares originate from central engine with low accretion rates and low disk masses \citep[e.g.,][]{Bernardini2011,Margutti2011,Mu2016}, most of flares can be covered by the power of the neutrino annihilation, however, ultra-long flares (UL flares) lasting about tens of thousand seconds may come from magnetars \citep{Mu2016}.

\subsection{Variability}

Rapid variability has been observed in GRB prompt emission lightcurve. Variability may originate from internal shock model \citep[e.g.,][]{Kobayashi1997}, relativistic mini-jets \citep[e.g.,][]{Lyutikov2003,Yamazaki2004,Zhang2014}, or relativistic turbulence \citep{Narayan2009,Kumar2009,Lazar2009,Lin2013}. More fundamentally variability arises from the central engine, which may be modified as the jet interacts with the stellar envelope \citep{Morsony2010}.

Variability can naturally arise from the hyperaccreting central BH engine. The thermal and viscously instabilities in NDAFs have been widely studied \citep[e.g.,][]{Janiuk2004,Janiuk2007,Lee2005,Lei2009,Kawanaka2013a,Kimura2015,Xie2016} to explain the variability. \citet{Carballido2011} characterized the time variability of energy release at small scales in NDAFs through shearing box MHD simulations. \citet{Masada2007} investigated the magnetorotational instability in the hyperaccretion disk. Once the dead zone gains a large amount of mass and becomes gravitationally unstable, the episodically intense mass accretion can cause short-term variabilities. \citet{Kawanaka2012} found that convective motion in the vertical direction can also trigger sporadic mass accretion, which causes variability in GRBs. Differently, \citet{Lin2016} showed that the propagating fluctuations mechanism \citep[e.g.,][]{Lyubarskii1997,King2004,Lin2012} that drives variabilities in BH binaries and AGNs may be responsible for the observed variability in prompt emission.

\subsection{Jet precession}

Another area of research in NDAF-GRB connection is the possibility of jet precession \citep[e.g.,][]{Blackman1996,PortegiesZwart1999,Reynoso2006,Lei2007,Liu2010b,Stone2013}. \cite{Blackman1996} first investigated a relativistic precessing blob-emitting NS jet in a binary pulsar. They considered Newtonian tidal torque and gravitomagnetic interaction between pulsar binary to account for precession and nutation. \cite{PortegiesZwart1999} proposed that the BH forces the accretion disk and jet to precess due to the Newtonian tidal torque, and fitted the observational data. Since the gravitomagnetic interaction between the BH and disk is, however, much stronger than the tidal force \citep{Thorne1986}, \cite{Reynoso2006} suggested that the gravitomagnetic interaction can precess jet from the central engine of GRBs. \cite{Lei2007} considered the effects of the half-opening angle of a precessing jet on shaping GRB lightcurves. In their model, since the whole disk precesses around the BH (i.e., the angular momentum of the disk is less than that of the BH), the required size of the disk needs to be small enough.

\begin{figure}
\centering
\includegraphics[angle=0,scale=0.40]{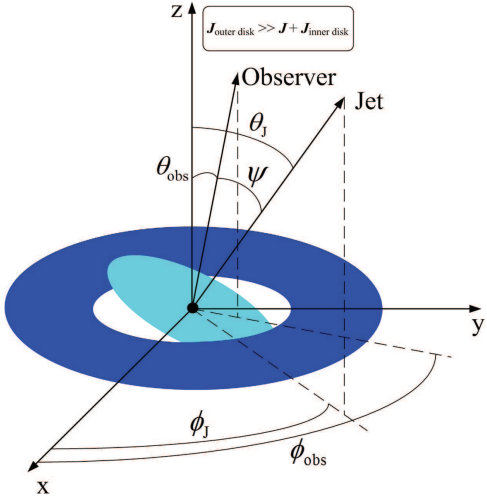}
\caption{Schematic picture of a precessing system (adapted from Figure 1 in \citet{Liu2010b}).}
\label{631}
\end{figure}

Naturally, the angular momentum of the disk is much larger than that of the BH in BH-NDAF systems, thus a new jet precession model should be built \citep{Liu2010b,Sun2012}. The basic picture (see Figure \ref{631}) is that a Kerr BH surrounded by a tilted NDAF whose initial orbital axis is misaligned with the BH spinning axis. In this framework, the outer disk whose angular momentum is sufficiently larger than that of the BH can maintain its orientation and force the BH to precess. Meanwhile, the inner disk whose angular momentum is significantly smaller than that of the BH, should be aligned with the BH spin axis \citep{Bardeen1975}.

The angular momentum per ring at radius $r$ with width $dr$ is $dJ_{\rm disk}=2\pi r^2\Sigma v_{\varphi}dr$, so the typical angular momentum of the disk is \citep[e.g.,][]{Sarazin1980}
\beq J_{\rm disk} = \frac{dJ_{\rm disk}}{d(\ln r)} = 2\pi r^3\Sigma v_{\varphi}. \eeq
Due to the Lense-Thirring effect \citep{Lense1918}, there exists a critical radius $r_{\rm{p}}$ where the typical angular momentum $J_{\rm disk}|_{r=r_{\rm p}}$ is equal to the BH angular momentum $J$, i.e.,
\beq J_{\rm disk}|_{r=r_{\rm p}}= J = \frac{a_*GM^2}{c}, \eeq
The outer disk ($r>r_{\rm{p}}$) will maintain its orientation and therefore force the BH and the inner disk ($r<r_{\rm{p}}$) to be a whole precessing system.
The precession rate is expressed as \citep[e.g.,][]{Sarazin1980,Lu1990}
\beq \Omega_{\rm p}=\frac{2GJ}{c^2r_{\rm p}^3}.\eeq
According to the mass conservation equation, the precession period can be derived as
\beq P = \frac{2\pi}{\Omega_{\rm p}} = \pi M (\frac{a_\ast}{G})^{\frac{1}{2}}(-\frac{cv_{r}|_{r=r_{\rm p}}} {\dot{M}v_{\varphi}|_{r=r_{\rm p}}})^{\frac{3}{2}}.\eeq

\begin{figure*}
\centering
\includegraphics[angle=0,scale=0.25]{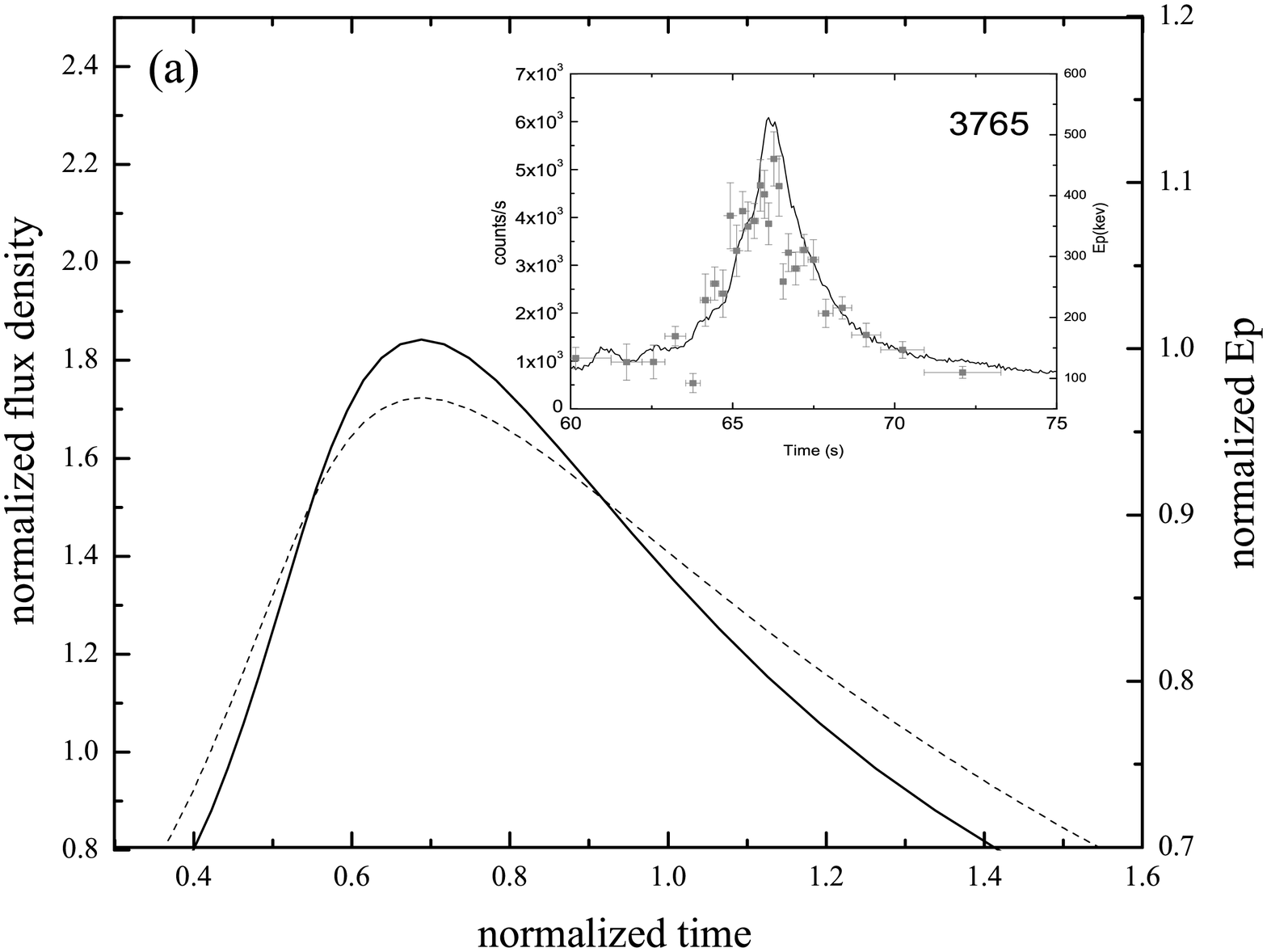}
\includegraphics[angle=0,scale=0.27]{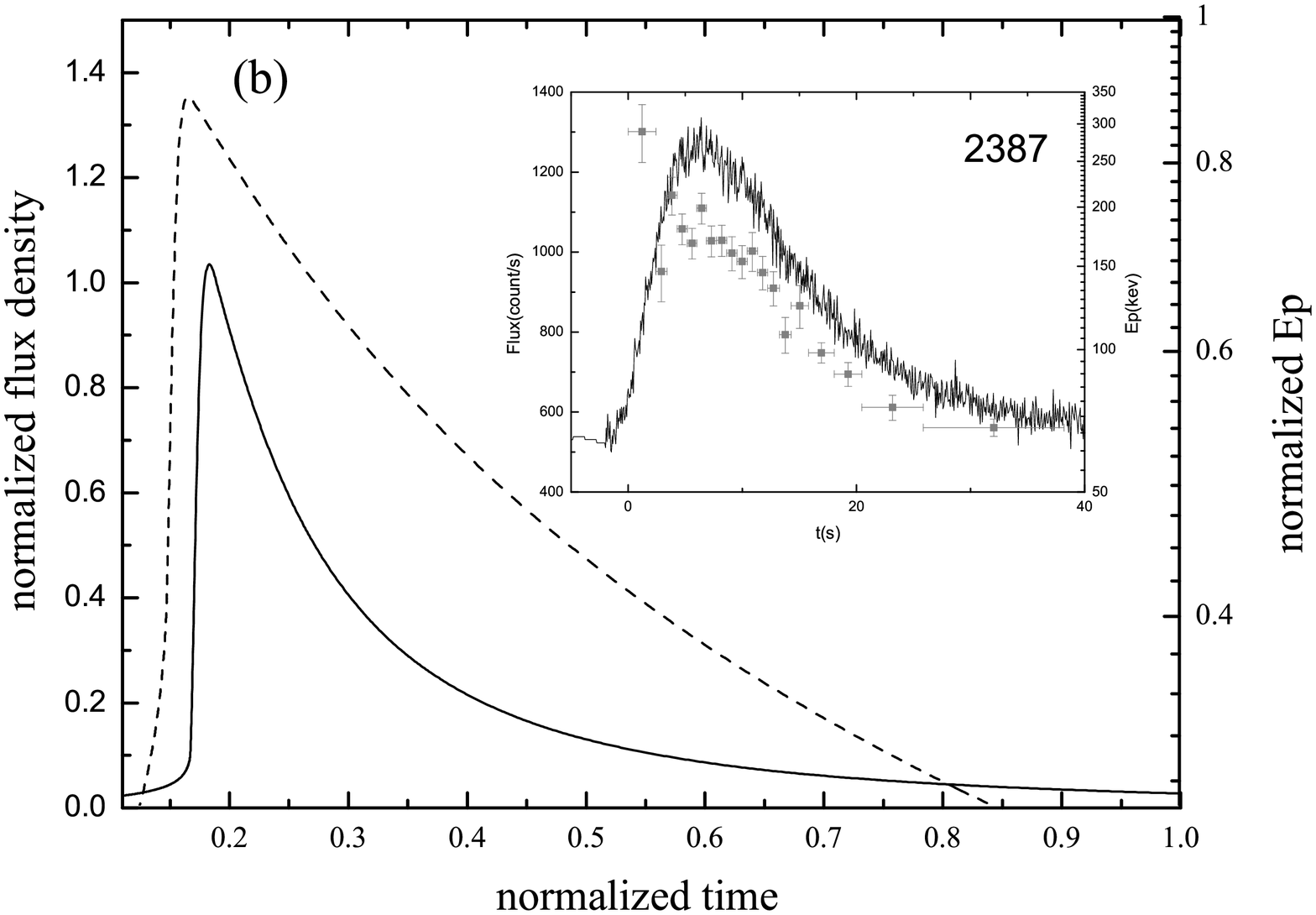}
\caption{Predicted flux $F$ (the solid line) and $E_p$ (the dashed line) with jet precession model for a symmetric pulse (panel a) and a FRED pulse (panel b) with comparisons to the observations (adapted from Figure 4 in \citet{Liu2010b}).}
\label{632}
\end{figure*}

Jet precession model can also interpret some lightcurves that show $E_p$ tracking behavior. Usually a GRB pulse shows a fast-rise-exponential-decay shape with the $\nu F_\nu$ spectrum $E_p$ evolves from hard to soft. This can be naturally explained by synchrotron radiation of an expanding shell \citep{Uhm2014,Uhm2016b}. However, some well-separated GRB pulses show a symmetric structure, and their $E_p$ traces the lightcurve behavior \citep[e.g.,][]{Liang1996,Liang2004b,Lu2010,Lu2012,Liang2015}. Both the temporal and spectral properties of these symmetric pulse are difficult to be explained with internal shocks. Figure \ref{632} shows the flux $F$ and $E_p$ in jet precession model as compared with the observations \citep{Liu2010b}.

Using the above equations and the analytic formulae of the structure of NDAFs derived from \citet{Popham1999}, the analytic expression of the precession period $P$ can be estimated by
\beq P = 2793~ a_\ast^{17/13} m^{7/13} \dot{m}^{-30/13} \alpha^{36/13} ~\rm s,\eeq
Assuming $\alpha$ as a constant, the time derivative of $P$ is expressed as
\beq \frac{1}{P} \frac{dP}{dt} = \frac{17}{13} \frac{1}{a_\ast} \frac{d a_\ast}{dt} + \frac{7}{13} \frac{\dot{m}}{m} e_{\rm ms} - \frac{30}{13} \frac{1}{\dot{m}} \frac{d \dot{m}}{dt}. \eeq

These equations suggest a quasi-periodic variability and its evolution of GRBs. \citet{Hou2014a} exhibited that a $86^{+5.9}_{-9.4}~\rm s$ periodic oscillation may exist in the data from about 5300 s to about 6100 s in the bump of GRB 121027A using the stepwise filter correlation method \citep{Gao2012} and the Lomb-Scargle method \citep{Scargle1982}, which can be interpreted by jet precession model \citep{Liu2010b}. \citet{Hou2014b} discovered that there is a remarkable time evolution in the data of the flares in GRB 130925A. It can be also explained by Equation (104) with reasonable initial BH mass and spin.

\subsection{Nucleosynthesis}

\begin{figure}
\centering
\includegraphics[angle=0,scale=0.45]{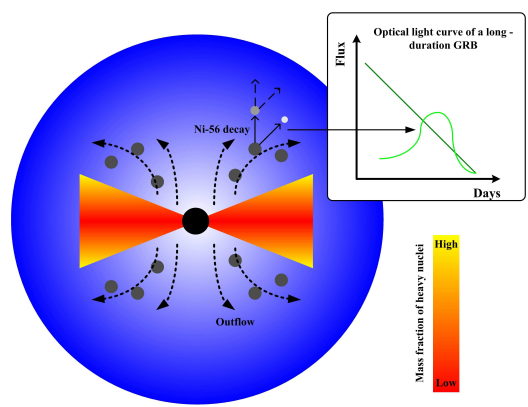}
\caption{Schematic picture of a Nickel factory in the collapsar (adapted from Figure 3 in \citet{Liu2014a}).}
\label{64}
\end{figure}

The central engine of GRBs is an ideal location to supply an extremely hot and dense environment to produce heavy elements. The radial and vertical components of NDAFs have been studied in detail in \citet{Liu2013} and \citet{Xue2013}, which can be connected to the origin of strong Fe K$\alpha$ emission lines and the SN bump observed in long GRB afterglow lightcurves (see Figure \ref{64}).

The SN bump may be driven by the decay of $^{56}\rm Ni$ \citep[e.g.,][]{Galama1999,Woosley2006}, but the product of massive $^{56}\rm Ni$ in SNe accompanied GRBs remains unsolved. The possibilities including $^{56}\rm Ni$ being originated from the central engine transported by the outflow \citep[e.g.,][]{MacFadyen1999,MacFadyen2003,Surman2011} or due to explosive burning \citep[e.g.,][]{Maeda2003,Maeda2009}. \citet{Reeves2002} reported on an \emph{XMM-Newton} observation of the X-ray afterglow of GRB 011211, which spectrum reveals evidence for emission lines of Magnesium, Silicon, Sulphur, Argon, Calcium, and possibly Nickel, arising in enriched material with an outflow velocity of order $\rm 0.1c$. However, there are no reported metal lines in GRBs from \emph{Swift} XRT observatory.

The production processes of the heavy nuclei from the central engine of GRBs are widely studied recently. As shown in Figures \ref{45} (c) and (d), for the low accretion rate of NDAFs exactly corresponding to LGRBs that are associated with SNe, $^{56}\rm Ni$ dominates near the disk surface \citep{Liu2013}. The decay of $^{56}\rm Ni$ in the outflows can produce the optical lightcurve SN bump. \citet{Surman2011} focused on nucleosynthesis, particularly for the progenitor of $^{56}\rm Ni$ in the hot outflows from GRB accretion disks. \citet{Metzger2011} suggested that the composition of ultrahigh energy cosmic rays becomes dominated by heavy nuclei at high energies forming GRB jets or outflows \citep[also see e.g.,][]{Sigl1995,Horiuchi2012,Hu2015}.

\subsection{Extended emission}

Significant extended emission (EE) after prompt emission of SGRBs has been observed up to $\sim$ 100 s by {\em CGRO}/BATSE \citep{Norris1995,Lazzati2001,Connaughton2002}, BeppoSAX \citep{Montanari2005}, and \emph{Swift}/BAT \citep{Norris2010}. The most prominent case is GRB 060614 lasting $\sim 110$ s, whose lightcurve is composed of some initial hard spikes and a long soft gamma-ray tail \citep{Gehrels2006}. There is no accompanied SNe detected for this nearby GRB \citep{DellaValle2006,Fynbo2006,Gal-Yam2006}, disfavoring the collapse of a massive star as the progenitor \citep{Zhang2007b}.

A magnetar model is proposed to explain the origin of SGRBs with EE \citep[e.g.,][]{Bucciantini2012,Gompertz2013,Gompertz2014}. \citet{Barkov2011} suggested that BZ jet and neutrino annihilation may coexist in BH-NDAF system, and GRBs with and without EE may depend on observer's viewing angle from the jet axis.

\begin{figure}
\centering
\includegraphics[angle=0,scale=0.42]{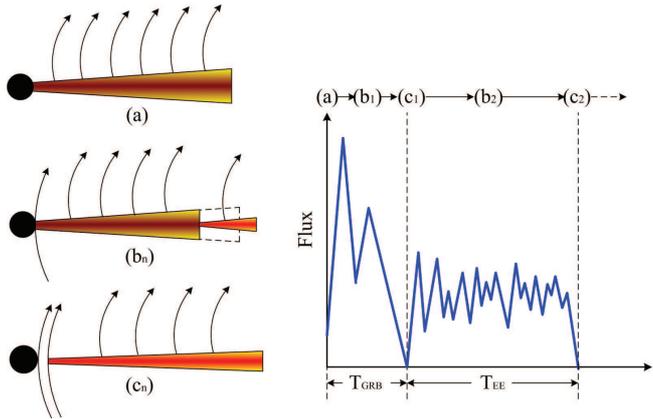}
\caption{Schematic illustration of our model and corresponding cartoon lightcurve: (a) the initial state of the central engine---the filled circle stands for the BH and the fuscous trapeziform region for the disk with magnetic field (curves); $(b_{\rm n})$ angular momentum transfer process and outward flow (light gray region); $(c_{\rm n})$ magnetic barrier in vicinity of the BH, where $n$ is for the $nth$ emission episode, $n>2$ may have no observable effects (adapted from Figure 1 in \citet{Liu2012b}).}
\label{65}
\end{figure}

In the framework of the BH-NS merger models, \citet{Liu2012b} studied the roles of radial angular momentum transfer in the disk and time evolution of the BH mass and spin as shown in Equations (92) and (93), then calculated the magnetic barrier around the BH, as Equation (53), to estimated the timescale to separate prompt emission and EE. The basic picture is shown in Figure \ref{65}, i.e., the radial angular momentum transfer may significantly prolong the lifetime of the accretion process and multiple episodes may be switched by the magnetic barrier. Our numerical calculations suggest that the model can fit most data of SGRBs with EE with a reasonable disk mass and BH characteristics.

\subsection{X-ray flares}

X-ray flares \citep[e.g.,][]{Burrows2005,Chincarini2007,Falcone2007,Margutti2011} are common in GRBs and occur at times well after the initial prompt emission, with a time lag of the order of hundreds or thousands of seconds, which might be related to the activities of the central engine \citep[e.g.,][]{Fan2005,Lazzati2007,Maxham2009,Luo2013,Yi2016}. Several mechanisms have been proposed to generally explain the episodic X-ray flares, including gravitational instability in the hyperaccretion disk \citep{Perna2006,Liu2014b}, fragmentation of a rapidly rotating core \citep{King2005}, a magnetic switch of the accretion flow \citep{Proga2006,Liu2012b,Cao2014}, differential rotation in a post-merger millisecond pulsar \citep{Dai2006}, transition from a thin to a thick disk \citep{Lazzati2008}, He-synthesis-driven wind \citep{Lee2009}, instability in the jet \citep{Lazzati2011}, outflow caused by the maximal and minimal possible mass accretion rates at each radius of NDAFs \citep{Liu2008}, episodic jets produced by the magnetohydrodynamic mechanism from the disk \citep{Yuan2012}, and MC-NDAFs \citep{Luo2013}.

\citet{Mu2016} investigated UL flares with duration $\gtrsim 10^4$ s, and argued that the corresponding central engine may not be associated with BHs, but a fast rotating NS with strong dipolar magnetic fields as shown in Figure \ref{614}.

\subsection{Constraint on the mass and spin of central engine BHs in GRBs}

BHs are mysterious and fascinating compact objects, which are generally related to multiband electromagnetic radiation, gravitational waves (GWs), neutrino emission, and cosmic rays. Two essential properties of BHs, i.e., mass and spin, are not easy to measure. Some dynamical or statistical methods have been introduced to constrain these parameters of supermassive BHs \citep[e.g.,][]{Natarajan1998,Brenneman2006,Tchekhovskoy2010,LeiZhang2011,Kormendy2013,Wang2013} and stellar-mass BHs \citep[e.g,][]{Bahcall1978,Zhang1997,McClintock2014}.

In NDAFs, neutrinos can tap the thermal energy gathered by the viscous dissipation and liberate a tremendous amount of binding energy, and neutrino-antineutrino annihilation above the disk would launch a hot fireball. A GRB powered by this mechanism is therefore thermally dominated. The launch site of the fireball should be above the typical neutrino annihilation radius, which can be constrained by the observed thermally spectral component \citep{Peer2007}.

\begin{figure}
\centering
\includegraphics[angle=0,scale=0.33]{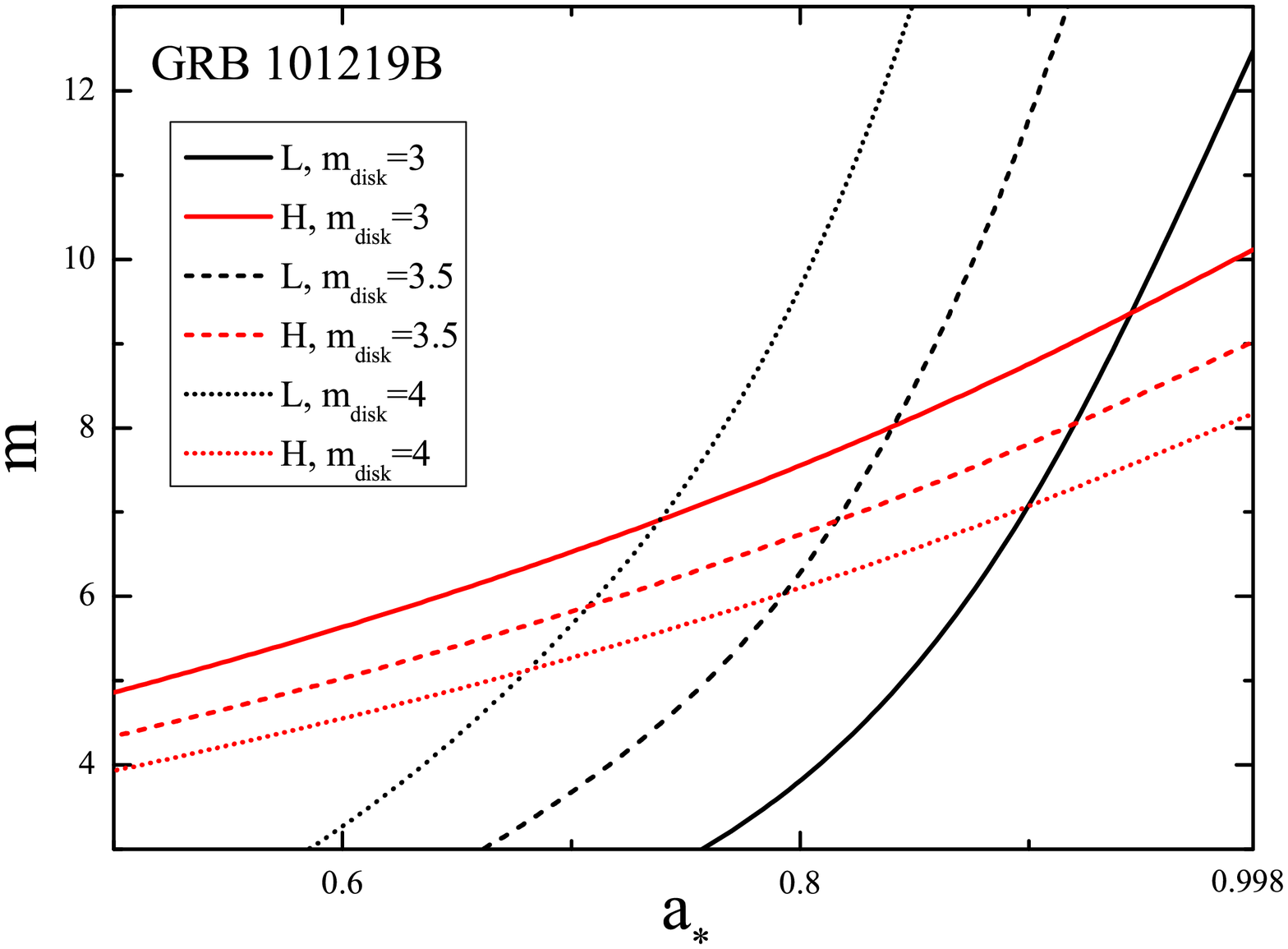}
\caption{The constrained mean BH mass and spin of GRB 101219B for different disk masses. The black and red lines correspond to the constraints from annihilation luminosity and height, respectively. The solid, dashed, and dotted lines correspond to $m_{\rm disk}$=3, 3.5, and 4, respectively (adapted from Figure 2 in \citet{Liu2016b}).}
\label{67}
\end{figure}

From the analytic solutions in \citet{Xue2013}, we obtain that the dimensionless annihilation height $h=H_{\rm ann}/r_{\rm g}$ is
\beq \log h  \approx  2.15 - 0.30 a_* - 0.53 \log m +  0.35 \log \dot{m},\eeq
Combing with Equation (49), using the GRB luminosity and fireball launch radius, one can constrain the mass and spin of a GRB central engine BH. As shown in Figure \ref{67}, \citet{Liu2016b} estimate the following constraints on the central engine BH of GRB 101219B: mass $M_{\rm BH} \sim 5-9~M_\odot$, spin parameter $a_* \gtrsim 0.6$, and disk mass $3~M_\odot \lesssim M_{\rm disk} \lesssim 4~M_\odot$. The results also suggest that the NDAF model is a competitive candidate for the central engine of GRBs with a strong thermal component.

\subsection{Kilonovae}

BH-BH, BH-NS and NS-NS mergers are the strong GWs transient sources, which may be associated with short GRBs \citep[e.g.,][]{Paczynski1986, Eichler1989, Narayan1992}, short-GRB-less X-ray transients \citep{Zhangb2013,Sun2017}, kilonovae/mergernovae \citep[e.g.,][]{Li1998,Metzger2010,Yu2013}, even fast radio bursts \citep[e.g.,][]{Liu2016a,Zhang2016,Wang2016}.

\citep{Li1998} first predicts a type of optical transient lasting a few days, which is powered by radioactive decay of the neutron-rich ejecta from an NS-NS or BH-NS merger system. The kilonova signature has been investigated in detail recently \citep[e.g.,][]{Metzger2010,Barnes2013,Metzger2014,Kasen2015,Metzger2017}, and the enhanced merger nova signature due to energy injection of a post-merger magnetar engine has been studied in detail \citep[e.g.,][]{Yu2013,Yu2015, MetzgerP2014,LiS2016,Gao2017}. Observationally, several kilonova/mergernova candidates have been discovered in association with some short GRBs, including typical kilonovae in GRB 130603B \citep{Tanvir2013,Berger2013}, GRB 060614 \citep{Jin2015,Yang2015}, and  GRB 050704 \citep{Jin2016}, and a few magnetar-powered merger nova \citep{Gao2016,Gao2017}. The applicability of NDAFs for some of these GRBs have been studied \citep{Shen2017}.

\begin{figure}
\centering
\includegraphics[angle=0,scale=0.5]{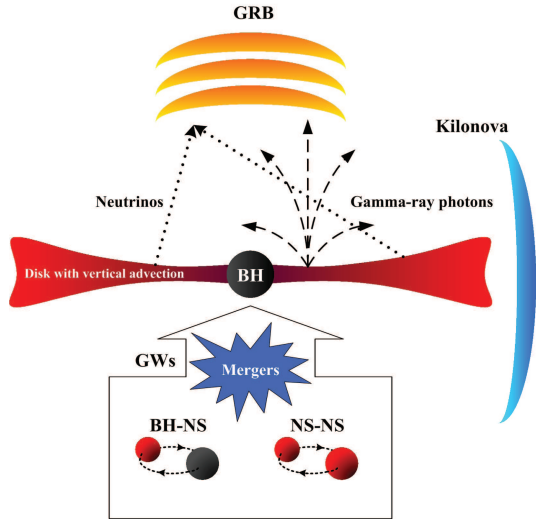}
\caption{Illustration of the association of short GRBs, kilonovae, and GW events. (adapted from Figure 7 in \citet{Yi2017}).}
\label{68}
\end{figure}

For NDAF models, \citet{Yi2017} studied the effects of vertical advection \citep{Jiang2014} on the structure and luminosity of NDAFs. Due to the strong cooling through the vertical advection, the neutrino luminosity from the disk is lower than that of NDAFs without vertical advection. Interestingly, the gamma-ray emission from the disk surface can be extremely super-Eddington, which may have significant contribution to the kilonovae as shown in Figure \ref{68}. Following radioactive decay from the outflow of the disk, this process may power a transient even in the direction off the jet. \citet{Song2017} proposed that a BH hyperaccretion disk with a strong outflow (about 99$\%$ accretion matter) via radioactive decay may trigger a kilonova-like transient as powerful as the one powered by a magnetar.

\subsection{Progenitor stars}

For the same settings of the accretion rate and BH characteristics, the BH hyperaccretion inflow-outflow model including BZ mechanism is more powerful than the NDAF model, which can be used to constrain the progenitor stars of GRBs \citep{Liu2017b,Song2017}.

Considering outflows from an NDAF, \citet{Liu2017b} showed tight constraints on the properties of progenitor stars from long GRB observations. Specifically, only the solar-metallicity, massive stars or parts of zero-metallicity stars can be as the progenitors of LGRBs lasting from several seconds to tens of seconds in rest frame. The fraction of bursts in the LGRB population is more than 40$\%$, which cannot be accounted for by these rare required progenitors. It implies that the activity timescale of central engine may be much longer than the duration of prompt gamma-ray emission, as indicated by the extended X-ray activities observed from the Swift data \citep{Zhangbb2014,Lv2014a}. Alternatively, LGRBs may be powered by the magnetars rather than NDAF systems.

\section{Multi-messenger signals from NDAFs}

There may be two observational messenger signals to probe the invisible central engine of GRBs and test the existence of NDAFs, i.e., MeV neutrinos and GWs.

\subsection{Detectable MeV neutrinos from NDAFs}

Whereas the leading candidates of MeV neutrino sources are SNe similar to SN 1987A. NDAFs around rotating BHs or NSs can be another source of cosmic MeV neutrinos. Assuming a distance of 10 kpc, the detectability of a nominal NDAF with a large accretion rate by Super-Kamiokande was discussed \citep[e.g.,][]{Nagataki2002,Caballero2012,Caballero2015}.

Based on the solutions of \citet{Xue2013}, \citet{Liu2016c} derived the fitting formulae for the mean cooling rate due to electron neutrino and antineutrino losses, $Q_{\nu_{\rm e}}$ and $Q_{\bar{\nu}_{\rm e}}$, and the temperature of the disk $T$, as the functions of the mean BH spin parameter, mean accretion rate, and radius ($m=3$ adopted), i.e.,
\beq \log Q_{\nu_{\rm e}}~({\rm erg~cm^{-2}~s^{-1}}) = 39.78 + 0.15 a_*\nonumber \\+ 1.19 \log \dot{m}- 3.46 \log r, \eeq
\beq \log Q_{\bar{\nu}_{\rm e}} ~({\rm erg~cm^{-2}~s^{-1}})= 40.02 + 0.29 a_* \nonumber \\+1.35 \log \dot{m}- 3.59 \log r, \eeq
\beq \log T ~({\rm K}) = 11.09 + 0.10 a_*+ 0.20 \log \dot{m} \nonumber \\- 0.59 \log r. \eeq

\begin{figure}
\centering
\includegraphics[angle=0,scale=0.5]{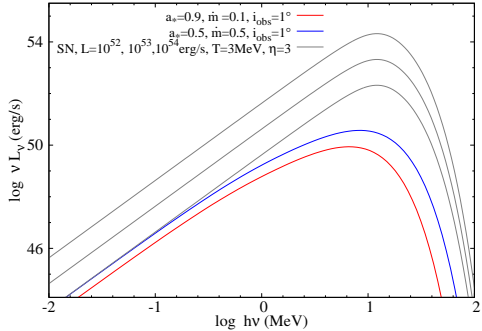}
\caption{Electron neutrino spectra of typical SGRBs (blue line), LGRBs (red line), and O-Ne-Mg core-collapse SNe (gray lines) (adapted from Figure 2 in \citet{Liu2016c}).}
\label{71}
\end{figure}

According to the above equations, \citet{Liu2016c} calculated the electron neutrino and antineutrino spectra of NDAFs by fully taking into account the general relativistic effects by using the null geodesic equation \citep{Carter1968}, and studied the effects of viewing angle, BH spin, and accretion rate on the spectra. Figure \ref{71} shows that even though a typical NDAF has a neutrino luminosity lower than that of a typical SN, it can reach $10^{50}-10^{51}~{\rm erg~s^{-1}}$ peaking at $\sim 10-20$ MeV, making them potentially detectable with the upcoming sensitive MeV neutrino detectors if they are close enough to Earth. Based on the detailed star formation rate and metallicity in Local Group, we estimate a detection rate up to $\sim$ (0.10-0.25) per century for GRB-related NDAFs by the Hyper-Kamiokande (Hyper-K) detector \citep{Abe2011,Mirizzi2016}, if one neglects neutrino oscillation and demands at least 3 neutrinos being detected. If one assumes that all Type Ib/c SNe have NDAFs, the Hyper-K detection rate would be $\sim$ (1-3) per century. By considering neutrino oscillations, the detection rate may decrease by a factor of 2-3. By the way, it can be expected that neutrinos from NDAFs may slightly effect on the universal neutrino background.

\subsection{GWs from NDAFs}

One of the methods to infer the existence of NDAFs may be through detections of the GW signals from NDAFs due to precession \citep{Romero2010,Sun2012} or anisotropic neutrino emission \citep{Epstein1978,Sago2004,Hiramatsu2005,Suwa2009,Kotake2012} in the disk. \citet{Liu2017a} compared the detectabilities of the GWs from NDAFs, BH hyperaccretion disks with BZ jets, and magnetars.

\begin{figure}
\includegraphics[angle=0,scale=0.2]{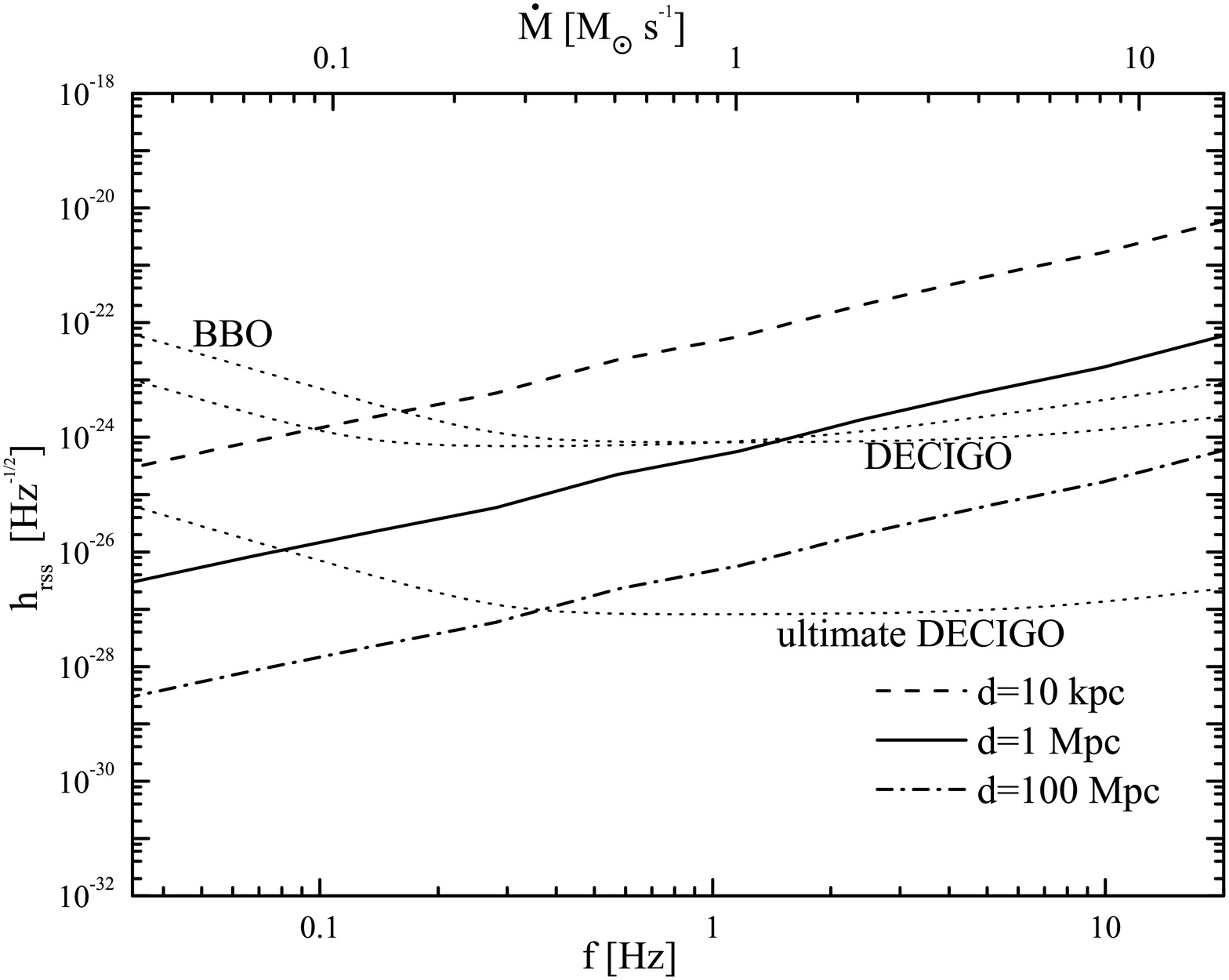}
\centering
\caption{The GW root-sum-square amplitude as a function of the frequency (or the accretion rate), for which $M=6 M_{\odot}$. The dashed, solid, and dash-dotted lines correspond to $d=10\,\rm{kpc}$, $1\,\rm{Mpc}$, and $100\,\rm{Mpc}$, respectively. The dotted lines represent the detectability of GW detectors (adapted from Figure 4 in \citet{Sun2012}).}
\label{72}
\end{figure}

The jet precession mechanism introduced by \citet{Liu2012b} leads to a non-axisymmetric mass distribution, which can cause gravitational radiation. \citet{Sun2012} estimated that the typical quadrupole power of NDAFs is about $10^{44}~\rm erg ~s^{-1}$ at 10 Hz, which is much less than the typical GRB luminosity. According to Figure \ref{72}, the GWs from NDAFs caused by jet precession may be detectable by the future GW detectors if the sources are in the Local Group. GW power and frequency from NDAFs may be different from these from compact objects mergers or collapsars \citep[e.g.,][]{Nakar2011,Ott2011,Gao2013,Messenger2014} and relativistic jet \citep[e.g.,][]{Akiba2013,Birnholtz2013}, but detailed studies are needed to model the wave forms of these events.

Two confirmed GW events and another candidate have been discovered by LIGO \citep{Abbott2016a,Abbott2016b}. The \emph{Fermi}/GBM recorded a possible gamma-ray transient 0.4 s after GW 150914 \citep{Connaughton2016}. Several models have been proposed to explain the possible electromagnetic counterpart of GW 150914 \citep[e.g.,][]{Li2016,Zhang2016,Loeb2016,Liu2016a,Perna2016,Woosley2016,Zhang2016a,Janiuk2017}. If such an association is true, the combined GW and electromagnetic signals can be used to test Einstein's Equivalence Principle \citep{Wu2016}. Detecting NDAF-related GWs and the electromagnetic counterparts can also serve the purpose.

\section{Summary and discussion}

As a theoretical concept, NDAFs are studies within the context of accretion systems with extremely high accretion rates such as cosmological GRBs. NDAFs are the naturally extension of slim disks for extreme accretion rates. As the main cooling mechanism, neutrino cooling process leads to a copious neutrinos that may power GRBs via neutrino annihilation. Due to the extreme density, temperature involved in such a system with extreme accretion rate, rich physics (gravitational, thermal dynamical, nuclear, and particle physics) is involved. We have reviewed the recent progress in studying the physical processes and properties of NDAFs, including the steady radial and vertical structure of NDAFs and the implications for calculating neutrino luminosity and annihilation luminosity, jet power due to neutrino-antineutrino annihilation and BZ mechanism and their dependences on parameters such as BH mass, spin, and accretion rate, time evolution of NDAFs, and effect of magnetic fields. The applications of NDAF theories to the GRB phenomenology such as lightcurve variability, precession, extended emission, X-ray flares, etc., are reviewed, and possible probes NDAFs using multi-messenger signals such as MeV neutrinos and gravitational waves are discussed.

Since GRB observations are directly related to the relativistic jets rather than the central engine, directly inferring the existence of NDAFs as well as their properties are not easy. The uncertainties lie in the unknown composition, energy dissipation mechanism, and particle acceleration and radiation mechanism of the jet \citep[][for recent reviews]{Zhang2011,Kumar2015}. Furthermore, the identity of the central engine, be it a hyperaccreting BH or a spinning down millisecond magnetars, is still uncertain. It is possible and even likely that both types of engines are operating in different GRBs. In any case, NDAFs must exist in BH systems, and may also exist and operate in magnetar systems as well. For BH engines, the BZ-dominated systems tend to produce Poynting-flux-dominated outflows whereas the neutrino annihilation dominated systems tend to produce hot fireballs. GRB spectral observations seem to suggest that both types of jet composition may exist in different GRBs \citep[e.g.,][]{Abdo2009a,Abdo2009b,Ryde2010,Zhang2011,Guiriec2011,Guiriec2013,Axelsson2012,Zhangb2009,Gao2015}. For thermally-dominated fireballs, the connection between the central engine and observational properties is closer, and interesting constraints can be placed to the properties of the BH and NDAFs \citep{Liu2016b}. Future more sensitive GRB spectral detectors and hopefully multi-messenger observations of some nearby NDAF sources may shed light into more detailed information about NDAFs and may even directly infer its existence.

\section*{Acknowledgements}

We thank Zi-Gao Dai, En-Wei Liang, Yi-Zhong Fan, Xue-Feng Wu, Da-Bin Lin and Li Xue for beneficial suggestions and discussion. This work was supported by the National Basic Research Program of China (973 Program) under grant 2014CB845800, the National Natural Science Foundation of China under grants 11473022, 11333004, and 11573023.

\clearpage

\end{document}